%% file: ms.tex
\documentclass[preprint] {aastex}
\input{lmod.tex} 

\begin{document}

\tightenlines 

\shorttitle{SEDs of $z>3$ quasars}
\shortauthors{}


\title{A search for signatures of quasar evolution: Comparison of the
shapes of the rest-frame optical/UV continua of quasars at 
$z>3$ and $z\sim0.1$\altaffilmark{1}}
\altaffiltext{1}{Work reported here was based on observations obtained with the Multiple 
Mirror Telescope, a joint facility of the Smithsonian Institution and the University 
of Arizona and at CTIO and KPNO which are operated by the Association of Universities
for Research in Astronomy Inc., under a cooperative agreement with the National 
Science Foundation as part of the National Optical Astronomy Observatories.} 
\author{Olga Kuhn\altaffilmark{2}, Martin Elvis\altaffilmark{3}, Jill Bechtold\altaffilmark{4}, and Richard Elston\altaffilmark{5}}
\altaffiltext{2}{Joint Astronomy Centre, 660 N. A`ohoku Place, University Park, Hilo, HI 96720; e-mail: o.kuhn@jach.hawaii.edu}
\altaffiltext{3}{Harvard-Smithsonian Center for Astrophysics, 60 Garden St., Cambridge, MA 02138; elvis@cfa.harvard.edu}
\altaffiltext{4}{Steward Observatory, University of Arizona, Tucson, AZ 85721; jill@as.arizona.edu}
\altaffiltext{5}{Department of Astronomy, University of Florida, Gainesville, FL 32611; elston@astro.ufl.edu}
\received{}
\accepted{}
\journalid{}
\articleid{}

\begin{abstract}

For 15 bright (V$<17.5$), high redshift ($z>3$) quasars, we have obtained infrared spectra and 
photometry, and optical spectrophotometry and photometry, which we use to construct their
spectral energy distributions (SEDs) from $\lambda_{rest}\sim1285-5100$\AA. 
High resolution spectroscopy for 7 enable measurements of their continua shortwards of Ly$\alpha$,
and L$^{\prime}$ detections of 4 of these extend their SEDs redwards to $\lambda_{rest}\sim7500$\AA.
We examine the optical/UV continuum shapes, and compare these to those of a set of 27
well-studied low redshift ($z\sim0.1$) quasars (Elvis \etal\ 1994a) which are matched to the high
redshift ones in {\em evolved} luminosity.
 
Single power law fits to the average fluxes within a set of narrow, line-free, windows between 
1285\AA\ and 5100\AA, but excluding the $2000-4000$\AA\ region of the FeII+BaC `small bump', are 
adequate for most of the objects. 
 
For both the high and low redshift samples, the distributions of spectral indices, $\alpha_{ouv}$
($F_{\nu} \sim \nu^{\alpha_{ouv}}$)
span a wide range, with $\Delta\alpha_{ouv}\sim1$. The cause of such diversity is investigated,
and our analysis is consistent with the conclusion of Rowan-Robinson (1995): that it arises from
differences in both the emitted continua themselves and in the amounts of intrinsic extinction 
undergone.

The mean (median) optical/UV spectral indices for the high and low redshift samples are $-0.32$($-0.29$) 
and $-0.38$($-0.40$), respectively. A Student's $t$-test indicates that these do not differ significantly, 
and a K-S test shows likewise for the distributions.
Assuming the optical/UV continuum derives from accretion, the similarity of the spectral 
indices at high and low redshift is inconsistent with models which interpret the statistical evolution 
as resulting from a single generation of slowly dimming quasars, and instead favors those 
involving multiple generations of short-lived quasars formed at successively lower 
luminosities.

A clear difference between the high and low redshift samples occurs in the region of
`small bump'. The power law fit residuals for the low redshift sample show a systematic excess 
from $\sim2200-3000$\AA; but this feature is weak or absent in the high redshift sample.  Further
study is needed to determine what is responsible for this contrast, but it could reflect differences
in iron abundance or FeII energy source, or alternatively, an 
intrinsic turnover in the continuum itself which is present at low but not at high redshift.

\end{abstract}


\section{Introduction}

It has been well established that the population of quasars undergoes strong 
evolution (Schmidt 1968, Schmidt \& Green 1983, Marshall 1985, and others).
However, not until the last 10-15 years, with the technological 
advances leading to larger or deeper surveys (\eg\ AAT, Boyle, Shanks \& Peterson 1988; 
LBQS, Hewett, Foltz \& Chaffee 1995; Edinburgh quasar survey, Goldschmidt \etal\ 1992; 
HES, Wisotzki \etal\ 2000; 2dF, Boyle \etal\ 2000), has it been possible to probe the manner 
of evolution. A key development was the discovery of a break in the luminosity function 
(LF; Koo \& Kron 1988, Boyle, Shanks \& Peterson 1988, but see Hawkins \& V\'eron 1993, 1995). 
By tracking this feature with redshift, it was determined that the evolution of
the luminosity function (at least up to $z\sim2$) was well described as pure 
luminosity evolution (PLE; Boyle, Shanks \& Peterson 1988, Boyle \etal\ 1991). Results from later surveys 
indicated the need for a more complex model such as luminosity dependent density evolution 
(LaFranca \& Cristiani 1997, K\"ohler \etal\ 1997, Goldschmidt \& Miller 1998, Wisotzki 2000, 
but see Londish, Boyle \& Schade 2000). But though the exact form of evolution may be uncertain,
the general model is of positive luminosity evolution up to $z\sim2$.
At redshifts between 2 and 3, the evolution slows or stops (Boyle \etal\ 1991, 
Hewett, Foltz \& Chaffee 1993), and at $z\gtrsim3$ the number density declines 
(Warren, Hewett \& Osmer 1994, Kennefick, Djorgovski \& de Carvalho 1995, Schmidt, 
Schneider \& Gunn 1995), perhaps signaling this as the initial epoch of quasar 
formation (Warren, Hewett \& Osmer 1994). Pei (1995a), by extrapolating a model of 
the luminosity function evolution, projects the initial formation redshift to 
$z\sim5.2-5.5$, though it may be significantly higher, as suggested by 
the spectrum of the recently discovered $z=5.8$ quasar, SDSS1044-0125 (Fan et al. 2000). 

One goal of these efforts to pin down the luminosity function evolution is a better 
understanding of the quasars' physical evolution; but as yet, the questions of how quasars 
form and evolve, what triggers the activity and how long it lasts remain. Cavaliere \etal\ (1988) 
outlined three modes of quasar evolution which can explain the statistical luminosity evolution: 
first, the most direct interpretation assumes continuous evolution of a single 
generation over timescales long compared to the Hubble time; second is recurrent 
activity of relatively short duration; and the third involves multiple generations of 
galaxies which undergo a brief active (quasar) phase. These are distinguished by their 
predictions for quasar activity or remnants of it in the nearby universe. The discoveries
of massive dark objects or black holes in the centers of many nearby ellipticals 
(Richstone \etal\ 1998, Magorrian \etal\ 1998) argue against continuous evolution, 
which would predict a smaller fraction of these (Salucci \etal\ 1999). 
 
Most of the physical evolution models proposed recently focus on the quasar phenomenon as a 
short-lived recurrent or single event (\eg\ Haehnelt \& Rees 1993, Haehnelt, Natarajan 
\& Rees 1998, Siemiginowska \& Elvis 1997, Haiman \& Menou 1999, Cavaliere \&  
Vittorini 1998, Kauffman \& Haehnelt 2000, though Choi, Yang \& Yi 1999, 2001 also 
consider a single generation of long lived quasars). 
They combine theories of structure formation with those for energy generation in 
quasars, assuming the accretion disk paradigm (\eg\ Shields 1979, Malkan \& Sargent 1982). 
But these models are not well constrained: Tytler (1999) lists the diverse characteristics of
several which can account for the luminosity function evolution. A needed constraint may 
come from the observed spectral evolution.
Accretion disk models make specific predictions for the continuum shape as a function of 
black hole mass, $M$, accretion rate, $\dot{M}$ ($\dot{m}=\dot{M}/\dot{M}_{Edd}$) and 
inclination of the disk axis to the line of sight ($\theta$, $\cos\theta=1$ is face-on), with the
peak frequency determined by the apparent disk temperature, which for optically thick/geometrically thin 
disks follows the relation:
$\log T \sim {1\over4} \log\dot{m}/M - 2.4(\cos\theta-1)$
(Sun \& Malkan 1989, McDowell \etal\ 1991). 
Though the greatest changes in $M$ and $\dot{M}$ are expected 
for continuous evolution of a single generation, some change would also be consistent with 
the multiple generation models, since the factor $\sim50$ drop in the characteristic luminosity
between $z\sim3$ and $z\sim0.1$ implies some difference in the central engines of high and 
low redshift quasars.

We focus here on the optical/UV, where the `blue bump' component, which contains 15-45\% of the 
bolometric luminosity and is commonly interpreted as primary emission from an accretion disk (\eg\ 
Shields 1979, Malkan \& Sargent 1982, Sun \& Malkan 1989, Laor \& Netzer 1989), is prominent.
To look for evidence of evolution in this region, we construct rest-frame optical/UV continua 
for a set of bright high redshift ($z>3$) quasars and compare these with the continua
of a set of 27 low redshift ($z\sim0.1$) ones which were chosen from the `Atlas of Quasar Energy 
Distributions' (Elvis \etal\ 1994a) to have {\em evolved} luminosities within the range spanned
by the high redshift sample.  The organization of this paper is as follows. 
The sample selection is discussed in \S 2, the observations and data reduction in \S 3 and \S 4, 
and the construction of the SEDs in \S 5. Power law fits to the rest-frame $1285-5100$\AA\ continua 
are made in \S6, and the distributions of resulting spectral indices for the high and low redshift 
samples are examined and compared in \S 7. In \S 8 and \S 9, the continuum shape is analyzed in more 
detail: \S 8 discusses our searches for previously reported redshift and luminosity trends, and \S 9 
focuses on deviations of the continuum shape from a power law in the UV and optical, particularly 
the 3000\AA\ region. Finally, \S 10 lists the conclusions.

\section{Sample Selection}
\label{samplesel}

We selected two samples of quasars with redshifts $z\sim 3$ and $z\sim0.1$. 
Because of the need for IR spectra to determine accurate continua for the $z\sim3$ 
quasars, and the relative scarcity of high redshift quasars bright enough 
to make the collection of such data feasible on 4-m class telescopes, 
we were forced to select the high redshift sample first. Our high redshift sample 
spans the range of $\Mv = -27.5 $ to $-29.5$, which corresponds to 
\llstarz $\sim 1-7$, where $L^{*}(z) = L^{*}_{o} (1+z)^{k}$ and the rate, $k=3.15$ 
($H_{o}=50$ km s$^{-1}$ Mpc$^{-1}$, $q_{o}=0.5$) and 
normalization are what Boyle, Shanks \& Peterson (1988) found best described the luminosity function 
(albeit for $z\lesssim2.2$) constructed largely from their AAT survey.
For comparison, the full range of \llstarz\ is approximately $\sim 0.05 - 25$ 
(Boyle \etal\ 1991).

Following the tracks of $L/L^{*} = 1$ and $7.4$ to $z\sim0.1$ delimits the absolute magnitude 
range $\Mv = -22.75$ to $-24.75$ for a low redshift sample which matches the high redshift one 
in {\em evolved} luminosity.
To minimize the number of new observations required, we selected the low redshift sample
from a set of quasars for which Elvis \etal\ (1994a) had previously 
constructed radio to X-ray spectral energy distributions.

\begin{figure}
\figcaption[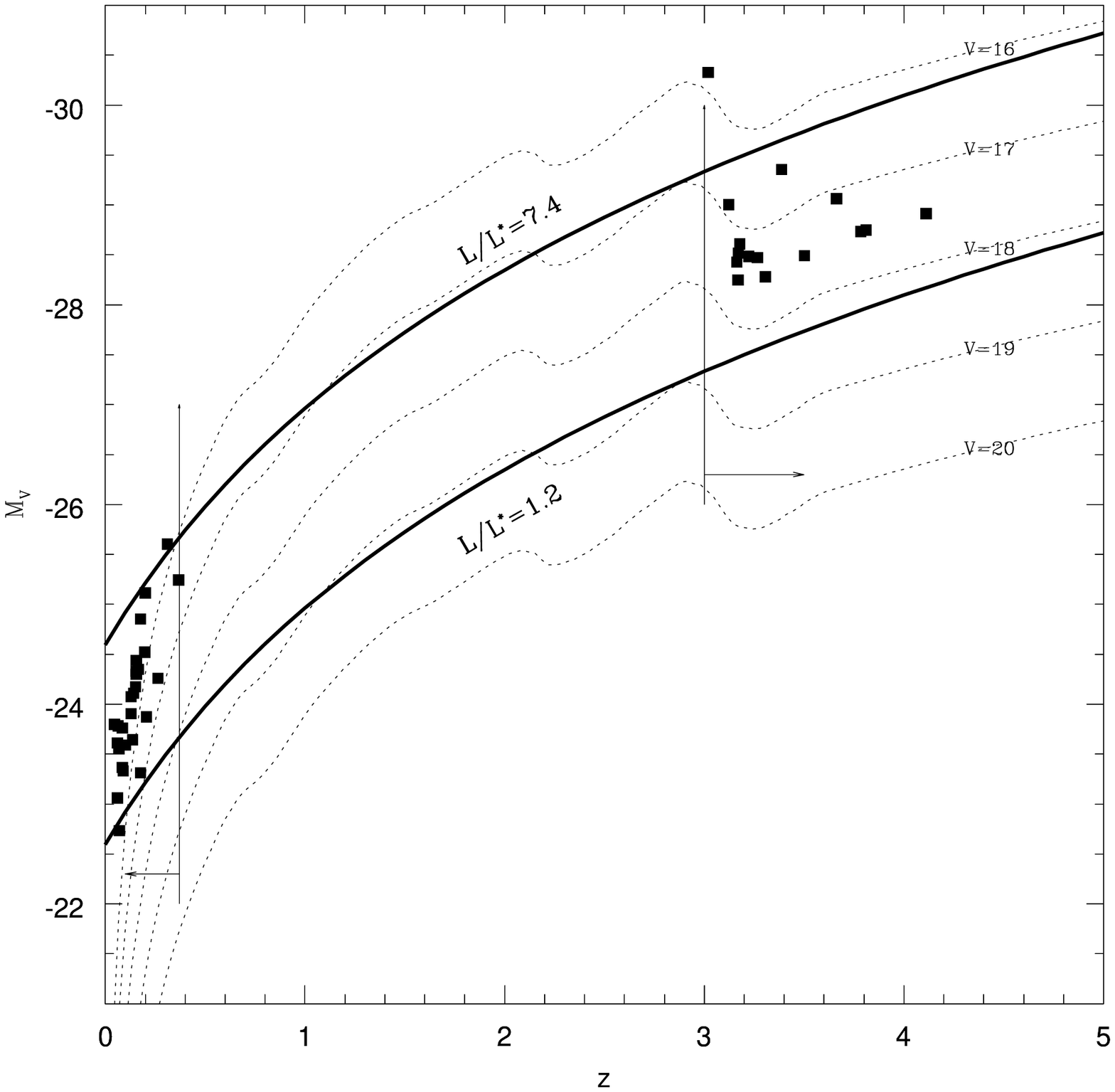]{
Absolute magnitudes, \Mv, and redshifts, $z$, of the quasars in the high and low redshift 
samples (solid squares). The \Mv\ are computed for $H_{o}=50$ km s$^{-1}$ Mpc$^{-1}$ and 
$q_{o}=0.5$ to match the parameters assumed by Boyle \etal\ (1988), whose model for luminosity 
evolution was referred to in the sample selection, although, as stated in the text, we 
subsequently adopt $H_{o}=75$ km s$^{-1}$ Mpc$^{-1}$ and $q_{o}$=0.1. 
The thick solid lines indicate tracks of pure luminosity evolution for $L/L^{*} = 1.2$ and $7.4$,
where these values span the range for all but one of the quasars in the high redshift sample and 
correspond to $M_{V} = -29.5$ and $-27.5$, respectively. $\lstar \sim (1+z)^{k}$ ($k=3.15$; Boyle \etal\ 
1988), so at $z\sim0.1$, these tracks delimit a range for the `matching' low redshift sample of 
$M_{V} = -22.75$ to $-24.75$. 
The thin dashed lines mark curves of constant apparent magnitude, \mv, where, following
V\'eron-Cetty and V\'eron (1993), the $k$-correction was made assuming an optical/UV spectral index, 
$\alpha=-0.7$ ($\Fnu \sim \nu^{\alpha}$) and the wiggles near redshifts 2 and 3 arise from strong 
emission lines of Ly$\alpha$ and CIV entering and leaving the V band.
\label{sample_dist}}
\end{figure}

Several objects had to be discarded from the original samples because we were unable
to obtain sufficient observations, or their data were of poor quality; so in the 
analysis, 15 high and 27 low redshift quasars were considered.
Tables \ref{hizsample} and \ref{lowzsample} list the properties of the quasars in 
the initial samples: radio-loudness, redshift, magnitude, $\alpha_{ox}$, and selection method;
and the final column indicates, by a $\surd$ or $X$, those which were used 
or had to be rejected from the sample.
In Figure \ref{sample_dist}, we plot the redshift-absolute magnitude distribution 
of the quasars in the final high and low redshift samples.
Note that while the sample selection was made assuming $H_{o}=50$ km s$^{-1}$ Mpc$^{-1}$, $q_{o}=0.5$,
to be consistent with Boyle, Shanks \& Peterson (1988), for the rest of the paper (\eg\ 
when converting data to the rest frame) we adopt a different set of cosmological
parameters: $H_{o}=75$ km s$^{-1}$ Mpc$^{-1}$, $q_{o}$=0.1, which are intermediate between an 
open, old and closed, young universe. $\Lambda=0$ is assumed throughout.

\placetable{table1} 
\placetable{table2} 

There is evidence that at redshifts greater than $2$, evolution 
slows or stops (Boyle \etal\ 1991; Hewett, Foltz \& Chaffee 1995, Warren, Hewett \& Osmer
1994). This would make a smooth extrapolation of the $L\sim(1+z)^{k}$ evolution out to 
$z\sim3$ invalid, and instead the curve would turn over, implying that the $z\sim3$ sample
spans yet higher values of $L/L^{*}$. The absolute magnitude range of a matching 
low redshift sample would then be $0.5$ to $1$ magnitude brighter. About 5-15 of the 
quasars in the sample that we chose would be too faint to satisfy this new constraint.  
However, selecting brighter quasars at low redshift is difficult since their low space 
density means there are few available.

Both the high and low redshift samples are inhomogeneous. Each is roughly evenly split 
between radio-loud and radio-quiet quasars and they contain quasars discovered in a variety 
of surveys: radio; UV excess; objective prism. A clear selection bias exists in the low 
redshift sample, which was drawn from the `Atlas' sample that is comprised of AGN with strong 
detections with {\em Einstein} (Wilkes \& Elvis 1987).  The high redshift sample contains 
some of the brightest objects in the universe.  One, Q$1208+101$, is a gravitational lens 
candidate (Maoz \etal\ 1992). The lens has not been detected, to H$\sim20$ in NICMOS images 
(L\'ehar \etal\ 2000), but magnification of the luminosity by a factor from $1-22$ is 
estimated from the distribution of Ly$\alpha$ absorption systems near the quasar (proximity 
effect; Bechtold 1994, Giallongo \etal\ 1999). None of the other high 
redshift sample quasars is considered a lens candidate, however Pei (1995b) suggests that 
from 8-80\% (depending on the lens model assumed) of bright ($M_{v}<-30$) $z\sim3$ quasars 
may be lensed and consequently, their apparent brightnesses enhanced. If true and a 
fraction of the high redshift sample quasars are indeed lensed, this would affect continuum shape-luminosity 
correlations, but not the primary goal to look for differences in the continuum shapes 
between $z>3$ and $z\sim0.1$.  We eliminated quasars known to be highly variable and those 
classified as BALQSOS (except Q$1426-015$, whose optical spectrum published by Wilkes \etal\ 
(1999) does not show strong absorption), and aside from the biases toward strong X-ray emitters 
in the low redshift sample and extremely luminous quasars in the high redshift one, there is 
no other obvious bias in either sample. 

\section{Radio Observations}

To ascertain the radio classification of the sample objects which
were not obviously radio bright (\eg\ discovered in a radio survey), we looked
in the literature and in the 87GB catalog (Gregory \& Condon 1991) which
has a flux limit of 25 mJy. We observed the 9 sample quasars whose radio
classifications were still not known at the VLA at 5 GHz and
1415 MHz in November 1992, and found all these to be radio quiet (radio
loudness, $\cal{R} \equiv \log$[S(5 GHz)/S($\lambda1450$)]
and $\cal{R}>$ 1 is radio loud, $\cal{R} \leq$ 1 is radio quiet; Bechtold \etal\
 1994).
Fluxes and upper limits for the 9 high redshift quasars observed at the VLA
are listed in Table \ref{vlaobslog}.

\placetable{table3} 

\section{IR and Optical Observations}

At $z\sim3$, the $1\mu$m to $\sim1000$\AA\ optical/UV `blue bump' is redshifted 
to optical and infrared wavelengths. To determine its shape accurately,
moderate resolution spectroscopy and broadband photometry are needed: the
spectroscopic data to distinguish
the continuum level in between the numerous broad emission lines,
and the photometry to provide an absolute flux calibration. 
But near-IR observations still do not reach the rest-frame continuum redwards of 
$\sim5500$\AA. To extend coverage of the `blue bump' beyond that,
we observed a subset of the sample at L$^{\prime}$(3.4$\mu$m) 
and N(10.6$\mu$m). Because of the difficulty of these observations, we 
only obtained, at L$^{\prime}$, fluxes for 5 of the high redshift quasars and upper limits 
for another 5 and, at N, upper limits for 4.

In this section, we report on the observations and the procedures 
followed to reduce the data. Because of the number of different observing runs
and instruments used, we summarize in Table \ref{obssum} the information 
relevant to each: 
the telescope and instrument, the instrument configuration 
and the typical integration times. Logs of the IR and optical spectroscopic 
observations are given in Tables \ref{irspec} and \ref{optspec}; and Tables
\ref{irphot} and \ref{optphot} list the results of the IR and optical photometry,
respectively.
\placetable{table4} 
\placetable{table5} 
\placetable{table6} 
\placetable{table7} 
\placetable{table8} 

\subsection{IR spectroscopy}

We used three instrument/telescope combinations to obtain the moderate resolution
(R$\sim500-700$) near-IR spectra presented here. Table \ref{obssum} lists the detectors,
slit widths, gratings used and resolutions obtained. 
With two of the instruments, Fspec (Williams \etal\ 1993) and CRSP (Joyce, Fowler \& Heim
1994), a single grating setting 
covered the entire J, H or K band, and with the third, OSIRIS (DePoy \etal\ 1993), a 
cross dispersing grism enabled the simultaneous acquisition of 1.1-2.5$\mu$m J, H and K 
spectra.
The observing procedures with Fspec, CRSP and OSIRIS were generally similar.
In each case, a series of short integrations (background limited in the case of CRSP
and OSIRIS) was accumulated, with the object stepped a few arcseconds along the slit 
between each or every other one, either progressively (Fspec) or in an ABBA pattern 
(CRSP and OSIRIS).
Total on-source integration times ranged from about 0.5-2 hours, with most between
$\twid45$ and $90$ minutes (see Table \ref{obssum}). 
To enable adequate removal of atmospheric absorption, the observations of each quasar 
were bracketed by those of a bright star with a nearly featureless near-IR continuum 
(early A or late F or early G dwarf) and within $\twid10^{\circ}$ of it. 
With CRSP and OSIRIS, the stars were stepped sequentially along the slit
rather than being placed in just two positions as were the quasars. 
For calibration, dome flats were taken with the lamps on and off, to enable the subtraction
of thermal radiation and the dark current, except in the case of Fspec, where only on-flats 
were obtained and then dark frames were taken to subtract from these.


We reduced the IR spectroscopic data using IRAF and following standard procedures.
Reduction of the Fspec and OSIRIS data used IRAF scripts specifically written for these 
instruments. Each of the individual images was corrected for non-linear response 
on a pixel-by-pixel basis (OSIRIS and CRSP only), and then sky subtracted, using for 
sky either the average of the two images which bracketed it or the single image taken 
immediately before or after it but with the object spectrum on a different part of the 
array.  The sky-subtracted images were then divided by the combined set of dark-subtracted 
flat fields. A residual bias, determined from an unexposed portion of the array was removed
from the Fspec images. The values of bad pixels were replaced with those determined by 
interpolating over (CRSP and OSIRIS) or median combining (Fspec) neighboring pixels. Usually 
the sky subtraction did not remove all traces of sky emission, so this residual background 
was fit and subtracted. 

The subsequent steps to extract, co-add and wavelength calibrate the 
one-dimensional spectra were carried out in a different order for the
data from CRSP, OSIRIS and Fspec. The Fspec images were rectified to align
the spatial axis along the rows and combined in sets of 4. Spectra were
then extracted, wavelength calibrated using a sky spectrum extracted
from the same set of images, and then all wavelength calibrated spectra of
the same object were combined. CRSP spectra were extracted from 
each reduced image, then wavelength calibrated using the sky spectrum that 
was extracted from the corresponding sky image within the aperture determined 
by the object. Next all the wavelength calibrated spectra for a 
given object were combined.
Finally, with the cross-dispersed OSIRIS data, the images were sky- and residual 
background- subtracted, then the array regions containing the individual orders were 
extracted, flat-fielded, and rectified. The reduced subimages were combined, one-dimensional 
spectra optimally extracted from these, and wavelength calibration done using the
OH lines in the spectra that were extracted from the sky images.

The spectroscopic standard stars were reduced in the same way as 
the quasars, though with CRSP and OSIRIS where the stars were stepped in several 
positions along the slit, a sky image was produced by median combining all the star 
integrations. The stellar spectrum
extracted from the sky-subtracted, flat-fielded image was wavelength calibrated
and then divided into the spectrum of the quasar for which it was taken, to 
remove telluric absorption lines and correct for spectral response. Before division, 
intrinsic absorption features in the standards, 
such as Pa$\beta$ at $1.28\mu$m and Br$\gamma$ at 2.16$\mu$m were removed.
Another feature, at 2.18$\mu$m, was treated as intrinsic when
reducing the Fspec and CRSP data, but for the later OSIRIS reductions, it was properly 
treated as atmospheric. 
The strength of this feature was comparable to the noise in the quasar spectra, and so 
removal of it did not significantly affect the spectra.

The quasar spectra were flux calibrated using the broad band IR magnitude of the standard 
star as determined from its V magnitude and the optical to IR color (Johnson 1966) 
corresponding to its spectral type. The stellar continuum shape was determined
from its effective temperature (Johnson 1966) and removed from the final quasar
spectrum.

\subsection{IR photometry}
\label{iphot}

We obtained the J, H, K and L$^{\prime}$ photometry presented in this paper
using the single channel IR photometer at the MMT (Rieke 1984) and OSIRIS 
in imaging mode at the CTIO 4-meter. The observations at N were also made at the MMT, 
using the single channel IR bolometer (Keller, Sabol \& Rieke 1990). 

For all of the MMT photometer and bolometer observations,
the secondaries were chopped at a frequency of 7.5 or 20 Hz, respectively, 
and the telescope was wobbled every 10 seconds to obtain a frequent sky measurements on 
either side of the source. The chop throw and wobble were both 15\asec, except for 
one run where they were 10\asec (Table \ref{irphot}). 
The observations were continued until sufficient signal-to-noise 
was achieved, typically 30-40 minutes for the quasar, and 5 minutes for the standard star.
Table \ref{obssum} lists the aperture sizes, chop rates and throw (in place of grating
and resolution) and typical integration times. The bolometer has just one aperture available, 
though the photometer has two | the choice for any given night depended on seeing conditions.
The aperture used for each quasar is given in Table \ref{irphot}.
For the observations with OSIRIS, we took series of 5 to 9 short (10 seconds times
3 co-adds) integrations between which the telescope was offset to position the 
object in a grid which covered the array.
Dome flats and dark frames were not taken, but flat fields were produced by 
median combining the all the science observations taken through the 
same filter and with similar integration times during the course of that night. 
The dark current could not be subtracted from these flats, but was negligible 
(1 e$^{-}$ s$^{-1}$; Cooper, Bui \& Bailey 1993).  


The IR photometer and bolometer data were reduced in the same way. 
After each wobble, the instrument output the total background-subtracted counts, 
where the background was taken to be the average of the two bracketing sky
observations. Instrumental magnitudes and errors were computed from the average 
sky-subtracted counts and the standard deviation in the average.

To reduce the images from OSIRIS, we first produced a blank sky image by median-combining
all the images taken within a single grid sequence and rejecting pixel values which deviated 
significantly from the median.  This sky image was then subtracted from each of the images in the 
sequence and the result was divided by the normalized flat field at the corresponding filter. 
Bad pixels were replaced with the value determined by interpolating over adjacent ones.  
Finally, all the reduced images from a single grid sequence were aligned and averaged together.
Aperture photometry was done on the individual and combined images using
a circular aperture with a 32 pixel (12\arcsec) diameter; and the mode of values
within an annulus that was centered on the object and had a radius and width 
of 25 pixels was taken as a measure of the residual background and subtracted from 
the aperture flux.

To flux calibrate the data from the MMT IR photometer and OSIRIS, we 
observed stars from the lists of bright standards (Elias \etal\ 1982) and 
faint UKIRT standards (Casali \& Hawarden 1992, Hawarden \etal\ 2001), respectively.
To calibrate the N band bolometer data, we used $\beta$And, $\alpha$Tau, 
$\alpha$Cyg and $\mu$Cep as standards (Tokunaga 1984; Rieke, Lebofsky \& Low 1985).
For all instruments, the instrumental magnitudes were converted to apparent magnitudes 
assuming extinction coefficients based on previous observing runs at the MMT 
(M. Rieke 1994, private communication): $k = 0.1$ mag/airmass at J, H and K and $k=0.2$ 
mag/airmass at L$^{\prime}$, though $k=0.08$ may be more appropriate for
the CTIO data. We made no extinction correction 
to the data at N since we had no detections, only upper limits, at this band.
No color transformations between the MMT and CIT systems were assumed except at
J where, to account for the difference between the peak wavelengths of
the transmission curves for the filter used by Elias and that used in
the MMT photometer, the following correction was applied to the Elias 
magnitudes: $J_{MMT} = 1.006(J-K)_{Elias} + K_{MMT}$ (Willner \etal\ 1985).
No color transformations between the UKIRT (to which the OSIRIS data were referenced)
and the CIT systems were assumed.
%
We placed the magnitudes on a flux scale using the following zero magnitude 
fluxes, also listed in Table \ref{flux0mag}: 
at J ($\lambda_{o}=1.25\mu$m), $F_{0}$=1603 Jy; 
H ($\lambda_{o}=1.60\mu$m), $F_{0}$ = 1075 Jy; and 
K ($\lambda_{o}=2.22\mu$m), $F_{0}$=667 Jy 
(Campins, Rieke and Lebofsky 1985). 
At L$^{\prime}$ ($\lambda_{o}=3.4\mu$m), the zero magnitude flux was taken 
to be 309.0 Jy, based on a blackbody fit to the spectrum of Vega (R. Cutri 1994, 
private communication) and at N ($\lambda_{o}=10.6\mu$m), we assumed $F_{0}=36$ 
Jy (Rieke, Lebofsky \& Low 1985).
\subsection{Optical spectrophotometry}

The optical spectrophotometry was obtained using four different 
instruments: the red channel spectrograph at the MMT; the Boller and 
Chivens spectrograph at the Steward Observatory 2.3-m; the FAST 
spectrograph (Fabricant \etal\ 1998) at the FLWO 1.5-m telescope; and 
the RC spectrograph at the CTIO 1.5 m telescope. Details of the various 
instrument set-ups, including slit widths, gratings, resolution, wavelength
coverage, as well as the range of integration times, are given in 
Tables \ref{obssum} and \ref{optspec}.

To reduce light losses, we used, for all of the objects, a 4-5\asec\ wide slit;
for the MMT and Steward Observatory observations, it was oriented 
parallel to the atmospheric dispersion, though for those made at 
FLWO and CTIO it was always oriented E-W. 
With the MMT and Steward 2.3 m, a single 20-30 minute integration gave 
sufficient signal-to-noise; at the other telescopes, we obtained a set of 3
15-20 minute integrations which we median-combined to increase signal-to-noise and
eliminate cosmic rays. An internal HeNeAr or HeAr lamp spectrum was obtained for
each position. 
Spectrophotometric standards from Massey \etal\ (1988), Massey \& Gronwall (1990) 
(except for the IIDS standard,
EG 129; Oke 1974), and Hamuy \etal\ (1992), for the southern hemisphere observations, 
were observed frequently throughout the night and at a range of airmasses.
Dome flats were taken during the day, and, at CTIO, twilight flats were also taken. 
Bias frames were obtained for each night.  We reduced these data in IRAF following 
the usual procedures for bias subtraction, flatfielding, spectral extraction 
with background subtraction, and flux calibration.

\subsection{Optical photometry}
\label{ophot}

Optical photometry was obtained at the 1.2 m telescope at FLWO on Mt. Hopkins 
and the 0.9 m telescope at CTIO.  The CTIO data were obtained in photometric 
conditions. There were cirrus or clouds during some of the observations 
made at the FLWO, as noted in Table \ref{optphot}, but we repeated the 
observations obtained in non-photometric conditions. During the course of the night, 
we observed groups of standard stars from Landolt (1992).

The images were reduced following the standard IRAF procedures. The data
were flat-fielded using dome (FLWO and CTIO) and twilight flats (FLWO).  
Aperture photometry was done where the diameter of the circular aperture,
$12$ to 17\arcsec, depending on the seeing, was 
chosen to include all the light yet not too much background noise.
The same aperture size was adopted for all observations taken during the night. 
The sky counts were measured in an annulus centered on the object.
The IRAF task {\em photcal} was used to convert instrumental to absolute magnitudes. 
It uses the standard star data to 
determine the extinction and color correction. For the October 1993 dataset, 
we did not include a color correction term in the transformation equations, 
but instead made a correction to the zero-magnitude fluxes at each filter, 
based on a comparison between an expected quasar spectrum (a power law, $\fnu 
\sim \nu^{\alpha}$) and the convolved CCD response and filter transmission 
curves (sixth column of Table \ref{flux0mag}). Finally, we placed the 
magnitudes on a flux scale, using the zero-magnitude fluxes in Table 
\ref{flux0mag}. 
\placetable{table9} 

\section{The Spectral Energy Distributions}

This section describes the procedures followed to piece together 
the optical and IR data to produce rest-frame optical/UV spectral
energy distributions. These consisted of 
(1) grey-scaling the IR and optical spectroscopy by constant factors 
to match the fluxes determined from our broad band photometry;
(2) correcting the spectra for Galactic extinction and extinction in 
line-of-sight damped Ly$\alpha$ systems when present; and 
(3) correcting the continua at wavelengths shorter than redshifted Ly$\alpha$ 
for the numerous HI absorption lines, the `Ly$\alpha$ forest'. 
Additionally, some re-construction of the rest-frame continua for the 
low redshift sample was necessary to make them 
consistent with those of the high redshift sample.

\subsection{Scaling and correction for Galactic extinction}
\label{hizcorr}

The observations of spectroscopic standards were not sufficient to
flux calibrate the IR spectra. The slit was narrow ($1-2\arcsec$), so 
centering and maintaining standards and quasars in it was difficult
and not uniformly done. Besides, conditions were sometimes not photometric. 
Therefore, we relied on our IR photometry to greyscale the spectra.
The scaling factor was determined by dividing the photometric flux by the 
spectroscopic flux averaged over the corresponding J, H or K bandpass.
Since the IR filter curves closely approximate a `tophat' shape, averaging 
over the bandpass is not a bad representation of the convolution of the 
spectrum with the filter curve. Finally, each spectrum was flux calibrated by
multiplying by the appropriate scaling factor. In the same way, the flux 
levels of our optical spectrophotometry were checked against the optical 
photometry and corrected when needed. The scaling factors are listed in Table 
\ref{scalfac}.
\placetable{table10} 

All of the data were corrected for Galactic extinction, using the Galactic 
extinction curve (optical/UV | Savage \& Mathis 1979, IR | Rieke \& Lebofsky 1985) 
and gas to dust ratio of $4.8 \times 10^{21}$ cm$^{-2}$ mag$^{-1}$ (Savage \& Mathis 1979).
When possible, the Galactic column densities were taken from Elvis, Lockman 
\& Wilkes (1989) and Murphy \etal\ (1996), who directly measured 
the column densities towards a number of the sample quasars to an accuracy of 
$<10^{19}$ cm$^{-2}$. For the southern quasars or those otherwise not targeted
by Elvis \etal\ or Murphy \etal, we used column densities from the Bell Labs HI 
survey, which have typical uncertainties of $\lesssim10^{20}$ cm$^{-2}$ 
(Stark \etal\ 1992). 
Table \ref{galext} lists the line-of-sight column densities for each of the 
high redshift sample quasars.
 
\placetable{table11} 

Three of the high redshift quasars: Q$0000-263$, Q$0347-388$ and Q$1946+7658$;
have damped Ly$\alpha$ systems in the line-of-sight. 
There is evidence that these contain dust (Pei, Fall \& Bechtold 1991, Carilli et al. 1998),
so we corrected for extinction in these systems. Their metallicity
and dust to gas ratios are assumed to be lower than
Galactic (10\%; Pei, Fall \& Bechtold 1991), and so to correct for reddening
by these, we used the SMC extinction curve and a dust to gas ratio 10 times 
smaller than the Galactic value. 
%

\subsection{Correction for Ly $\alpha$ forest}
\label{lyforcorr}

Along a given line of sight, the number density of Ly$\alpha$ forest clouds increases with 
redshift (Peebles 1993). Absorption by these is clearly evident in the $\lambda < 
\lambda$(Ly$\alpha$) spectra of the high redshift sample quasars.
Since the accretion disk spectrum predicted for parameters typical of quasars: \eg,
$M_{BH}\sim10^{8-9}$ \msun\ and $\dot{m}\sim1$; peaks in the UV at wavelengths $\lambda\lesssim
1000$\AA, knowledge of the intrinsic continuum shape bluewards of Ly$\alpha$ would be 
useful in constraining disk models.

\begin{figure}
\figcaption[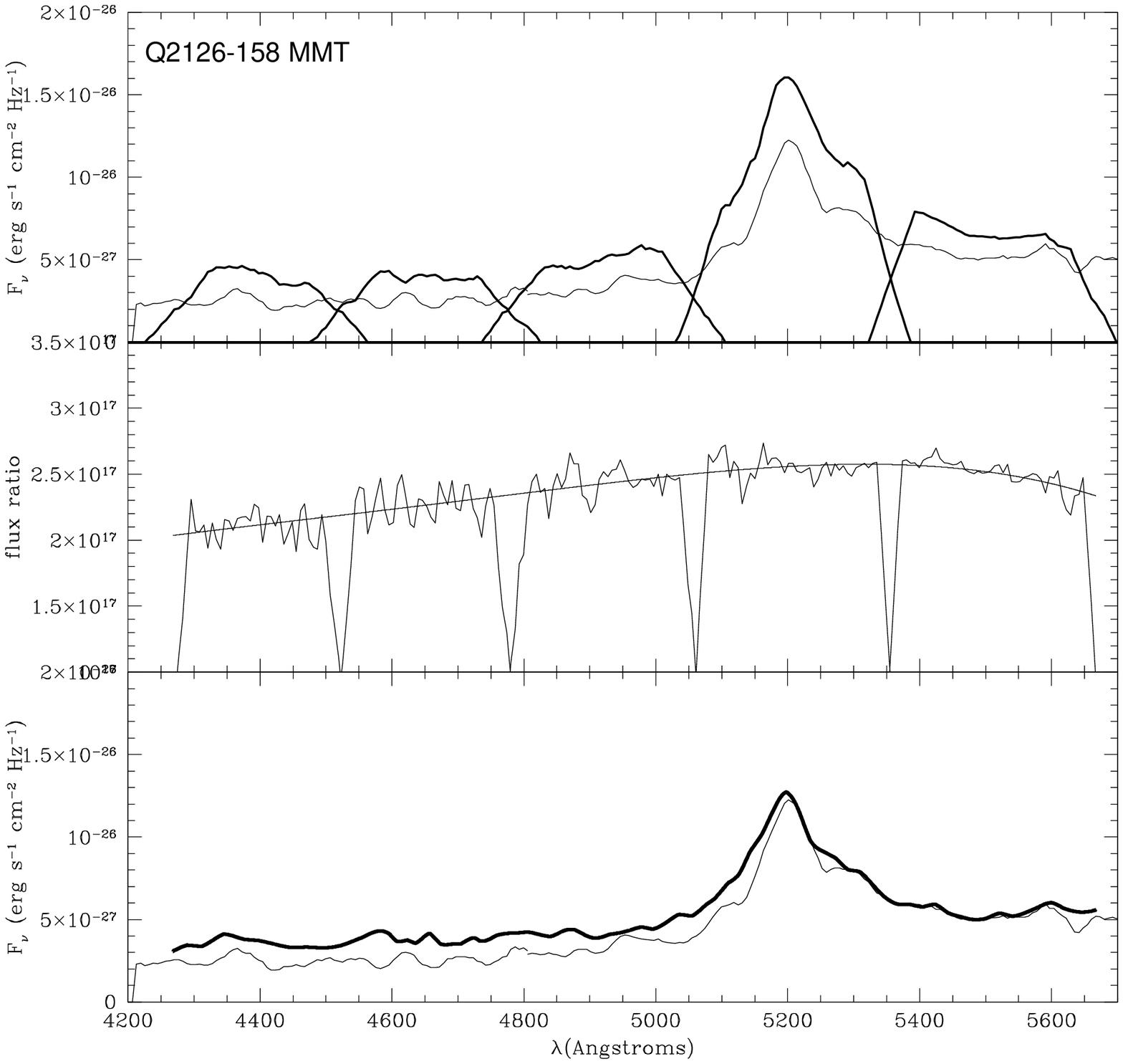]{
The three panels, from top to bottom, illustrate, for Q$2126-158$, the steps taken 
to correct for suppression of the UV continuum by `Ly~$\alpha$ forest' absorption.
In the top panel, the high resolution spectrum is binned and smoothed (thick 
curve) to approximate the resolution of the spectrophotometry (thin curve), 
and the spectrophotometry shifted to match the
wavelength calibration of the high resolution data. The middle panel
shows the ratio of the binned and smoothed high resolution spectrum to the
shifted spectrophotometry and the best-fitting polynomial, termed the 
`correction function'.
In the bottom panel, the corrected continuum (thick curve), \ie\ the product of the 
continuum fit to the high resolution spectrum and the `correction function', is plotted 
with the shifted spectrophotometry (thin curve) for comparison. Note that the levels 
match well to the red of Ly$\alpha$, but on the blue side, the corrected continuum flux 
is larger (by a factor of $\sim 1.5$ for this quasar) than the level of the spectrophotometry.
\label{lysteps}}
\end{figure}

To correct the UV continua for flux suppression by superposition of numerous Ly$\alpha$ forest
lines, we used high resolution spectra that had been obtained for 7 of the quasars from the 
$z>3$ sample (Bechtold 1994; Dobrzycki \& Bechtold 1996; Table \ref{lymatch}).  The resolution 
of these spectra, FWHM $\twid 50-100$ km s$^{-1}$ (19 km s$^{-1}$ for 2 of the objects), is 
sufficient to distinguish the absorption lines and permit a measurement of the continuum level 
in between these (Bechtold 1994). 
However, since these spectra are of necessity taken through a narrow slit, the continuum 
determined from them is not photometric. We have corrected it using our low resolution 
spectrophotometry, following the procedures described below and illustrated in Figure 
\ref{lysteps}. 
\placetable{table12} 
%

(1) We binned and smoothed the high resolution spectra to match the resolution
of the spectrophotometry. (2) This was then divided into the spectrophotometry, 
which had been shifted by a few pixels to match the more accurate wavelength 
calibration of the high resolution data, to produce a `correction function'. 
(3) The `correction function' was fit with a low order polynomial. The smoothing lengths,
shifts and fit orders varied for each pair of spectra and are listed in
Table \ref{lymatch}.  Because our smoothing
procedure included zeroes outside the data range, it produced artificial gradients
at the edges. 
When several high resolution spectra were used to expand the wavelength 
coverage, these spectra were treated separately until making the polynomial
fit to the ratios. We tried both fitting all the sections at once and
fitting each individually. As Table \ref{lymatch} shows, for Q$0420-388$ we 
chose to splice together a number of best fit polynomials while for Q$2126-158$,
we used the single polynomial which, as we judged, simultaneously fit all of the ratios 
well. (4) Finally, we multiplied the continuum that had been fit through the high resolution
spectra by this polynomial to produce the final Ly$\alpha$ forest corrected continuum. 
In some cases, there was ringing in the fitted polynomial near the edges of 
the spectra. The last column of Table \ref{lymatch} notes the regions where the corrected 
continuum is valid, not being affected by either smoothing or ringing.

\subsection{The SEDs of the high redshift quasars}

The flux calibrated, dereddened and blueshifted optical/UV spectral energy 
distributions of the 15 high redshift sample quasars are plotted in 
Figure \ref{seds}. Corrected continua bluewards of Ly$\alpha$ are plotted
using thick lines. The mean SED determined from the radio-loud
or radio-quiet quasars from the Elvis \etal\ (1994a) database of low 
redshift AGN is superposed for reference.

Four of the high redshift quasars were detected at 
L$^{\prime}$ or have upper limits at N which constrain their continua to 
be close to the mean low redshift continuum. Figure \ref{lseds} shows
their SEDs over the rest-frame spectral region which includes the L$^{\prime}$
and N datapoints.

\subsection{Modifications to the low redshift SEDs}
\label{lowzcorr}

For many of the low redshift quasars there were multi-epoch observations, 
and some of the quasars showed significant variability. Thus, when possible, we 
selected the multi-wavelength observations that were obtained closest in time and 
reconstructed SEDs using only these.  Table \ref{simultan} lists the observation dates 
for the data we selected.  For four quasars, Q$0121-590$, Q$0804+761$, Q$1211+143$ and
Q$2130+099$, more than one SED resulted. The one which was judged by eye to look 
cleanest was used in the final analysis. 

\placetable{table13} 

The Galactic gas-to-dust ratio and cosmological parameters that Elvis \etal\ 
(1994a) had assumed to deredden and blueshift the data on the low redshift
`Atlas' quasars (N$_{H}$/E(B-V) = $5 \times 10^{21}$ cm$^{-2}$ mag$^{-1}$; $H_{o}$ = 50 
km s$^{-1}$ Mpc$^{-1}$, $q_{o}=0.5$) differ from the values we adopt in the paper.
Therefore, before comparing the low and high redshift SEDs, it was necessary to 
convert the low redshift SEDs back to the observed frame using their values and
then deredden and blueshift the data using ours. 

\subsection{Computation of continuum fluxes}

We characterize the shapes of the rest-frame optical/UV continua by 
luminosities measured within a set of twelve narrow band windows which we chose
upon examination of the composite QSO spectra constructed by Francis \etal\ 
(1991; LBQS) and Boyle (1990; AAT). The central wavelengths and widths of
these bands are listed in Table \ref{windows} and indicated on the SED of one of the
sample quasars (Figure \ref{fwindows}).

Initially, eleven of these bands were considered continuum windows. 
Three are centered at the wavelengths through which Francis \etal\ (1991) 
fit a continuum to the composite LBQS spectrum, and five were added at wavelengths 
where the continuum fit closely approached the LBQS composite. A ninth band 
was centered at $\lambda_{rest}=7500$\AA\ to measure the red end of the optical
continuum, but the flux there can only be estimated for the
low redshift quasars and the $z>3$ ones detected at L$^{\prime}$. Finally,
two bands centered at 1160\AA\ and 1115\AA\ sample the flux below
Ly$\alpha$. The twelfth band, centered at 2500\AA, where there are
several strong FeII multiplets (Joly 1993), was selected to gauge FeII strength. 

Around 3000\AA\ there is a small bump in the continua of many low redshift
quasars (Oke, Shields \& Korycansky 1984). It is modelled as blended FeII and Balmer
continuum emission (Wills, Netzer \& Wills 1985). Though its strength varies from 
object to object, between 2000\AA\ and 4000\AA\ it may contribute as much as the 
continuum itself to the total flux (see Figure 2 of Neugebauer \etal\ 1987).
The LBQS composite, however, does not show significant excess over
the continuum fit within this region, and four of the `line-free' bands lie between
2000\AA\ and 4000\AA. In our analysis, we treat the fluxes within these bands with caution,
looking for indications of non-continuum emission (see Table \ref{aouv}, \eg, which lists
spectral indices computed with and without these points). Only the spectral indices determined
without the $2000-4000$\AA\ data are used in the analysis, and the $2000-4000$\AA\ residuals
are discussed in \S\ref{l3000bump}.

\placetable{table14} 

%
The luminosities within the narrow bands (Tables \ref{hiz_cfluxes}, \ref{lowz_cfluxes} 
and \ref{uvfluxes}) were determined as follows.  The monochromatic luminosity emitted 
at the central wavelength of each band, $\nu F_{\nu}$, was taken to be the average of the 
data points (scaled, dereddened, and blueshifted) within that band. The narrow widths 
of the bands ($\Delta\lambda=20-40$\AA) and the smoothness of the energy distribution 
within them make this a good approximation to the integral of flux over frequency.
Within the shortest wavelength bands (the bluest 2 for the high redshift
quasars and the bluest 5 for the low redshift ones), where the datapoints
are from moderate resolution (R$\sim100-500$) optical spectrophotometry, there were 
abundant points over which to average. But at longer wavelengths the data are sparser
and there are gaps between the datasets, which are mostly unavoidable because they are
where atmospheric transmission is poor.  Furthermore, 
the dispersion of the near-IR spectra is lower than the optical | about $11-17$\AA/pxl 
rather than $\twid7$. Thus, the average fluxes within the redder bands were often 
computed either from relatively few data points or if none, by interpolation between
the average fluxes computed within 50\AA\ wide bands on either side of the empty
window.  The width of 50\AA\ was chosen since it usually included sufficient data points
to provide good statistics, yet did not extend to emission line features.
%

There were various sources of error in these narrow-band average luminosities:
among them (1) errors in normalization, either from grey-scaling or
flux calibration; (2) rms errors in the individual spectra; (3) errors in
individual photometric points; and (4) scatter caused by having more than one 
epoch of data within a band. 
For the high redshift quasars, the normalization error was taken to be the error 
propagated from the errors in the photometry used to grey scale the IR spectra and
optical spectrophotometry.
For those quasars with Ly$\alpha$ forest corrected spectra, the normalization error
of these was taken to be 10\%, which was the typical scatter about the polynomial
fit to the ratio of the binned, smoothed high resolution data to the spectrophotometric
data.
For the low redshift quasars, the normalization error in the IUE spectra was
taken to be 10\% (Bohlin \etal\ 1980).
The errors that are given alongside the luminosities in Tables \ref{hiz_cfluxes} 
$-$ \ref{uvfluxes} both include the normalization and photometry errors, but
incorporate the rms scatter in different ways. The first, $\sigma_{avg}$,
uses a quadrature sum of the scatter in the individual spectra, which is an accurate
reflection of the quality of the data and valid when variability is not an issue.
In contrast, the second, $\sigma_{var}$, accounts for the rms scatter, \ie\
$\sigma_{N-1}$, for all datapoints, spectroscopic and photometric, within the 
band.

Tables \ref{hiz_cfluxes} and \ref{lowz_cfluxes} list the monochromatic luminosities
at each of the ten bands redwards of Ly$\alpha$, for the high and low redshift sample objects,
respectively. They also give the number of points used in the average, either within the 
band itself or within the two 50\AA\ regions used in interpolating the flux, and the 
confidence in the computed luminosity: 3 signifies that it was directly
computed from points within the band and well represents the flux level; 2, that it
was interpolated; 1, that it does not represent the flux level well; and 0, that it
could not be determined even by interpolation because of insufficient data coverage
to the red.

The continuum luminosities in the two windows at $\lambda < \lambda$Ly$\alpha$ can
be computed only for a subset of objects from the low and high redshift samples:
of the low redshift sample, only those objects with redshifts large enough to shift the
1115\AA\ and 1160\AA\ bands beyond the geo-coronal Ly$\alpha$; and of the high
redshift one, only those 7 for which high resolution spectra enabled the correction 
of the $\lambda<\lambda$Ly$\alpha$ continuum for flux suppression by Ly$\alpha$ forest.
Consequently, luminosities at 1160\AA\ were computed for 26 of the 27 low redshift quasars 
(all but Q$0121-590$) and 7 of the 15 high redshift ones, and luminosities at 1115\AA,
for 20 low and 7 high redshift objects.  Table \ref{uvfluxes} lists these luminosities.
For comparison, for the high redshift quasars, luminosities determined from uncorrected data 
are also given. 

\placetable{table15} 
\placetable{table16} 
\placetable{table17} 

\section{The optical/UV continuum shapes}

\subsection{Power law fits}

Our first approach towards characterizing the optical/UV continuum shape
was to fit a single power law through the set of average narrow band 
continuum fluxes. Since our coverage of the red and blue ends of the `blue
bump' was poor | only 4 of the 15 quasars in the high redshift sample were detected 
at $L^{\prime}$ and only 7 of the 15 had high resolution spectroscopy bluewards of 
Ly$\alpha$ | we focused on fitting the continuum from $\lambda_{rest}=1285-5100$\AA\
only.  This region included 9 narrow band windows, four of which lie between 2000\AA\ and
4000\AA, in the `3000\AA\ bump' region. To insure that the power law fits are 
not influenced by the blended line emission, we tried both fitting through all 9 bands from 
1285\AA$-5100$\AA\ and through the 5 bands which remained after excluding those between 
2000\AA\ and 4000\AA.
The spectral indices determined from both sets of fits are listed in 
Table \ref{aouv}. As the last column in the table shows, the differences between
the the spectral indices computed with and without the $2000-4000$\AA\ data are small,
especially for the high redshift quasars.
K-S tests on the sets of spectral indices determined with and without the $2000-4000$\AA\ 
data give probabilities of 7\% for the low and 0\% for the high redshift 
sample that they differ. 
The residuals from the fits to just the 5 bands around the `3000\AA\ bump'  are shown
in Figure \ref{plres}. For the low redshift quasars, the $2000-4000$\AA\ residuals are
systematically positive, suggesting either the presence of an extra component 
or that the continuum has an intrinsic curvature at low but not at high redshift.
This will be discussed further in \S \ref{l3000bump}, and for the analysis of continuum
shapes we use the optical/UV spectral indices obtained without the data between 2000 and 
4000\AA. 

\placetable{table18} 

\section{The distributions of optical/UV spectral indices}

The histograms in Figure \ref{hist_aouv} show the distributions 
of the $1285-5100$\AA\ spectral indices, $\alpha_{ouv}$ for the high and low redshift
samples.
The mean (median) spectral indices for the 15 high and 27 low redshift quasars and for
the 42 objects in the combined high and low redshift samples are $-0.32\pm0.07$ ($-0.29$), 
$-0.38\pm0.07$ ($-0.40$) and $-0.36\pm0.05$ ($-0.40$), respectively.
The latter is similar to the median determined for the set of 688 LBQS quasars 
(Francis \etal\ 1991), $-0.33$, and also to the average optical/UV spectral index
of $-0.33$ that Natali \etal\ (1998) find for a sample of 62 quasars, though they 
point out that the continuum is better described by a double power law, with a turnover
at 3000\AA. We examine the shape of the continuum around 3000\AA\ further in \S 
\ref{l3000bump}.
The mean values, medians and standard deviations of the distributions of optical/UV spectral
indices are summarized in Table \ref{astats}.
\placetable{table19} 

\subsection{Diversity of continuum shapes}

Even a cursory glance at the optical/UV continua and the distributions of spectral
indices reveals a broad diversity. 
For both samples, the values of $\alpha_{ouv}$ range from $\sim -1$ to $0.5$. The 
standard deviations about the means, $\sigma_{N-1}$, are $0.28$ and $0.35$, for the 
high and low redshift samples, respectively.
The interquartile ranges are from $-0.57$ to $-0.26$ 
for the low and $-0.51$ to $-0.22$ for the high redshift samples, so
one-half of the objects in each sample lie within a strip of width 0.3 about the median. 
This width is about 10 times larger than the measurement 
errors (0.04 and 0.03 for the low and high redshift samples, respectively | see 
Table \ref{aouv}), and thus indicates that the spread in optical/UV continuum 
shapes is real and not due solely to measurement errors. 


The broad range of optical/UV spectral indices in our samples is not unusual. 
Such diversity has been previously noted for many different
samples (\eg\ Richstone \& Schmidt 1980, Neugebauer \etal\ 1987, Cheng, Gaskell \& Koratkar 1991, 
Elvis 1992).  Elvis \etal\ (1994a) computed 
the mean and dispersion of the SEDs of the 47 low redshift `Atlas' AGN and
concluded that, while the mean well-represented the sample, with the continua of 
two-thirds of the quasars within a factor of 2-3 of it, the spread in continuum 
shapes was significant. And, Francis \etal\ (1991) plot a histogram (their Figure 1) of 
optical/UV spectral indices for 688 LBQS quasars which shows that they span a range from 
about $-1.5$ to 1.

Various explanations for the wide range of optical/UV continuum shapes have
been proposed, among these: that it may arise from measurement errors (Cheng et al. 1991 | though they discuss only the UV continuum slope); 
variability; different amounts of unknown, and thus not corrected for, intrinsic 
reddening (Cheng et al. 1991);  a range of shapes in the emitted continuum 
itself; or from combination of orientation and intrinsic reddening (Rowan-Robinson 1995).

As mentioned at the beginning of this section, the measurement errors are too small to 
account for the full range in our samples. To reduce the contribution of variability to 
the uncertainties, we have tried to obtain the data for the high $z$ quasars close in time 
and, as far as possible, to use quasi-simultaneous and simultaneous datasets for the 
low $z$ quasars (see \S\ref{lowzcorr} and Table \ref{simultan}). 
The observed range in spectral index cannot be explained by errors in the Galactic column 
densities, which are known to better than $\twid 10^{20}$ cm$^{-2}$ (Stark \etal\ 1992), 
and most to $\twid 1-3 \times 10^{19}$ cm$^{-2}$ (Elvis, Lockman \& Wilkes 1989,
Murphy \etal\ 1996). 

We have considered whether differing amounts of intrinsic extinction, applied to
similarly shaped emitted continua, might produce the observed distribution of optical/UV
continuum shapes. To account for the spread of $\Delta\alpha_{ouv} \approx 1$ in $1285-5100$\AA\ 
spectral index requires intrinsic reddenings up to $E_{B-V}$ = 0.11 or 0.25,
depending on which extinction law: Small Magellanic Cloud (SMC; Bouchet \etal\ 1985, 
Pr\'evot \etal\ 1984) or Galactic (MW; Savage and Mathis 1979, Rieke \& Lebofsky 1985);
is assumed (see the right hand panels, b and d, of Figure \ref{hist_aouv}).
The absence of the 10$\mu$m silicate feature and the non-universality of absorption at
2175\AA\ in quasar spectra argue for a dust composition in AGN unlike that in the Milky
Way and more similar to that in the SMC. Hence the SMC, rather than Milky Way, extinction 
curve may better represent the AGN one (Czerny \etal\ 1995). Assuming both the SMC 
and Milky Way extinction laws and gas-to-dust ratios ($4.45\times10^{22}$cm$^{-2}$ for the
SMC, Bouchet \etal\ 1985, Pr\'{e}vot \etal\ 1984 and $4.8\times10^{21}$ cm$^{-2}$ for the
MW, Savage \& Mathis 1979), we find that the color excesses needed to produce the observed 
range in spectral index correspond to intrinsic column densities of $4.9\times10^{21}$ 
cm$^{-2}$ (SMC) and $1.2\times 10^{21}$ cm$^{-2}$ (MW). 

These are about 100 times larger than the column densities of less than 
$5\times10^{19}$ cm$^{-2}$ which Laor \etal\ (1997) infer from fits to high S/N ROSAT PSPC 
spectra of 23 quasars from the BQS sample with $z<0.4$ and low Galactic column densities.
But, they are consistent with those for the RIXOS X-ray AGN ($0-4\times10^{21}$ cm$^{-2}$; 
Puchnarewicz \etal\ 1996).
Finally, they are about 10 times smaller than the intrinsic columns implied by fits to 
X-ray spectra of several high redshift, radio-loud quasars, three of which are in the high $z$ 
sample: Q$0014+813$ ($N_{H, int}$ = $57.8^{+22.2}_{-20.4} \times 10^{21}$ cm$^{-2}$ from ASCA; 
Reeves \etal\ 1997); Q$0636+680$ ($N_{H,int} = 20^{+10}_{-8}\times 10^{21}$ cm$^{-2}$, 
ROSAT PSPC; Fiore \etal\ 1998; and Q$2126-158$ ($N_{H,int} = 12.9^{+7.2}_{-3.8} \times 10^{21}$ 
cm$^{-2}$, ROSAT PSPC; Elvis \etal\ 1994b or $12.6^{6.6}_{-5.9}\times10^{21}$ cm$^{-2}$, 
ASCA; Reeves \etal\ 1997). However, while there is evidence that optical reddening and X-ray 
absorption are paired at low redshift (Puchnarewicz \etal\ 1996, Elvis \etal\ 1998), Elvis 
\etal\ (1998) find that at high redshift there is no such correlation. In line with their 
result, we find no correlation between intrinsic column density and optical/UV spectral 
index for the three high redshift radio-loud quasars with strong X-ray absorption
($\alpha_{ouv} = -0.14, -0.57$ and $-0.41$ for Q$0014+813$, Q$0636+680$ and Q$2126-158$). So
the gas-to-dust ratios in the intrinsic hydrogen columns at high redshift may differ
substantially from those at low redshift or in the SMC or Milky Way.
A better understanding of intrinsic absorption at high redshift is anticipated from XMM 
and Chandra.
%


The emitted optical/UV continuum might itself span a range of shapes. 
Thin disk spectra predicted for typical quasar masses and accretion rates 
are expected to show some curvature between 1285\AA\ and 5100\AA\ region, in which case 
the best-fitting power law slope within this spectral region would be flatter 
than the canonical $\nu^{1/3}$ dependence (Frank, King \& Raine 1995).
The predicted disk spectrum is also anisotropic, so the turnover depends on 
viewing angle. A range of orientations within the sample may cause some scatter
even if the black hole masses and accretion rates are identical.
In addition to changes caused by shifts in turnover frequency, more complex accretion
disk models which include scattering in the disk atmosphere (Czerny \& Elvis 1987), 
Compton scattering by a corona (Czerny \& Elvis 1987, Kurpiewski, Kuraszkiewicz \&  
Czerny 1997), and external X-ray irradiation (Matt, Fabian \& Ross 1993, Sincell \&  
Krolik 1997), predict a range of spectral indices down to $\sim -1$ (Malkan 1991), near the lower
end of the distribution, so accretion models can account for nearly the entire
range of continuum slopes.

The slope of the continuum generated by free-free emission depends on the temperature 
of the emitting clouds and can range from $\sim-0.6$ to $-0.2$, for T=$10^{5}$ to $10^{6}$K,
the range of expected temperatures in quasars (Barvainis 1993). This is not enough to account 
for the full distribution of optical/UV continuum shapes. 



\subsection{Continuum evolution?}
\label{no_evol}

The distributions of spectral indices within the high and low redshift samples, though
broad, are also similar, at a 96\% significance as given by a K-S test. The mean spectral
index is bluer for the high than for the low redshift sample, as would be expected for
continuous evolution of a single generation, but a $t$-test to compare the means indicates
the difference is not significant | the probability of randomly obtaining a larger 
$t$ is 60\%. Hence, our data do not give any strong evidence for evolution
of the optical/UV continua between $z>3$ and $z\sim0.1$. 

The similarity between the distributions of continuum shapes at high and low redshift 
is inconsistent with the evolution of only a single generation of quasars, 
whether the activity is continuous or recurrent, though the latter, since it allows for high
accretion rates at low redshift and less mass growth, may be reconcilable with the data.
The lack of significant evolution clearly favors models which involve the formation and evolution 
of multiple generations of quasars. Haiman \& Menou (1999) fit the observed luminosity function 
evolution with multiple generations of quasars in which the characteristic accretion rate 
increases with redshift, while Haehnelt, Natarajan \& Rees (1998) focus instead on evolution
of the mass in their model. Referring back to the dependence of the disk temperature 
on mass and accretion rate, evolution of both $\dot{M}/\dot{M}_{Edd}$ and $M$, at roughly
the same rate, is implied by the lack of significant evolution in the continuum shape.

In this simple interpretation, we have assumed first, that the high and low redshift quasars 
lie at ends of a single continuum and, second, that the structures of 
the accretion disks (thin, slim, thick) are the same at high and low redshift (luminosity).
The former assumption appears valid. The 27 low redshift objects have absolute magnitudes
large enough to be classified as quasars, and even quasars and the lower luminosity
Seyfert 1 galaxies are linked, as implied by the smooth continuity between
their luminosity functions (K\"ohler \etal\ 1997). The greater concern is probably at
high redshift, where several of our sample quasars are among the most luminous objects
in the universe. To move away from these extremes, we are beginning to survey the optical/UV 
continuum shapes of a larger, fainter sample of $z>3$ quasars.
Fits to some of these high redshift/luminosity quasars imply that they may be 
undergoing near- or super-Eddington accretion (Kuhn \etal\ 1995, Bechtold \etal\ 1994),
which is inconsistent with thin disks. A comparison between the SEDs presented 
here and the predictions of various accretion models is planned.

\section{Comparison with previous results}

Though we find no significant difference between the optical/UV continuum shapes at
low and high redshift (or low and high luminosity, given the strong $z-L$ correlation 
in the combined samples), prior studies have indicated that the shape of the UV continuum 
does evolve (O'Brien, Gondhalekar \& Wilson 1988, Cheng et al. 1991) and 
that the optical/UV continuum is luminosity dependent (Wandel 1987, Mushotzky \& Wandel 1989, 
Zheng \& Malkan 1993). Our samples were not selected for a comprehensive correlation analysis: 
there are too few objects, and these are not evenly distributed in redshift and luminosity.
We have looked for evidence of these previously reported correlations in our dataset, however, 
and this section discusses the outcome of these searches.


\subsection{$\alpha_{uv}$ vs. redshift, luminosity}

A trend toward harder (flatter) UV spectral indices at higher redshifts was reported both
by Cheng, Gaskell \& Koratkar (1991) and O'Brien, Gondhalekar \& Wilson (1988).
Multivariate linear regression fits to the UV spectral index as a function of both redshift,
log$(1+z)$, and UV luminosity, $L$(1450\AA), showed the primary correlation to be
between UV spectral index and redshift.

We looked in our samples for similar correlations between the UV continuum shape, log$(1+z)$ and 
log $L$(1460\AA), using spectral indices measured between two pairs of the narrow-band monochromatic 
luminosities listed in Tables \ref{hiz_cfluxes} and \ref{lowz_cfluxes}: (1285\AA, 2200\AA) and 
(1285\AA,1460\AA); to characterize the UV continuum shapes. These and other point-to-point spectral 
indices are given in Table \ref{uvao}.
The 1285\AA $-$ 2200\AA\ spectral index matches the definition used by Cheng et al.,
while the 1285$-$1460\AA\ one is deliberately cut off at a wavelength short enough
that there is no possibility of continuum contamination from the 3000\AA\ bump, the rationale 
of O'Brien et al. in measuring the UV spectral index from $\lambda$Ly$\alpha$ to 1900\AA.
(We had not defined a narrow-band window between 1460\AA\ and 2200\AA, so cannot readily approximate
their measure any better). 
Although the $1285-1460$\AA\ spectral index represents the true continuum better than
the $1285-2200$\AA\ one, its smaller baseline yields larger errors and increased scatter, 
especially for the low redshift sample where the UV data are from the IUE and, for some objects, 
are relatively noisy. 

We found positive correlations between the $1285-2200$\AA\ spectral index and both redshift and
luminosity (Figure \ref{uvslopecorrs}). The linear regression fits and two-tailed significances of the correlations,
\ie, one minus the probability of a chance correlation as given by the Kendall-tau test,
are summarized in Table \ref{uvcorr}. When the UV continuum shape is represented by the $1285-1460$\AA\
spectral index instead, the trends persist (the results of the linear regression fits
are similar, albeit with larger errors), but the correlations disappear (significance levels
$<82$\%). 
From this analysis, it is difficult to determine whether the correlations are stronger for
the $1285-2200$\AA\ spectral indices because of their smaller errors and scatter, or 
instead reflect a luminosity or redshift dependence of the 3000\AA\ bump,
which may be contaminating the continuum luminosity at 2200\AA\ (see section \ref{l3000bump}); 
but the similarity of the trends and linear fits for both measures of UV spectral index favors the
former explanation.
\placetable{table21}  

To compare our results directly with O'Brien \etal\ and Cheng \etal\ requires 
re-computing the luminosities under the same cosmological assumptions as they made. 
Table \ref{uvcorr} lists the parameters of the best-fitting lines alongside those that they determined.
Our results from fitting spectral index against luminosity are in good agreement with 
theirs, though the slope of the spectral index vs. redshift relation that we derive is shallower, but 
still consistent within 1$\sigma$, than their result based on simultaneous fits of 
spectral index against luminosity and redshift.  We tried a multivariate fit, but the 
strong luminosity-redshift correlation in our combined samples yields results very 
different from those of either group.

When our samples are broken down by redshift, the trend of increasing spectral index with
luminosity or redshift remains, though is weaker. Only for the low redshift sample is the
correlation between $1285-2200$\AA\ spectral index and luminosity marginally significant,
95\% by the Kendall-tau test.  Neither the trend of UV spectral index against UV luminosity
within the high redshift sample nor of UV spectral index against redshift within each sample is
significant: the Kendall-tau significance level is $<84$\% in each case.
We investigated whether the correlation between $\alpha$(1285\AA,2200\AA) and luminosity holds 
for both UV and optical luminosities, and find it is stronger for the UV than for optical ones. 
A similar result is found for the optical/UV 
continuum shape (discussed in the next section): it is correlated with UV but not with the optical 
luminosity.

In sum, we find a positive correlation between the UV spectral index and UV luminosity (or
redshift) in our combined high and low redshift samples, in good agreement with what Cheng et al. 
and O'Brien et al. had previously found for objects covering a similar range of redshifts. The 
strong luminosity-redshift correlation in our combined samples does not allow us
to test their conclusion that the primary correlation is with redshift. The trend
towards harder UV continua for higher luminosity or redshift exists for both measures
of UV slope, but its increased significance for the $1285-2200$\AA\ slope raises some doubt
as to whether the trend is intrinsic or instead due to a luminosity or redshift dependent contribution of
the 3000\AA\ bump at 2200\AA. Better fits to the UV continua would be needed to make stronger 
conclusions.


\subsection{$\alpha_{ouv}$ vs. luminosity}

As discussed in \S \ref{no_evol}, we find no evidence for significant evolution of the
optical/UV continuum shape. Given the strong redshift-luminosity correlation present in
our combined samples, this would imply no significant correlation between the
optical/UV continuum shape and luminosity, consistent with the results that Neugebauer \etal\ (1987)
obtain from near-IR and optical observations of 105 PG quasars.
Other groups, however, have reported trends between the rest-frame optical/UV continuum shapes
and luminosities. Wandel (1987) and Mushotzky \& Wandel (1989) found a correlation
between the optical spectral index, measured between 4200 and 7500\AA, and the rest-frame
luminosity at 4200\AA. And Zheng \& Malkan (1993) found a reasonably significant anti-correlation 
(correlation coefficient, $r=-0.59$) between the $1250-6500$\AA\ continuum slope and
$M_{4800\AA}$ for 145 QSOs and Seyfert 1 galaxies. 
We checked for correlations between optical luminosity, taken at $\lambda_{rest} = 4200$\AA\ 
and 4750\AA, and the $\alpha_{ouv}$ listed in column 7 of Table \ref{aouv} but found
none for the combined samples or the high or low redshift one individually (Figure \ref{aouvcorr}).
The results are given in Table \ref{uvcorr}.

Although the spectrum of starlight from the host galaxy peaks in the near-IR, its
contribution to the optical continuum, especially for the low luminosity AGN, may
be non-negligible (Elvis \etal\ 1994a). An increasing fraction of the starlight to AGN continuum
with decreasing luminosity could account for the reported correlations. Wandel (1987) and
Mushotzky \& Wandel (1989) considered the role of host-galaxy starlight in the correlation
and acknowledged that it may explain the correlation for luminosities, $L<10^{45}$ erg s$^{-1}$, 
though to contribute
significantly to the continua of higher luminosity quasars would require an implausibly high
starlight flux (Mushotzky \& Wandel 1989). Zheng \& Malkan (1993) also questioned whether the
contribution of host-galaxy starlight at 6500\AA\ could account for the observed
correlation; but, pointing to the persistence of the correlation when only the optical
or UV spectral index or the brightest AGN (with $M<-23$) are used, concluded that, while
important, the host galaxy contribution does not fully account for the correlation.
To gauge the importance of host galaxy starlight to the optical slope$-$luminosity
correlations, we measured the spectral indices from 1250\AA$-6500$\AA\ and absolute
magnitudes at 4800\AA\ of 46 AGN in the Atlas, with and without the host galaxy contribution
subtracted\footnote{To improve statistics, the full `Atlas' sample, and not just the subset
of 27 studied here, was used to examine the host galaxy contribution. But
since we have not computed narrow band luminosities for all these, we used the TIGER
software package written by J. McDowell (Wilkes \& McDowell 1995) to determine the
$1250-6500$\AA\ spectral indices and M$_{4800\AA}$.}.
As Figure \ref{hostgal} illustrates, the host galaxy starlight is most important at low
luminosity and does effect a slope-luminosity correlation. When the starlight is not subtracted 
the spectral index and magnitude are weakly anti-correlated (Spearman correlation coefficient,
$r_{s}=-0.37$, and the probability of a chance correlation, $p_{s}=3$\%).
With the host galaxy starlight spectrum subtracted, however, there is no evidence for a correlation 
between spectral index and absolute magnitude ($r_{s}=0.13$ and $p_{s}$=37\%). 
%

We checked whether the optical/UV continuum slope was also uncorrelated with UV luminosity, but
found instead a positive correlation (99.5\% significance) between $\alpha_{ouv}$ and the UV 
luminosity measured at 1285\AA\ (see figure \ref{aouvcorr}; there
is also a correlation when UV luminosity is measured at 1460\AA, but it is weaker). 
Though present for the combined samples (97\% significance), it is strongest within the low redshift 
one (99.5\%) and absent in the high redshift one, echoing the $\alpha_{uv}-$ UV luminosity correlation 
reported in the previous section. 
Therefore, while our analysis shows that host galaxy starlight can effect a correlation between 
$\alpha_{ouv}$ and optical luminosity, and this explains the absence of a correlation between
optical/UV slope and optical luminosity in our host-galaxy corrected data, it does not explain
why we should find a correlation between $\alpha_{ouv}$ and UV luminosity. A similar trend for 
variable quasars to have bluer continua in their higher luminosity states (Giveon \etal\ 1999)
suggests an intrinsic origin, but further investigation is clearly needed.

\section{Comparison continued | Continuum deviations from a single power law}

Whereas the single power law fits to the optical/UV continuum energy distributions 
are adequate in most cases, a closer look at several continua (\eg\ Q$0055-269$) 
shows that they steepen between the optical and UV, consistent with prior
consensus that the overall optical/UV shape is more complex than a single power law
(Neugebauer \etal\ 1987). This section examines the UV continuum shape shortwards of 
Ly$\alpha$ and around 3000\AA.

\subsection{UV turnover}
\label{uvturnover}

In the extreme ultraviolet (EUV), the continuum must turnover to match up with the X-ray flux. 
Quasar accretion 
disk models predict a turnover in the EUV, so an accurate determination of the continuum 
shape in that region is important for testing these. Unfortunately, the EUV data are 
sparse and, for high redshift quasars, complicated by Ly$\alpha$ forest absorption. However, 
HST observations of over 100 quasars with $z>0.33$ have provided valuable information on the 
rest-frame EUV continuum (Zheng \etal\ 1997).  The composite spectrum formed by these data,
after they were statistically corrected for Ly$\alpha$ forest absorption, shows a turndown 
at 1050\AA, believed to be intrinsic (Zheng \etal\ 1997).  

Although we do not have data at $\lambda_{rest}\sim1050$\AA, we can use the 1115\AA\ and 
1160\AA\ fluxes (Table \ref{uvfluxes}) to investigate the continuum behavior at $\lambda < 
\lambda$ Ly$\alpha$. Spectral indices were computed from 1115\AA$-1285$\AA\ and 1160\AA$-1285$\AA,
1285\AA\ being the closest narrow band window to Ly$\alpha$ and well representative of
the continuum level around the line. These are given in Table \ref{uvao}. 
Average spectral indices on either side of Ly$\alpha$ and statistics on the inflection
at 1285\AA\ are listed in Table \ref{dauv}.  The first column indicates which sample is 
considered, the second lists which spectral indices are used in computing the inflection 
at 1285\AA, and the third through sixth columns give the number of objects in the sample, 
the average slope change, the scatter ($\sigma_{N-1}$) and the error in the mean.  
In the final three columns, the numbers of objects showing, at a level 
greater than 1$\sigma$, a turn over, no change or a turn up, are tabulated.
The average slope change is, in general, positive (a turnover), except for the average 
change in slope from 1160\AA\ to 1285\AA\ to 2200\AA\ for the low redshift and combined samples. 
The turn up for these might be explained if the 3000\AA\ bump influences the luminosity 
measured at 2200\AA\ for the low redshift AGN. Since there are over 3 times as many low 
redshift AGN as high redshift ones in the combined sample of objects for which $1160-1285$\AA\ 
and $1285-2200$\AA\ spectral indices can both be measured, the overall average is heavily 
weighted by the low redshift sample. 
Among the average values that are positive, several are within $1\sigma$ of 
the mean and hence consistent with no slope change. The numbers in the last three columns 
indicate that for most of the sample objects, there is no compelling evidence for a turnover 
out to 1160\AA\ or 1115\AA, and thereby suggest that it must occur at shorter wavelengths, 
consistent with the HST results of Zheng \etal\ (1997).

\placetable{table22} 

\subsection{The 3000 \AA\ region | excess or continuum turnover?}
\label{l3000bump}

\subsubsection{3000\AA\ excess}

There is one clear difference between the high and low z samples. The residuals from a 
single power law fit (Figure \ref{plres}) show a systematic excess between 2000 and 4000\AA\
($\log\nu = 14.88 - 15.18$) 
for the low redshift sample, which is weak or absent in the high redshift one. This region
is that of the `small bump' made up of blended FeII emission ($\lambda\sim2000-3000$\AA) 
and Balmer continuum ($\lambda\sim2500-3800$\AA; Wills, Netzer \& Wills 1985). This result 
is robust; although the data coverage is somewhat sparse,
we have tried multiple ways of quantifying the difference and they all agree.
%

We use four parameters to gauge the strength of the 3000 \AA\ excess. 
Two are the residuals at 2500\AA\ and 2660\AA. Recall that the 2500\AA\ 
window is centered on a strong FeII multiplet (Joly 1993), and the 2660\AA\ one is
the only other window of the four between 2000 and 4000\AA\ within which luminosities 
could be directly measured for a significant part of the sample.
Another measure is the luminosity of the excess relative to the continuum between 
2200\AA\ and 3023\AA, 
\begin{equation}
{L(excess) \over L(cont)} = {{\int_{2200\AA}^{3023\AA} {(F_{\nu,total} - F_{\nu,fit})}
d\nu} \over {\int_{2200\AA}^{3023\AA}{ {F_{\nu,fit}} d\nu}}}. \label{ewidth1} 
\end{equation}
where the continuum level, $F_{\nu,fit}$, is given by the power law fit through 
all bands between 1285\AA\ and 5100\AA\ excluding those four between 2000\AA\ and 4000\AA.
Luminosity excesses were computed by trapezoidal integration over both 
directly measured and the interpolated narrow band fluxes, $F_{\nu,total}$, at 2200\AA, 
2500\AA, 2660\AA\ and 3023\AA.
Finally, the fourth parameter is the $2200-3023$\AA\ equivalent width, which was computed
as follows:
%
\begin{equation} {EW(2200-3023\AA)} = {{\int_{2200\AA}^{3023\AA} {(F_{\nu,total} -
F_{\nu,fit})}/F_{\nu,fit} d\nu}}. \label{ewidth2} 
\end{equation}
Errors on the residuals are simply taken to be the errors on
the measured narrow band fluxes (Table \ref{hiz_cfluxes} and \ref{lowz_cfluxes}), and
errors on the fractional luminosity excesses and equivalent widths are computed by 
propagating the sources of error in equations \ref{ewidth1} and \ref{ewidth2}. 
The 2500\AA\ and 2660\AA\
residuals, ratios of excess to total luminosity and the EW($2200-3023$\AA) are 
listed in Table \ref{fe2}. The last column indicates the quality of the four narrow
band average luminosities from 2500\AA\ to 3023\AA\ | whether they are computed directly (3), 
interpolated (2) or if they do not represent what is believed to be the true flux level, 
regardless of how they are determined (1). The values that are listed in parentheses in the 
table were determined solely from interpolated fluxes or from one or more discrepant points, 
and were not used to compute the statistics, however the results would not be significantly 
different if they were.

\placetable{table23}

The distributions of the residuals, luminosity ratios and equivalent widths are plotted 
in Figure \ref{fe2hist}. 
The K-S probabilities that the 2500\AA\ residuals, 2660\AA\ residuals, $2200-3023$\AA\ excesses 
and equivalent widths of the high and low redshift samples differ are 99.95\%, 99.97\%, 99.75\% 
and 99.75\%, respectively.  Equivalent widths for the high redshift sample range from $-10$\AA\ 
to 300\AA, while for the low redshift one, they are 100\AA$-700$\AA. The average equivalent 
width for the high redshift sample is 100\AA, but the median is 35\AA. These average and median 
values are 3 and 8 times lower than those for the low redshift sample. 

Though the 3000\AA\ excesses for most of the high redshift quasars are consistent with
zero, five show some evidence for excess emission. Q$0055-269$ has the clearest indication of 
a 3000\AA\ bump. Both its 2500\AA\ and 2660\AA\ residuals were
directly measured, and they and the fractional luminosity excess and equivalent width are 
significant to greater than 3$\sigma$. 
Q$0347-383$, Q$2204-408$ and Q$1935-692$ also show evidence for a 3000\AA\ bump, 
as does Q$1208+101$, though its residuals and equivalent width had to be determined from 
interpolated fluxes. 
The $2200-3023$\AA\ equivalent widths of these five `3000\AA\ excess' quasars are consistent 
with or greater than the $2000-3000$\AA\ equivalent width of 130\AA$-157$\AA\ (depending
on the assumed line profile) that Thompson, Hill \& Elston (1999) measured from a composite 
spectrum of six quasars with $<z>=3.35$. They also lie within the range covered by our
$z\sim0.1$ sample, though are all less than the average for it ($\sim300$\AA). 
It is interesting that two of our $z>3$ sample quasars which are considered to be strong 
iron emitters based on their FeII(opt)/H$\beta$ ratios: Q$0014+813$ and Q$0636+680$
(Elston, Thompson \& Hill 1994, but see Murayama \etal\ 1996); show little or no evidence 
for a 3000\AA\ excess. Direct measurements of the UV FeII multiplets in the J-band spectra, 
and comparison of these to the 3000\AA\ bump and optical FeII strengths would shed some
light on this apparent discrepancy.

These five $z>3$ quasars notwithstanding, we find that the 3000\AA\ excess is on the whole
much weaker (or absent) in the SEDs of the $z>3$ sample quasars than in those of the $z\sim0.1$ 
ones. If it is primarily blended FeII emission, and the iron abundance does not evolve 
significantly as Thompson, Hill \& Elston (1999) concluded, then the explanation must lie 
with other parameters which affect the strength of the FeII emission, such as its
energy source. 

There is abundant evidence for a link between the strength of FeII emission and the X-ray properties, 
though the origin of this is uncertain. In general, the AGN with flat X-ray spectra or strong X-ray 
emission tend not to have strong optical FeII (Wang, Brinkmann \& Bergeron 1996, Laor \etal\ 1997, 
Lawrence \etal\ 1997, Wilkes \etal\ 1999).
Green et al. (1995), however, found a positive 
correlation between the UV FeII and X-ray strengths: the LBQS quasars with the strongest UV 
FeII emission have soft X-ray (ROSAT) fluxes 2 times greater than the sample average. 
The discrepant results and inconsistency of most with the expectations of standard photoionization 
models (\eg\ Krolik \& Kallman 1988) are confusing, and other mechanisms such as collisional excitation 
may play a large role in FeII production (Kwan \etal\ 1995, Collin \& Joly 2001). 
Kwan \etal\ (1995) suggest that FeII may be collisionally excited within a well defined region 
of the accretion disk, and this dependence of FeII on the accretion flow has been invoked to 
explain the strong FeII emission seen in the spectra of narrow line Seyfert 1s and $z>4$ quasars,
both of which are assumed to be young and accreting at high rates (Mathur 2000). 
The strong evolution that we find in the 3000\AA\ excess does not 
support this analogy between NLSy1 and high redshift quasars, though a more rigorous test, 
using measures of UV and optical FeII features, would be interesting, especially given that 
our low redshift sample contains several objects classified as NLSy1s, including the prototype, 
IZw1. 

We looked in our data for any trends between the strength of the 3000\AA\ excess and the 
quasars' X-ray properties. All the quasars in our low redshift sample have X-ray data 
from the Einstein IPC, and 8 of the 15 in the high redshift one were detected with the 
ROSAT PSPC (but these include only 1 of the 5 high redshift quasars with a relatively 
strong 3000\AA\ excess). We found no significant correlations between the 3000\AA\ bump 
strength (any of the 4 measures) and the rest-frame 2500\AA\ to 2keV flux ratio, 
$\alpha_{ox}$ (see Tables \ref{hizsample} and \ref{lowzsample}) or X-ray spectral index, 
$\alpha_{x}$ (case A in Table 1 of Bechtold \etal\ 1994 and Table 2 of the `Atlas'), neither 
in the combined high and low redshift samples nor in each individually. 
Though the number of objects considered is small for such statistical tests, the lack 
of any trend between the strength of the 3000\AA\ excess and X-ray emission is 
puzzling, especially given that the samples are well separated in $\alpha_{ox}$. 

In sum, while the difference in strength of the 3000\AA\ excess between the high and low 
redshift samples is clear, its origin remains uncertain. First, we have assumed, based on
models at low redshift, that it is primarily due to FeII, though other factors could 
contribute. A comparison of its strength with direct measurements of UV and optical FeII 
multiplets, which are apparent in the J and K spectra of some of the high redshift objects 
and in the UV and optical for the low redshift ones (Wilkes \etal\ 1999), 
would help resolve this issue. Second, the models to explain FeII production in quasars
are themselves uncertain. However, the evidence for an X-ray-FeII link argues for the collection
of X-ray spectra for all the high redshift sample quasars; a comparison of these with 
the 3000\AA\ excess and UV and optical FeII strengths could yield some clues to the evolution.

\subsubsection{Optical continuum turnover}

Recently, several groups have reported evidence for an intrinsic turnover in the continuum
at 3000\AA\ (Zheng \& Malkan 1993, Natali \etal\ 1998, Carballo \etal\ 1999). 
The average spectral indices measured on either side of $\sim3000$\AA\ differ by 
$\sim 0.8-0.9$. The immediate suspicion is that the 3000\AA\ FeII and BaC emission bump
is responsible for this turnover. However, Zheng \& Malkan, using Malkan's (1988)  estimate 
that emission from the small bump contributes about 20\% of the total flux at 3100\AA, 
determine that, while consideration of it could reduce the intrinsic slope change to 0.45, 
it would not eliminate it.  Similarly, Carballo \etal\ have examined the role of FeII and BaC 
emission, but conclude that it alone cannot explain the turnover.

To test these results and assay the role of the small bump in producing the turnover, we have 
computed point-to-point spectral
indices from the SEDs of the high and low redshift quasars. Spectral indices computed from
$1285-3023$\AA\ and $3023-5100$\AA\ have been used to measure the continuum slopes below
(UV) and above (optical) 3000\AA, respectively. These closely match Zheng \& 
Malkan's (1993) definitions of $\alpha_{uv}$ and $\alpha_{opt}$. To move away from the
influence of the 3000\AA\ bump we also measured the UV spectral index from 1285\AA\ to 
1460\AA\ and 1285\AA\ to 2200\AA, and the optical one starting from 4200\AA\ rather than 3023\AA.  
Table \ref{uvao} lists the spectral indices, and Table \ref{optstats} lists the average
spectral index, scatter and standard deviation in the mean for the high and low redshift 
and the combined samples in addition to the results of statistical tests ($t$-tests and K-S tests)
to compare the various optical and UV slopes.  
As the table shows, there is a clear slope change when the spectral 
indices are measured to 3023\AA: the average UV and optical spectral indices are 
$-0.68\pm0.07$ and $0.18\pm0.07$, respectively. Both the $t$-test and K-S tests strongly indicate
a difference in optical and UV slopes when measured to 3023\AA.
\placetable{table24} 

When the spectral indices are computed without using data within the 2000-4000\AA\
small bump region, a difference between those computed shortwards and longwards of 3000\AA\
still exists, though is less significant. 
The average UV spectral indices are $-0.61\pm0.07$ (1285-2200\AA) and $-0.33\pm0.18$ (1285-1460\AA);
both redder than the average optical (4200-5100\AA) spectral index of $-0.09\pm0.13$.  
But these averages are dominated by the low redshift sample, for which the slope change appears
to be stronger. The statistics suggest that the 3000\AA\ excess is responsible for the 
observed optical/UV inflection, but they do not rule out an intrinsic continuum turnover. 
To discern the relative contributions of blended line and continuum emission to the 3000\AA\ bump 
would require a thorough treatment of the FeII+BaC emission and better measures of the spectral 
indices, especially over short baselines.

\section{Conclusions}

To look for signatures of physical evolution in quasars, which are expected given
their strong statistical evolution, we have constructed the rest-frame optical/UV 
spectral energy distributions of 15 $z>3$ quasars, using near-IR spectroscopy, 
optical spectrophotometry, and near-IR and optical photometry. We compare these
with the SEDs of the set of 27 low redshift ($z\sim0.1$) quasars from the `Atlas' (Elvis
\etal\ 1994a) whose luminosities with respect to $L^{*}$ lie within the same range
spanned by the high redshift sample quasars.
The SEDs of the $z>3$ quasars are presented, and their shapes are characterized 
by single power law fits to the average continuum fluxes measured within line-free
narrow windows between 1285\AA\ and 5100\AA. 
We list the conclusions drawn from an examination and comparison of the 
distributions of the optical/UV spectral indices of the high and low redshift samples.

1. {\em The optical/UV continuum shapes at high and low redshift span a broad range.}
The distributions of spectral indices for both the high and low redshift samples
are broad, spanning a full range of $\sim1$ in spectral index.
In both cases the interquartile spread is $\sim0.30$, too large to be explained
by measurement errors. It is also unlikely that variability or uncertainties in Galactic
extinction can account for it.  Intrinsic reddening and the
diversity in the shapes of the emitted spectra produced by accretion disk models may both
contribute to this spread, as Rowan-Robinson (1995) concluded in an earlier study. 

2. {\em The shapes of the optical/UV continua at high and low redshift do not
differ significantly.} 
Half of the high redshift sample objects have $\alpha_{ouv}$ between 
$-0.51$ to $-0.22$, and half of the low redshift ones have $\alpha_{ouv}$ 
between $-0.57$ to $-0.26$. 
The mean (median) spectral indices for the high and low redshift samples are 
$-0.32$($-0.29$) and $-0.38$ ($-0.40$), respectively. Though these are bluer for the 
high redshift sample, the difference in the means is significant to only 60\%, as measured
by the $t$-test.
A K-S test indicates that the spectral index distributions are similar to a high
confidence level, 96\%. This similarity disfavors the hypothesis that the observed luminosity
evolution derives from the continuous long-timescale evolution of a single 
generation of quasars, and instead is easier to reconcile with models which invoke multiple 
generations.

3. {\em The 3000\AA\ excess is stronger in the low redshift sample than in the high redshift
one, where it is generally weak or absent.}
While there is little evidence for evolution in the optical/UV continuum slopes, the SEDs
of the high and low redshift samples do differ significantly in the 3000\AA\ small bump
region. 
Residuals from the single power law fits, measured at 2500\AA, 2660\AA, and between
2200\AA\ and 3023\AA, are all generally stronger at low redshift than at high. 
The K-S significance of this result is $>99$\% for all four measures. Assuming the 3000\AA\
excess is primarily blended FeII emission, the relative weakness or absence of the 3000\AA\ 
bump at high redshift indicates either a smaller abundance of iron or differences in other
parameters affecting FeII strength. Recent results of Thompson \etal\ show no change in the 
UV FeII strength from $z\sim4$ to $z\sim0.88$, though they did not compare with a set of lower 
redshift AGN. The $z\sim0.1$ `Atlas' AGN could have greater iron abundances. A better explanation,
though, might focus on the shape of the photoionizing soft X-ray continuum, which is relatively 
strong in the `Atlas' AGN, since these were selected based on counts measured with the 
{\em Einstein} Observatory, or on other parameters which affect the FeII strength.  
Further work is needed to determine the origin of the 
differences between the $\sim3000$\AA\ continua of the high and low redshift quasars.
The planned emission line study, which will involve fitting the individual FeII complexes, 
should enable a more thorough analysis of the 3000\AA\ excess. To look for links with the 
X-ray properties, X-ray spectra are needed for all the high $z$ sample quasars. 

We looked in our dataset for previously reported correlations between UV spectral index
and redshift or luminosity and between optical/UV spectral index and luminosity.

4. {\em The UV spectral indices in our dataset become flatter (hotter) with increasing
luminosity or redshift, consistent with previous results.} The correlations are significant
for $\alpha$(1285\AA, 2200\AA) but much weaker for $\alpha$(1285\AA, 1460\AA); either
the large errors and scatter in the spectral indices measured over such a short baseline 
dissolve any correlation which may be present, or 
FeII and BaC emission affect the 2200\AA\ flux in which case the the correlation is 
not intrinsic to the continuum but reflects the stronger 3000\AA\ excess at low than 
at high redshifts (or luminosities, given the strong redshift-luminosity correlation in the
combined samples).

5. {\em The optical/UV spectral index, as measured by a power law fit, is correlated
with UV, but not optical, luminosity for the low redshift sample.}
Starlight from the the host galaxy does contribute to the red optical flux (Elvis \etal\ 1994a)
and must be considered. However, the correlation with UV luminosity exists even in our data 
where the host galaxy has been carefully removed from the SEDs of the low redshift quasars. 
It may be related to the energy generation mechanism, and we plan to investigate its origin more
carefully. 

The primary goal of this paper is to present the spectral energy distributions of 
the 15 quasars in the high redshift sample and contrast them with the SEDs of a 
comparison sample at low redshift. Future work will concentrate on fitting these with 
accretion disk and other models and analyzing the emission line properties. 
The samples discussed here suffered from (1) the strong redshift-luminosity correlation 
which resulted from the luminosity matching selection criteria, and (2) a bias to the bright end 
of the luminosity function (several of the high redshift quasars were at one time considered the 
most luminous objects in the universe) which arose from the need for IR spectroscopy of these.
We are beginning a survey of the continuum shapes of a set of $z>3$ radio loud quasars 
which should extend the present sample by 2-3 magnitudes. This will have the advantage of 
including less extreme quasars and enabling us to begin to disentangle luminosity from 
redshift trends.

\acknowledgements

Since this project involved many observing runs, using a range of different telescopes and
instruments, it relied on the help of many people. We thank M. and G. Rieke for 
setting up and helping us obtain the observations with Fspec. G. Rieke and R. Cutri helped
O.P.K. obtain the IR bolometer and photometer observations, and R. Cutri and S. Kenyon 
obtained the IR photometry that is presented here for 6 of the quasars.
We are grateful also to D. Joyce for his help with the CRSP observations at KPNO.  
P. Berlind obtained the optical spectrophotometry of Q$0956+122$ with FAST at the FLWO 1.5-m. 
We gratefully acknowledge the assistance of the staffs at the FLWO, MMTO, KPNO and CTIO. 
The Ohio State Infrared Imager/Spectrometer, OSIRIS, was built using funds from NSF awards 
AST 90-161112 and AST 92-18449 by the OSU Astronomical Instrumentation Facility.
We also thank those who helped with the data reduction.
The VLA data were reduced by C. Fassnacht.
K. Kearns and M. Davisson reduced some of the optical spectrophotometry presented here.
The Fspec data were reduced up to the stage of flux calibration using IRAF
tasks and scripts written at University of Arizona, Steward Observatory, by M. Rieke, C.
Engelbracht, L. Shier and D. Williams. 
B. Wilkes and J. McDowell were consulted throughout the course of the project
and we thank them for many helpful discussions key to its progress. Finally, we thank
the anonymous referee for suggestions which helped to improve the presentation.
O.P.K. acknowledges support from a NASA Graduate Student Researchers Program grant 
during a part of this work. M.E. acknowledges NASA grants NAGW-2201 
and NAG5-6078, and J.B. acknowledges support from NASA ADP NAG5-3691 and NSF AST-9617060.

\include{references}
\clearpage

\begin{figure}
\figcaption[fig3.eps]{Optical/UV spectral energy distributions of the
15 high redshift sample quasars. The optical 
spectrophotometry and IR spectra, scaled by the photometric data as
described in \S \ref{hizcorr}, are plotted, together with the average luminosities computed 
within each of the 9 narrow bands between 1285\AA\ and 5100\AA.
The symbols indicate how well the computed luminosity is believed to represent the continuum: 
solid squares mark the luminosities computed directly from the data and well representative
of the continuum level (Q=3 in Tables \ref{hiz_cfluxes} and \ref{lowz_cfluxes}); 
empty squares mark those which, though interpolated, seem to represent the continuum well
(Q=2); and x's mark those that are not well representative of the continuum (Q=1) and those 
determined at 2500\AA, which was not defined as a `continuum' window. 
At wavelengths shorter than Ly $\alpha$, the corrected continuum is plotted 
over the spectrophotometric data with a thicker line, and average fluxes
at 1115\AA\ and 1160\AA, computed for the corrected and uncorrected data, are
plotted (open circles for corrected continua, x's for uncorrected). 
The power law fits made through the 5 narrow band fluxes on either side of the $2000-4000$\AA\ 
small bump region, are plotted as dashed lines. Note that the corrected spectra nearly reach the 
extrapolated power law for some objects (Q$0000-263$, Q$0420-388$, Q$0636+680$) but fall short 
for others (Q$0014+813$, Q$1208+101$, Q$1946+7658$ and Q$2126-158$), as discussed in \S \ref{uvturnover}.
For comparison, the average radio-quiet or radio-loud low redshift spectral energy distribution 
(Elvis \etal\ 1994), normalized to match the quasar's 1460\AA\ luminosity, is overplotted 
(dotted line). 
\label{seds}}
\end{figure}
\begin{figure}
\figcaption[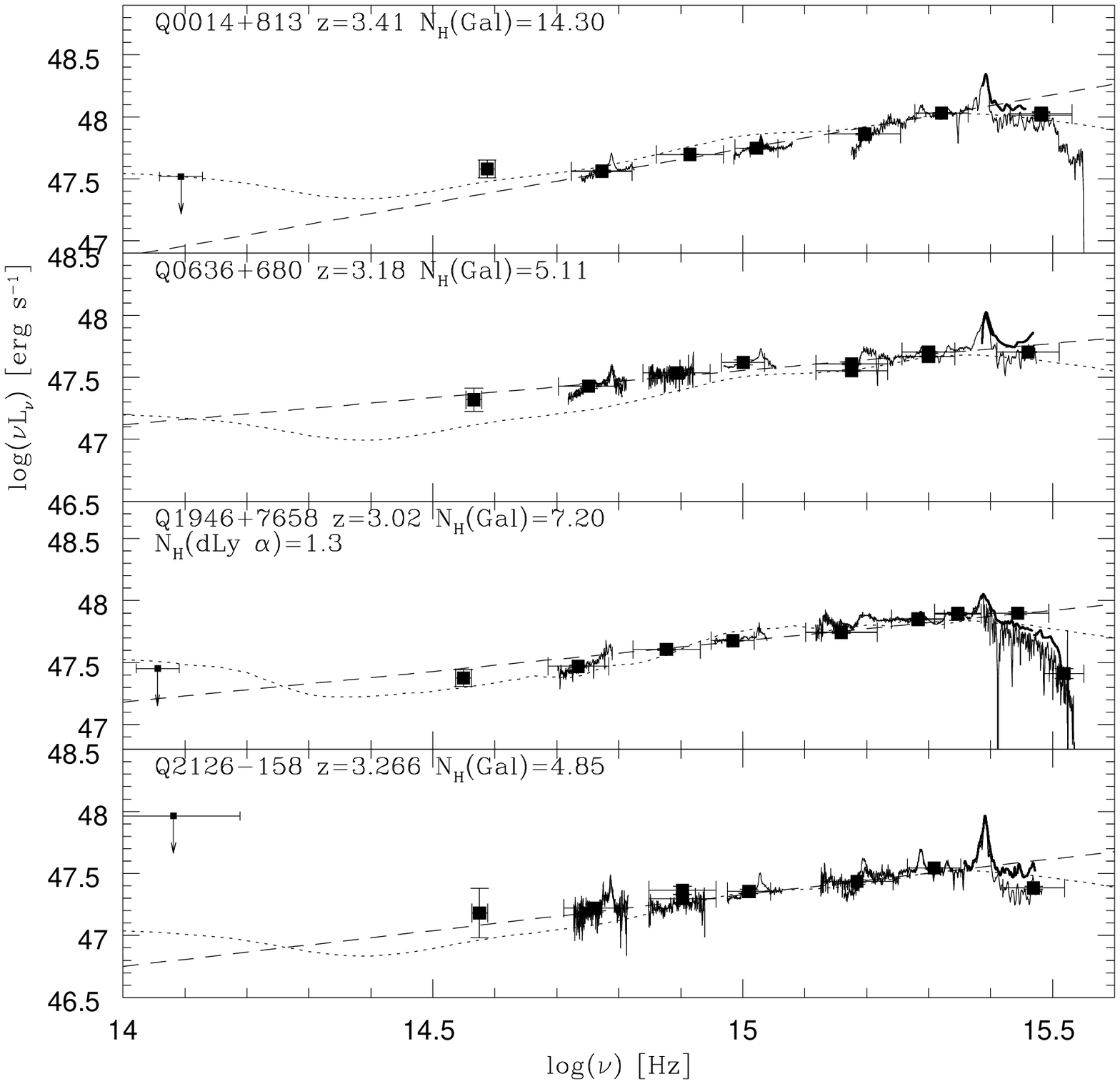]{Spectral energy distributions for 4 
high redshift sample quasars detected at L$^{\prime}$, 2 of which, Q$0014+813$ and Q$1946+7658$, 
also have interesting upper limits at N. Optical 
spectrophotometry, IR spectra and corrected $\lambda<\lambda$(Ly$\alpha$) continua are plotted as 
in Figure \ref{seds}, with the addition here of optical and IR photometry and upper 
limits (solid squares).  As in Figure \ref{seds}, the appropriate (radio-loud or radio-quiet) 
mean, low redshift, SED is overplotted for comparison (dotted line) as well as the power law 
fit (dashed line). 
\label{lseds}}
\end{figure}

\begin{figure}
\figcaption[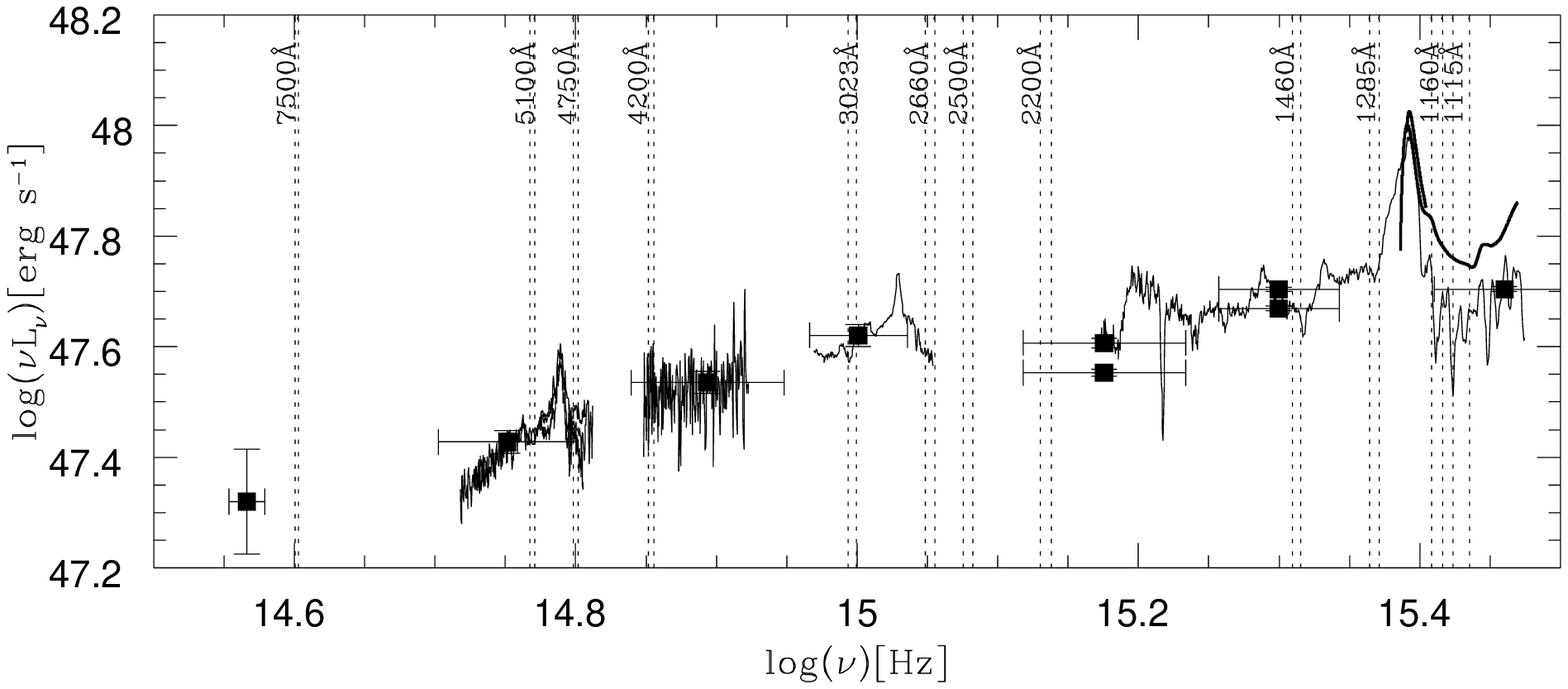]{The locations of the 12 narrow bands chosen 
for this study (defined in Table \ref{windows}) are marked with reference to the SED of 
one of the high redshift sample quasars, Q$0636+680$ at $z=3.18$.
\label{fwindows}}
\end{figure}

\begin{figure}
\figcaption[fig6.eps]{Residuals of single power law fits to the 
narrow band average continuum luminosities between 1285-5100\AA, excluding
the four between 2000\AA\ and 4000\AA\ where FeII+BaC emission (the 3000\AA\ small bump) 
contributes. Symbols are as in Figure \ref{seds} except that all 4 bands between 2000 and
4000\AA\ are indicated by an `x' in addition to the symbol used to signify how the luminosity
was determined (solid square | directly or open squares | by interpolation) and its quality (x |
poorly determined). 
For most of the low redshift AGN, the 2000-4000\AA\ residuals indicate excess emission within
this region, but this feature is weak or absent among the high redshift quasars, as discussed
in \S\ref{l3000bump}. 
\label{plres}}
\end{figure}

\begin{figure}
\figcaption[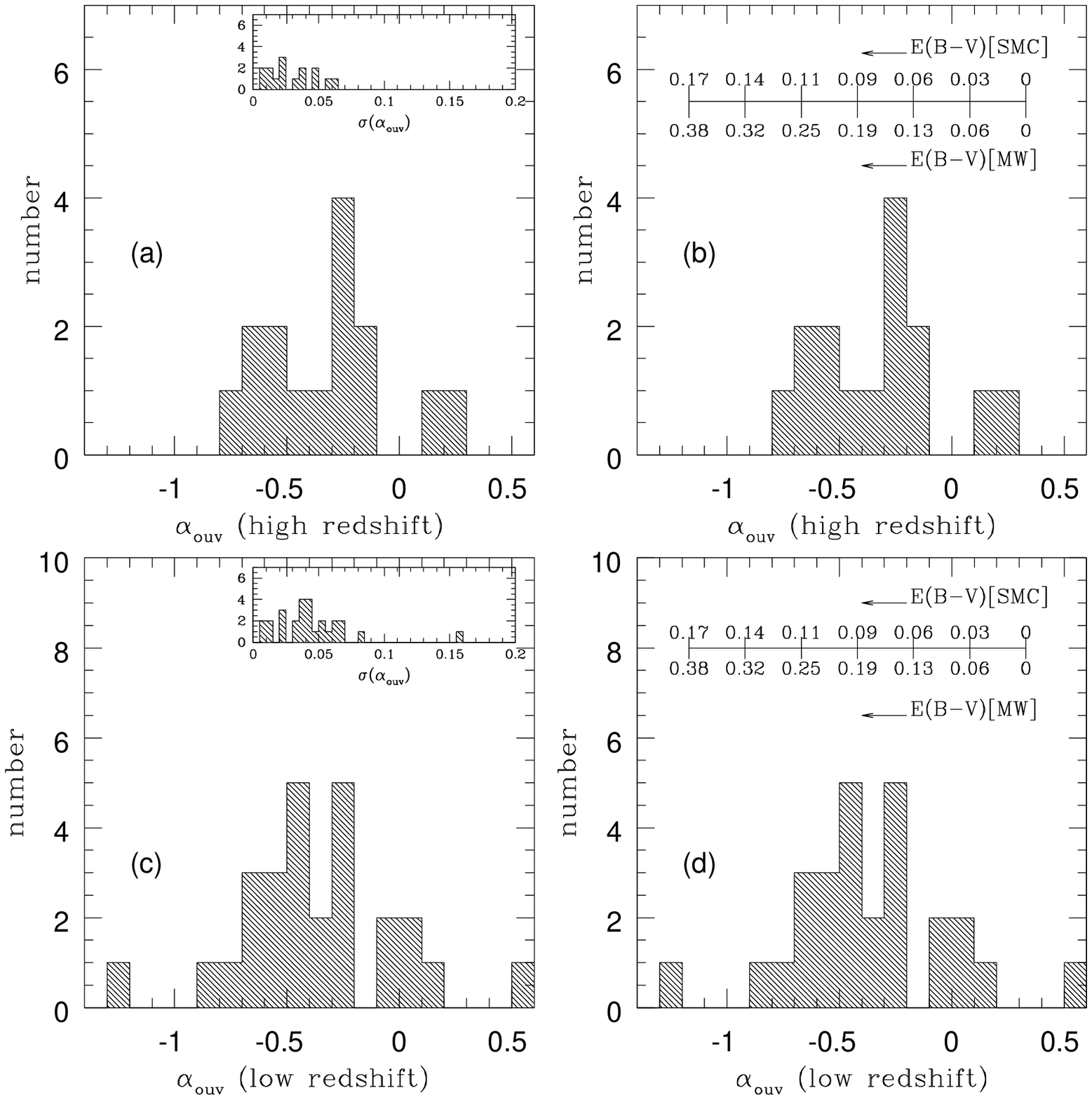]{
The histograms show the distributions of optical/UV spectral indices 
for the high and low redshift samples, computed using the narrow band luminosities
between 1285\AA\ and 5100\AA\ but excluding those 4 within the `small bump' region. 
In (a) and (c), error histograms are inset.
In (b) and (d), the scale indicates the E(B-V) needed, assuming both Milky Way
and SMC extinction laws, to redden a $\nu^{1/3}$ power law by a given amount.
To cover the range of spectral indices ($\Delta \alpha_{ouv} \approx 1$) would require 
E(B-V) $\approx$ 0.11 (SMC) or 0.25 (MW).
\label{hist_aouv}}
\end{figure}

\begin{figure}
\figcaption[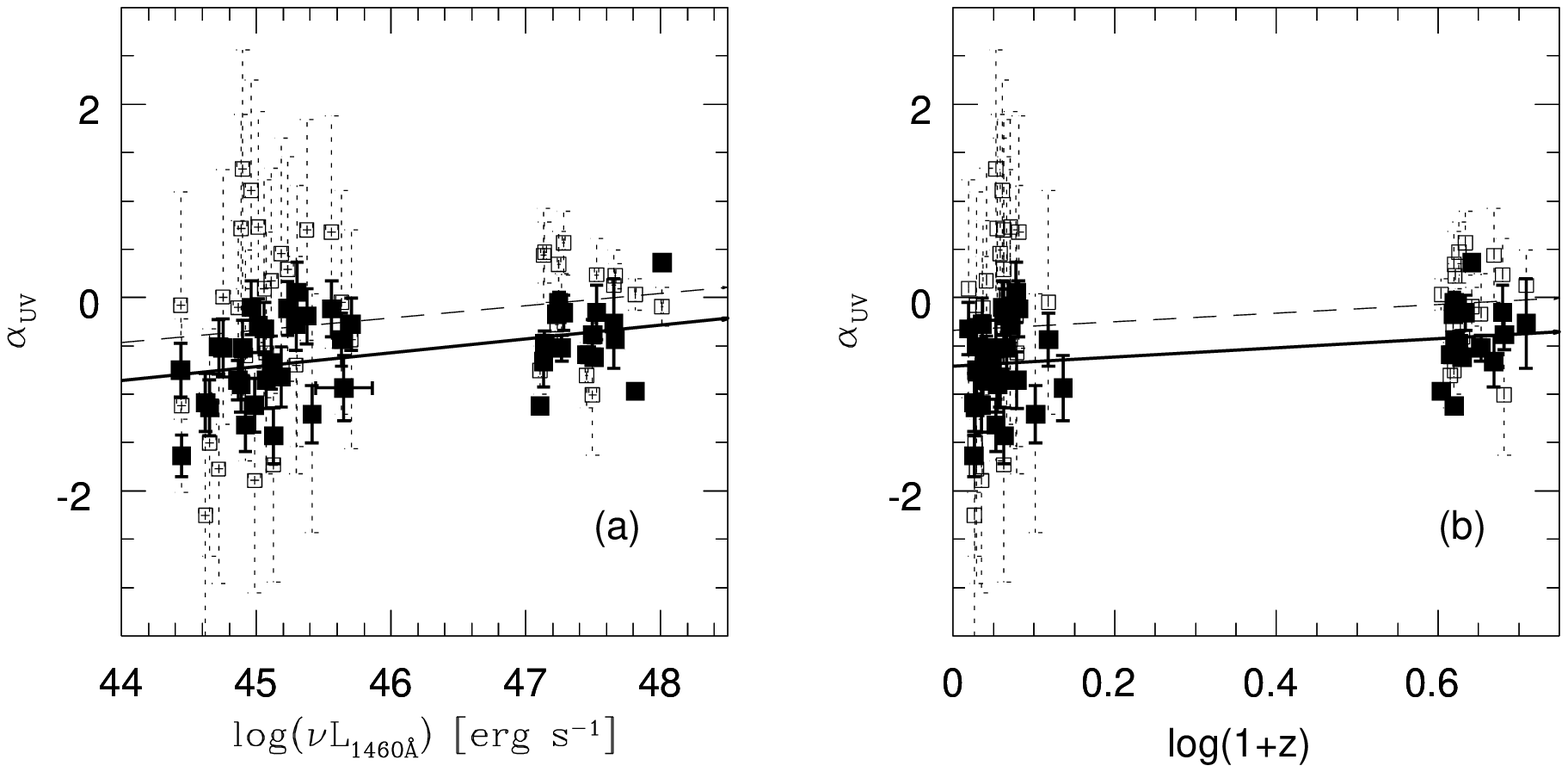]{
UV spectral indices, $\alpha_{uv}$, plotted against luminosity at 1460\AA, log$\nu L_{\nu}$(1460\AA) (a) 
and redshift, log$(1+z)$ (b). In both figures the filled symbols and solid thick error bars 
are used for the $\alpha_{uv}$(1285-2200\AA), and the open symbols and dotted error bars, 
for the $\alpha_{uv}$(1285-1460\AA). The solid lines show the best linear fit between the 
1285-2200\AA\ spectral indices and log $L$ or log$(1+z)$ while the dotted lines mark the 
best fits determined using the 1285-1460\AA\ spectral indices instead.
\label{uvslopecorrs}}
\end{figure}

\begin{figure}
\figcaption[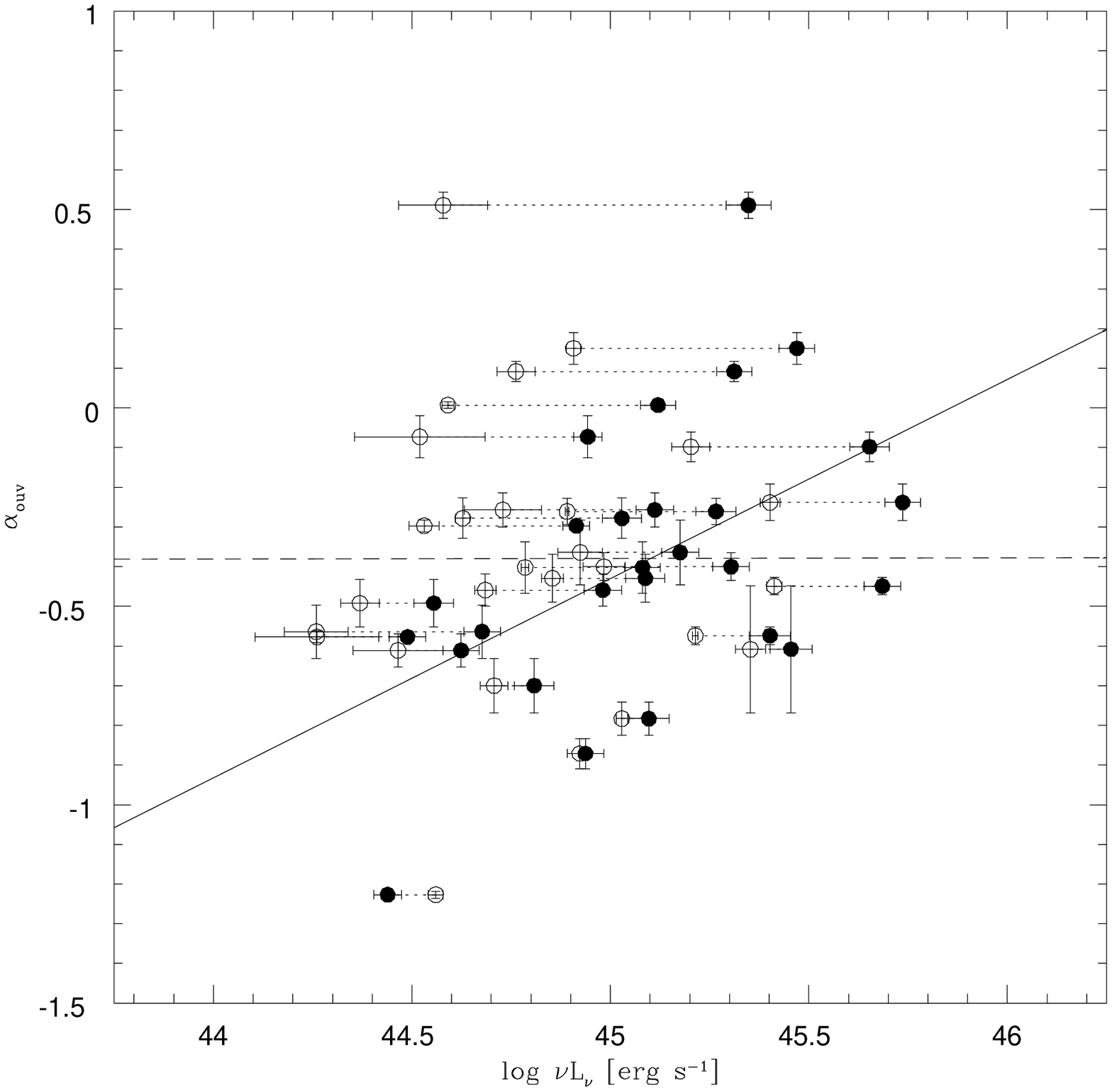]{
Optical/UV 1285-5100\AA\ spectral indices, $\alpha_{ouv}$, plotted against 
UV ($\log \nu L_{\nu}$(1285\AA); solid circles) and optical ($\log \nu L_{\nu}$(4200\AA); 
open circles) luminosities for the low redshift sample. The solid line indicates the best 
fit between $\alpha_{ouv}$ and $\log \nu L_{\nu}$(1285\AA) and the long-dashed line, the best 
fit between $\alpha_{ouv}$ and $\log \nu L_{\nu}$(4200\AA). The short-dashed lines connect 
optical and UV luminosities for
the same object, and illustrate the movement away from a slope-luminosity correlation when 
the luminosity is measured in the optical rather than the UV.
There is good evidence for a correlation between the optical/UV continuum shape and 
UV, but not optical, luminosity for the low redshift sample and marginal evidence that
it is present also in the combined low and high redshift samples, though not within the high 
redshift sample alone. 
\label{aouvcorr}}
\end{figure}

\begin{figure}
\figcaption[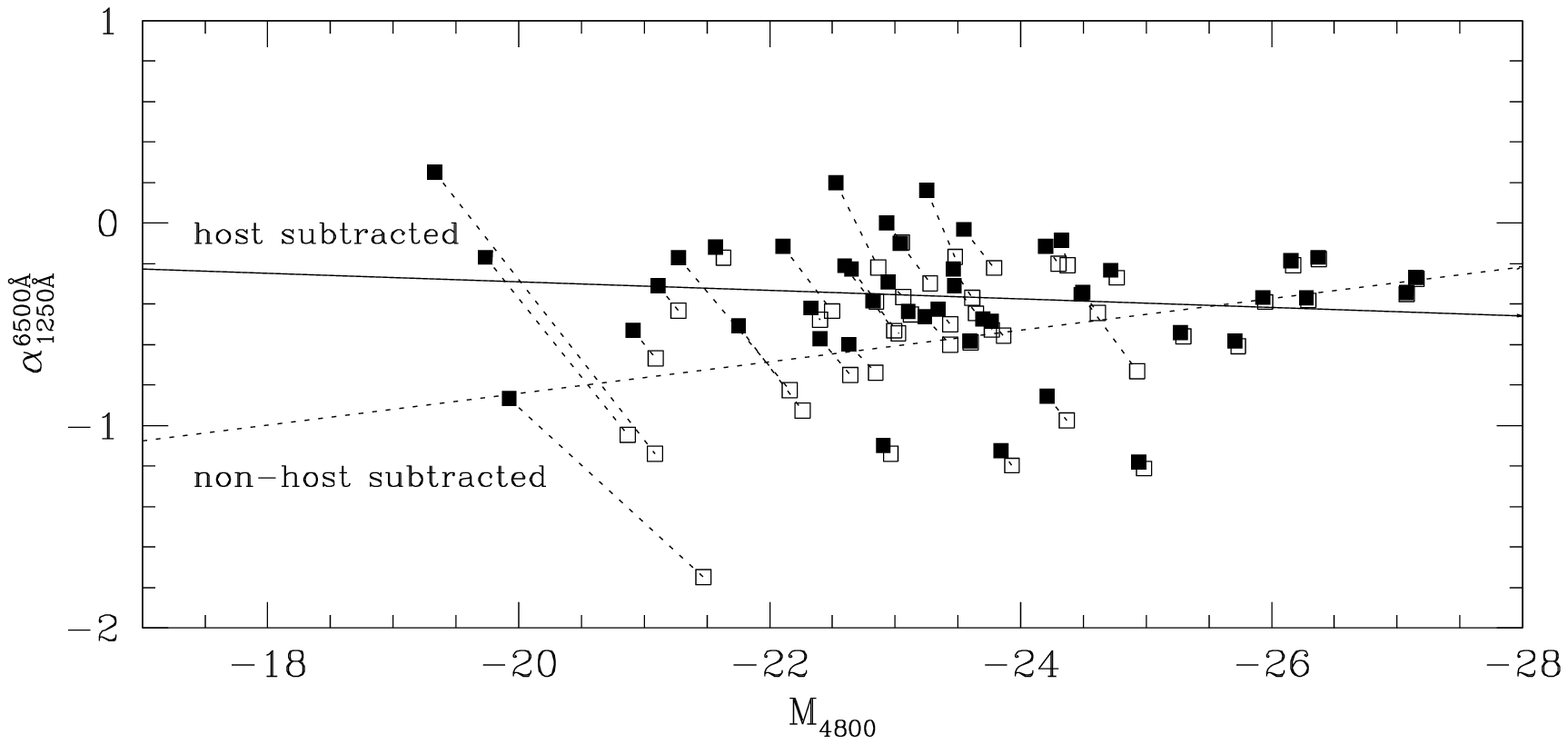]{The influence of host galaxy starlight
in the correlation between optical/UV slope and optical luminosity is illustrated 
in this figure where the spectral indices from 1285-6500\AA\ and absolute magnitudes 
at 4800\AA\ are plotted for 46 low redshift AGN from the `Atlas' (Elvis \etal\ 1994a), 
before (open squares) and after (solid squares) subtracting the host galaxy starlight 
contribution. There is a weak correlation between spectral index and luminosity when the 
host galaxy contribution is not removed (Spearman correlation coefficient, $r_{s} = -0.32$ 
with a 3\% probability, $p_{s}$, of a chance correlation). Subtracting it effects the lowest 
luminosity AGN most, and there is no evidence for slope-luminosity correlation in the 
resulting distribution ($r_{s}=0.13$, $p_{s}=37$\%). Linear least squares fits to the
data before (dotted) and after (solid) host galaxy subtraction are shown.
\label{hostgal}}
\end{figure}

\begin{figure}
\figcaption[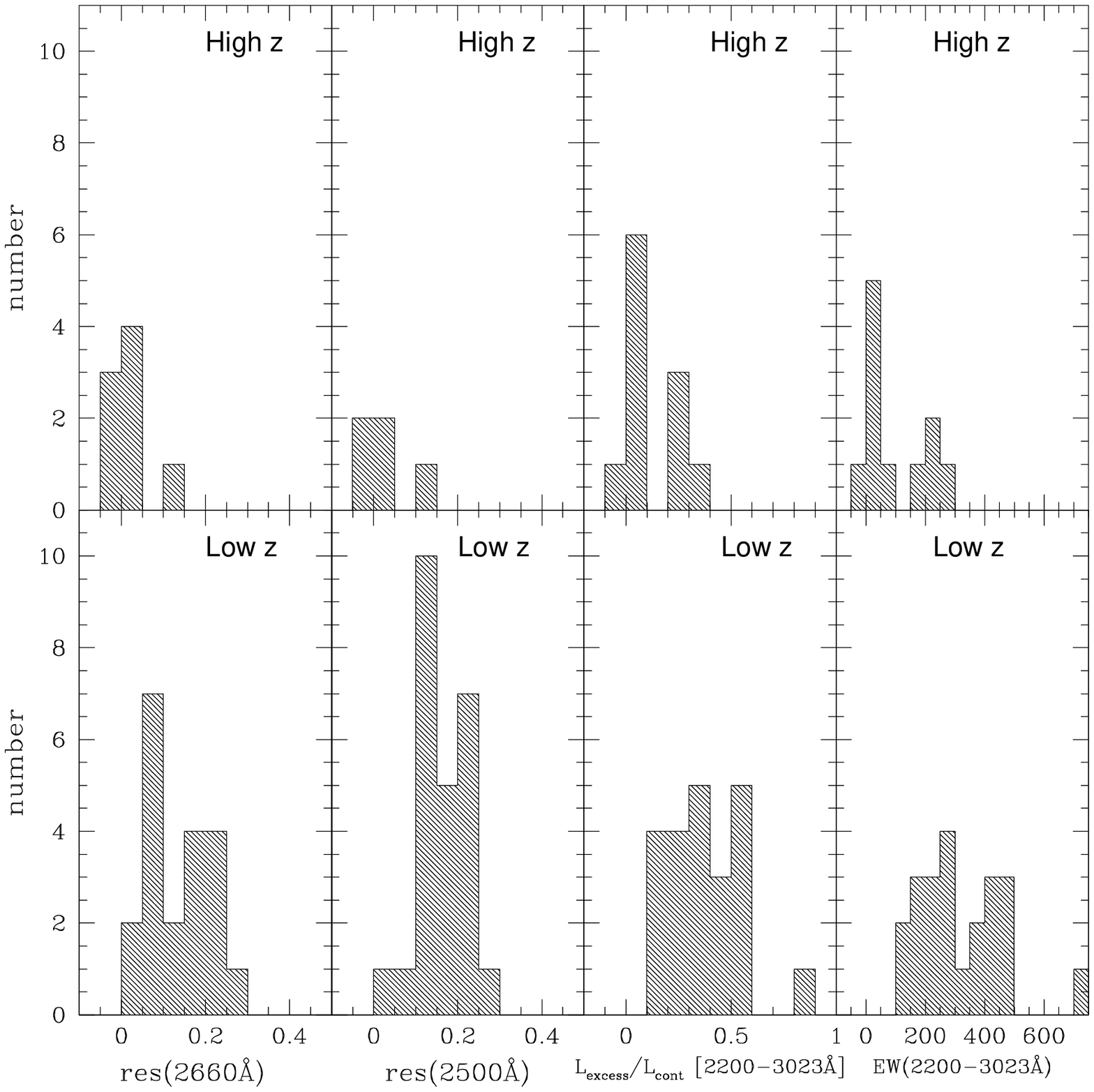]{
Evolution of the 3000\AA\ bump with redshift. The histograms show the distributions 
of the residuals at 2500\AA\ and 2660\AA, the ratios of the excess to continuum luminosity between 
2200\AA\ and 3023\AA, and the $2200-3023$\AA\ equivalent widths (as defined in the text) 
for the objects from the high and low redshift samples. Only those measures which
do not rely solely on interpolated fluxes or include discrepant flux points 
(\ie\ all values not in parentheses in Table \ref{fe2}) are plotted and used in the
sample comparisons, but the distributions which include the measures for all objects 
do not differ significantly from these. The high redshift quasars have significantly 
weaker 3000\AA\ bumps than the low redshift ones, as evidenced by all four measures. 
\label{fe2hist}}
\end{figure}


\clearpage
%
\plotone{fig1.eps}
\clearpage

\plotone{fig2.eps}
\clearpage
%
\plotone{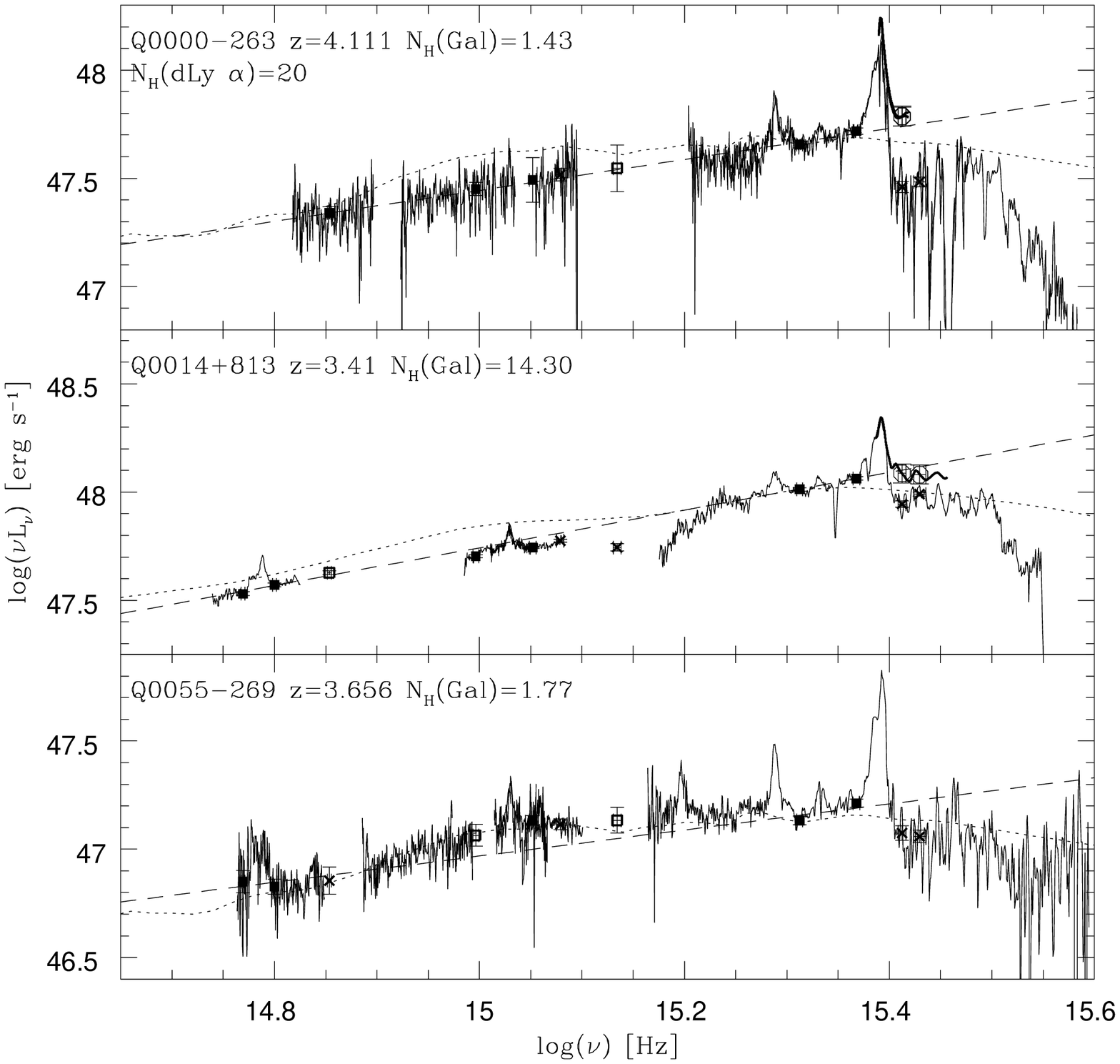}
\clearpage
\plotone{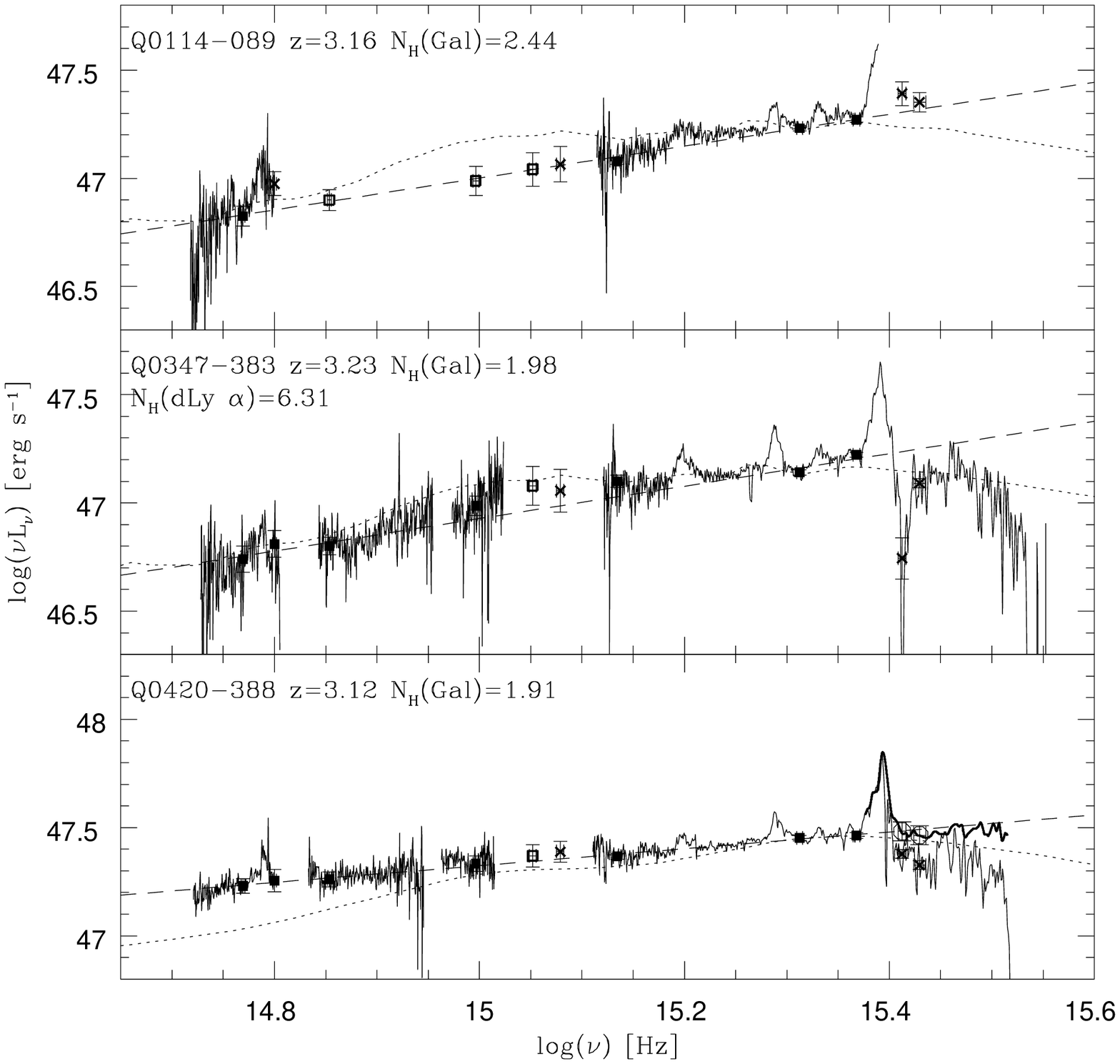}
\clearpage
\plotone{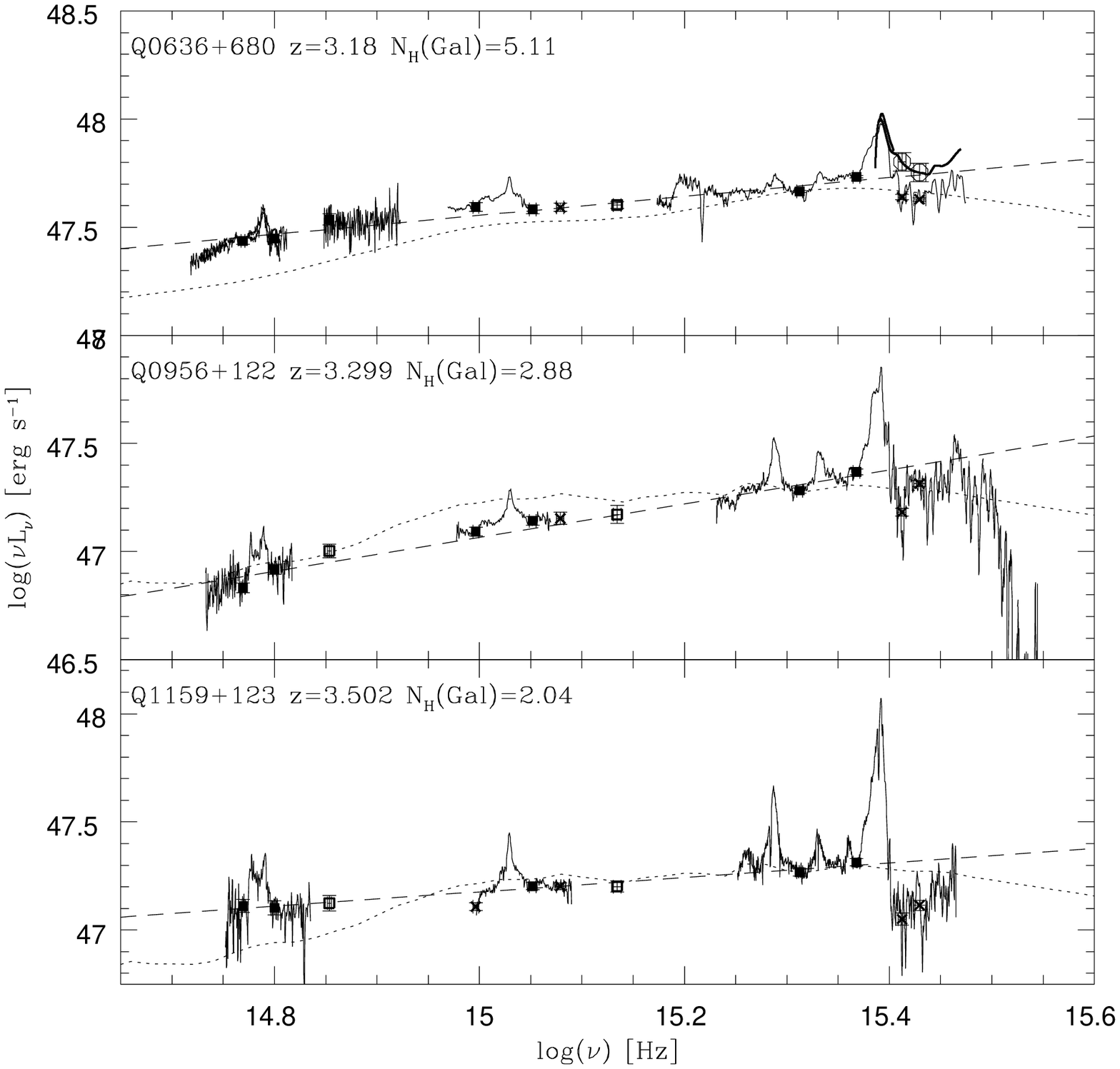}
\clearpage
\plotone{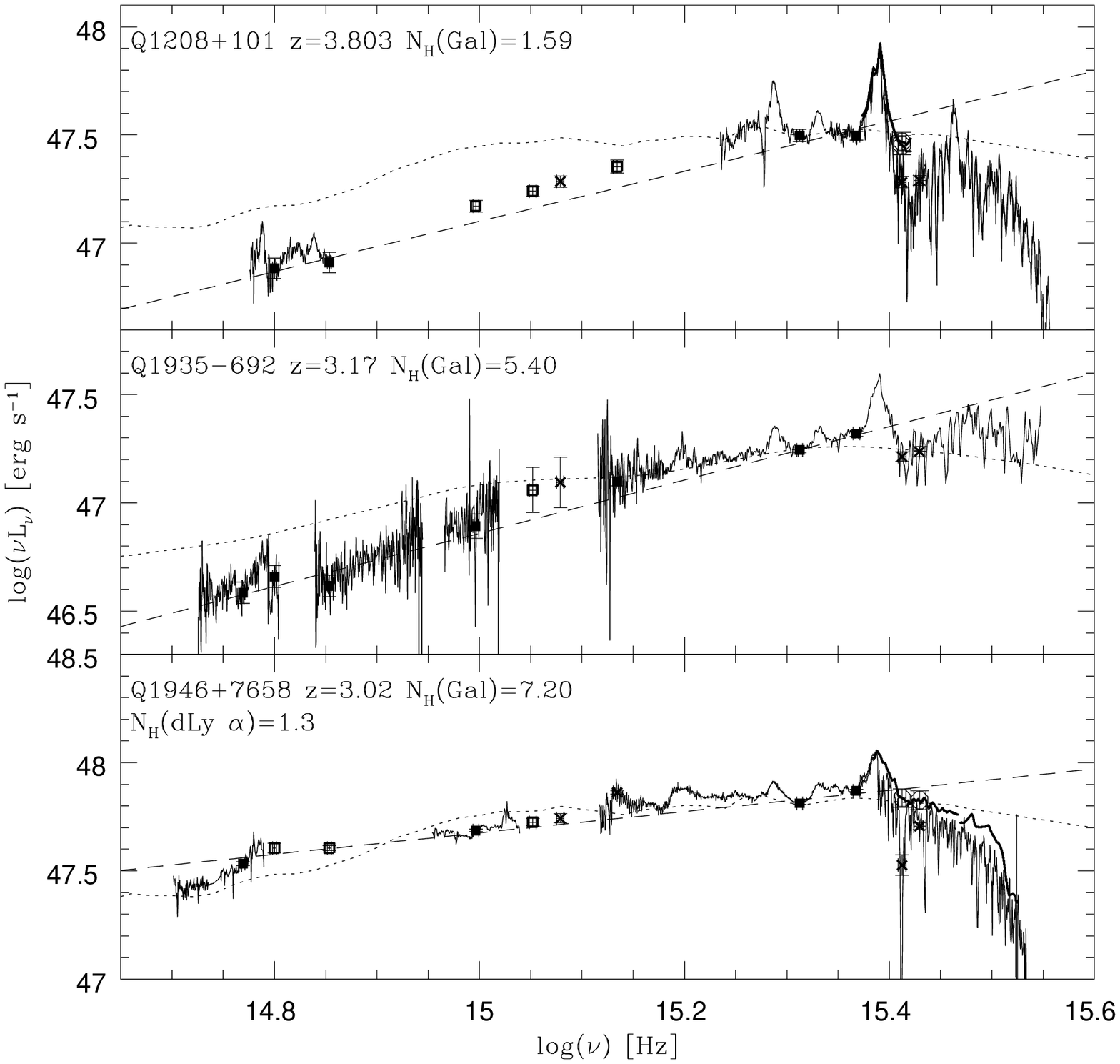}
\clearpage
\plotone{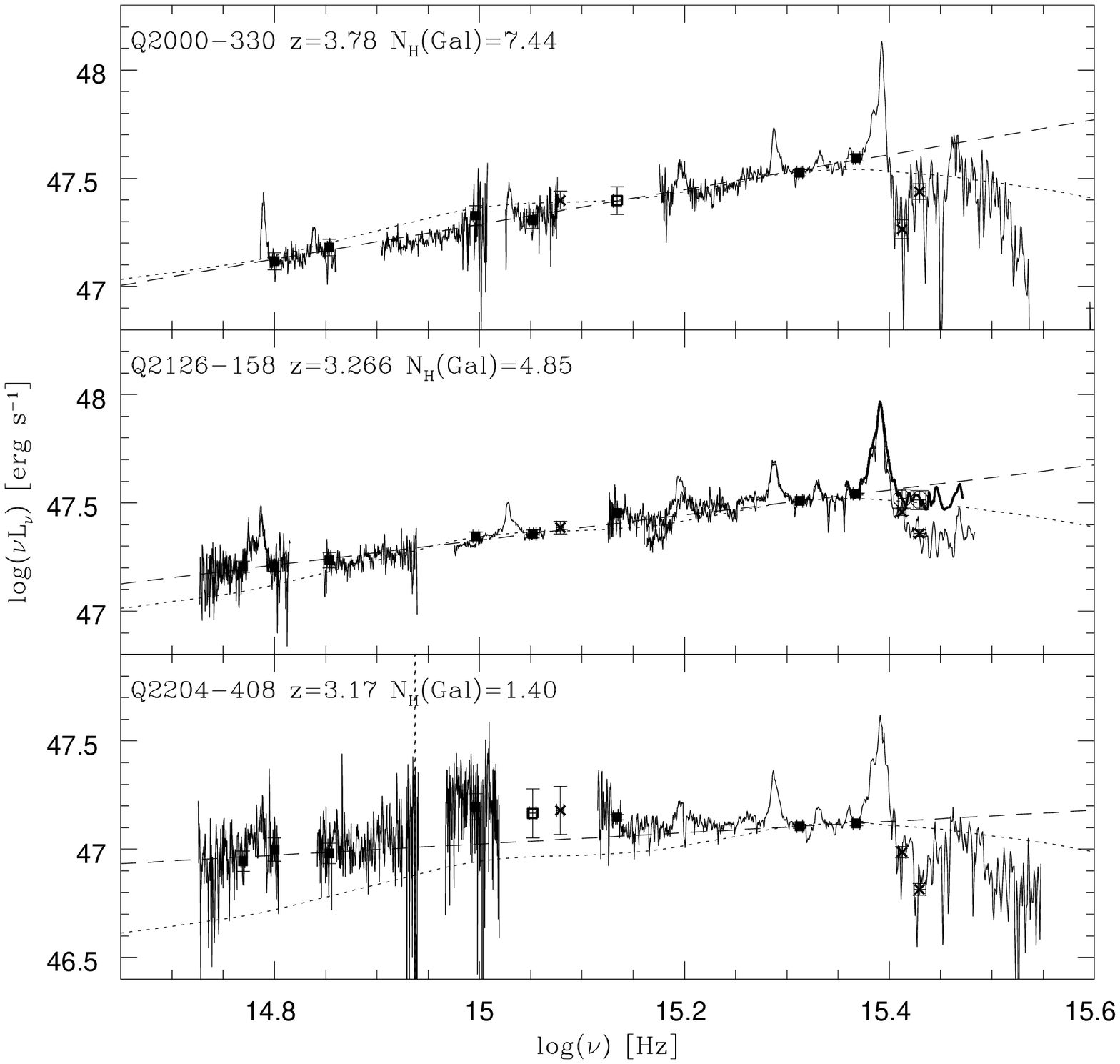}
\clearpage
%
\plotone{fig4.eps}
\clearpage
%
\plotone{fig5.eps}

%
\plotone{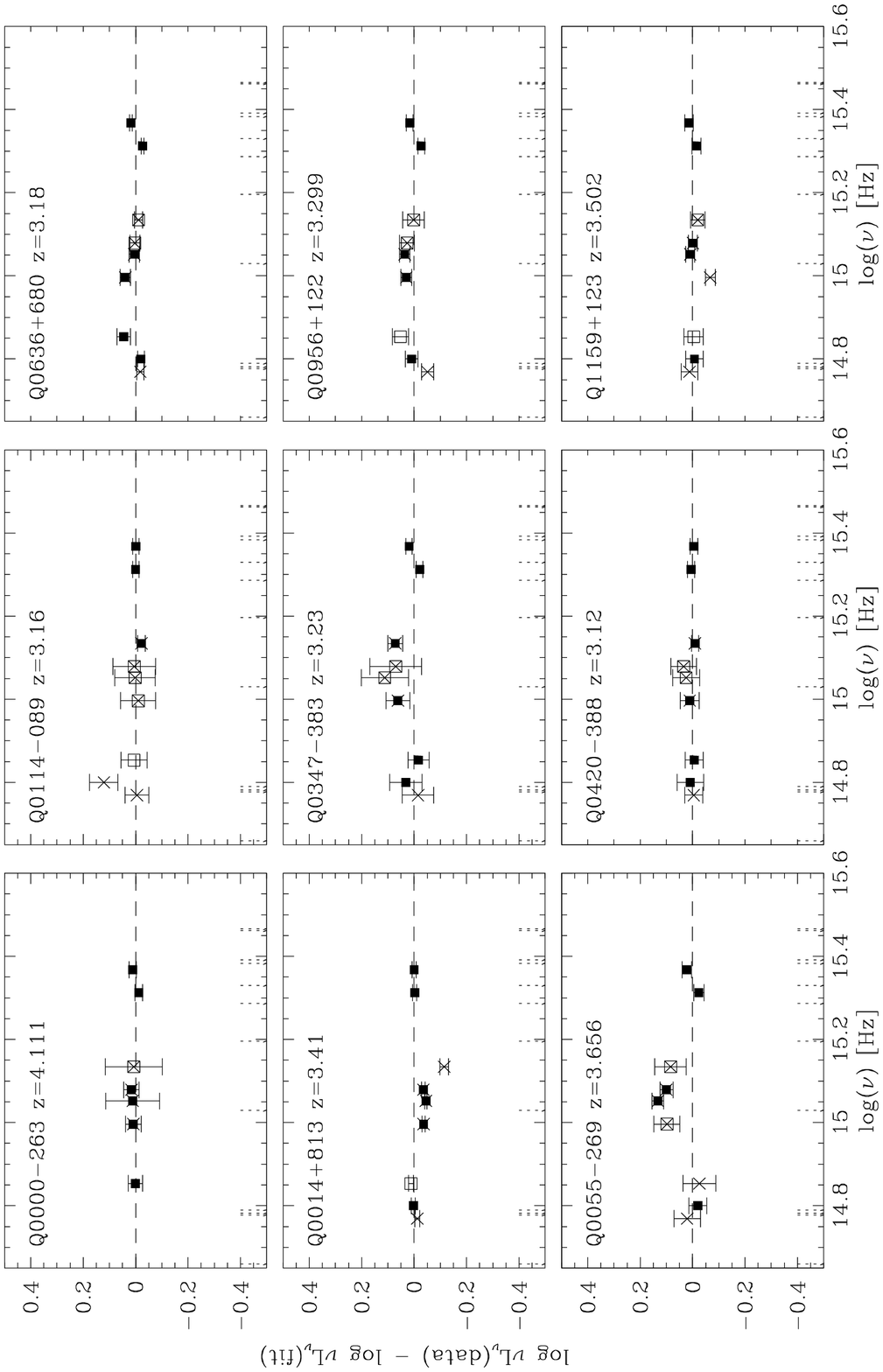}
\plotone{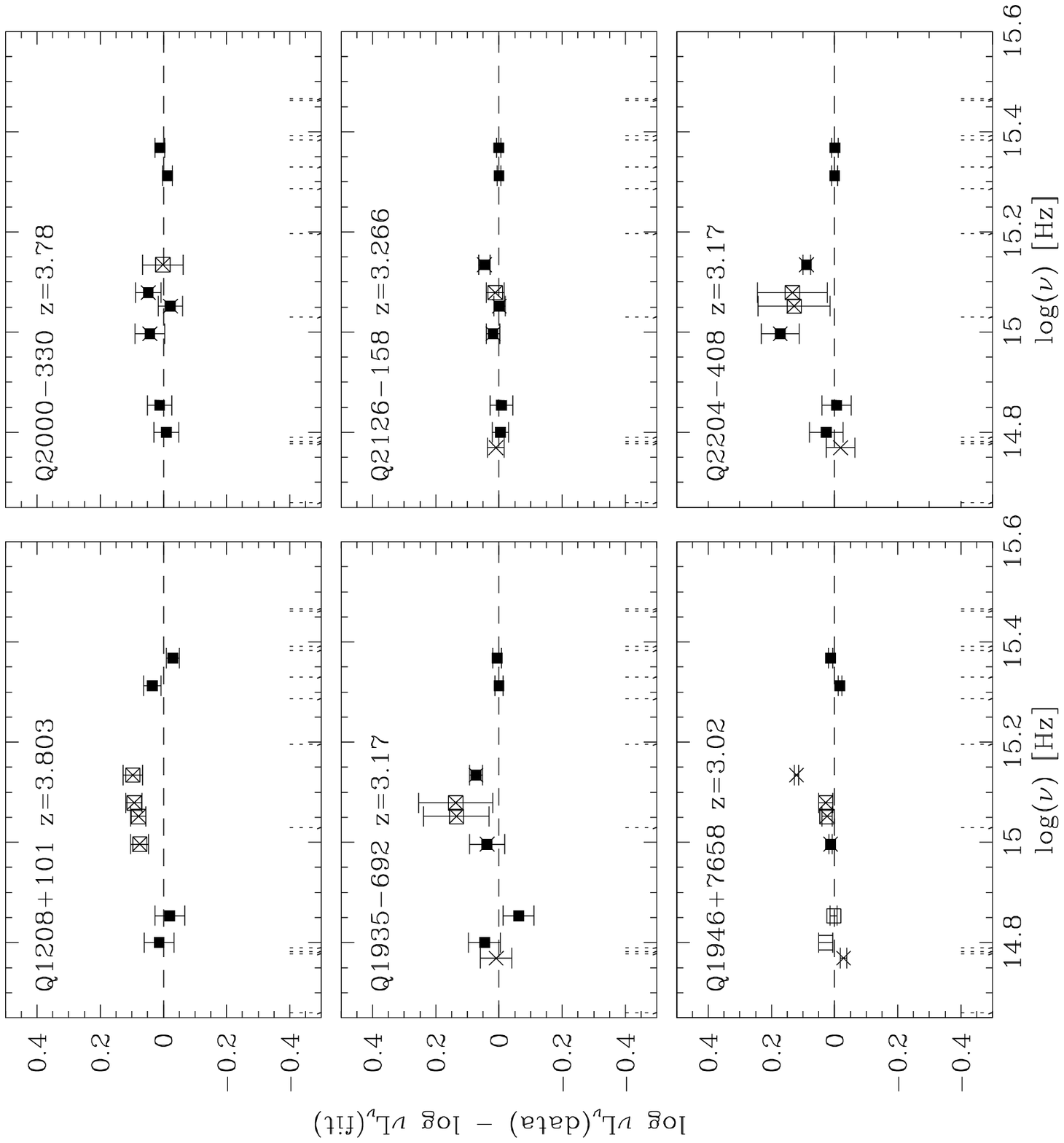}
\plotone{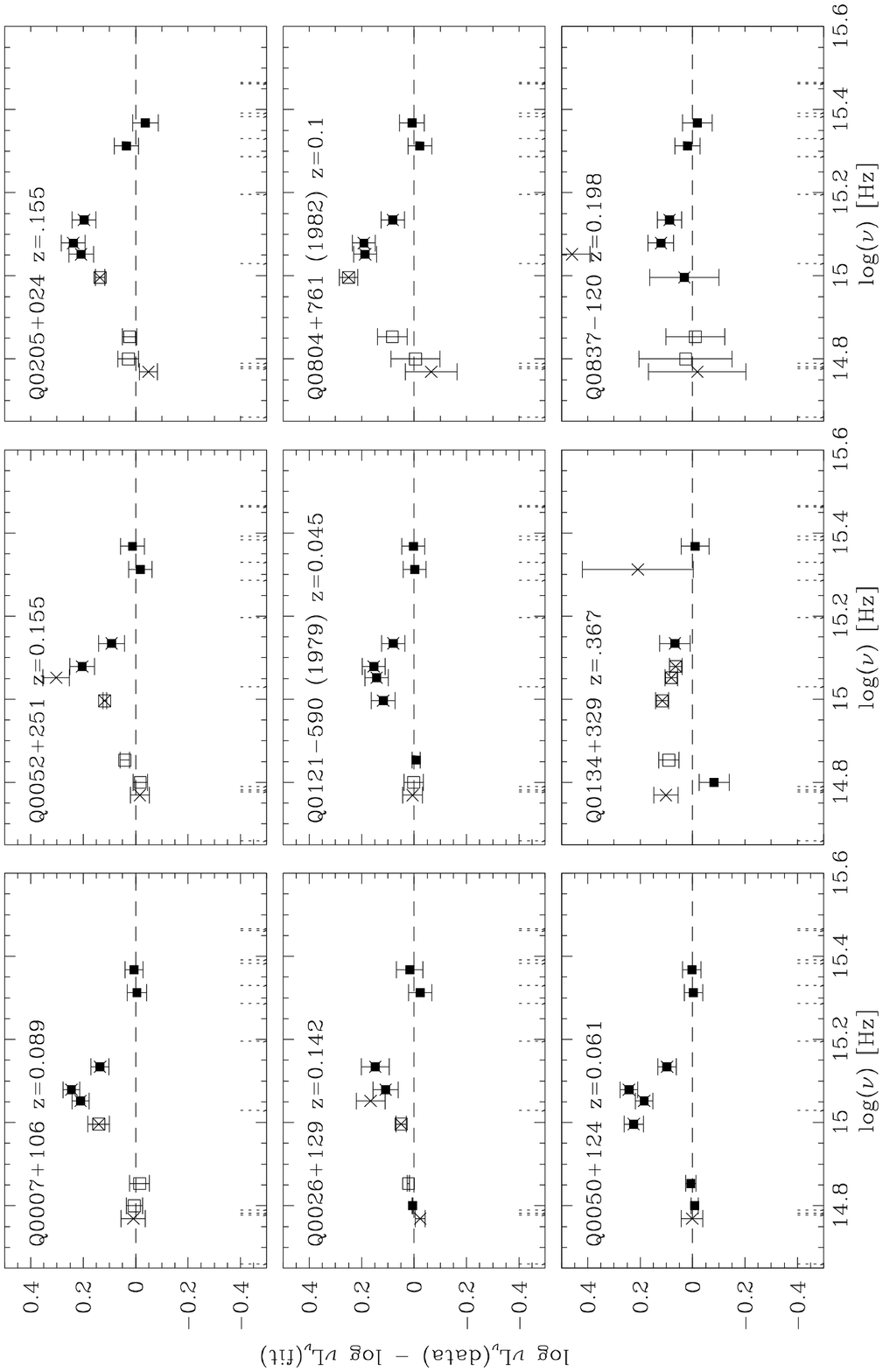}
\plotone{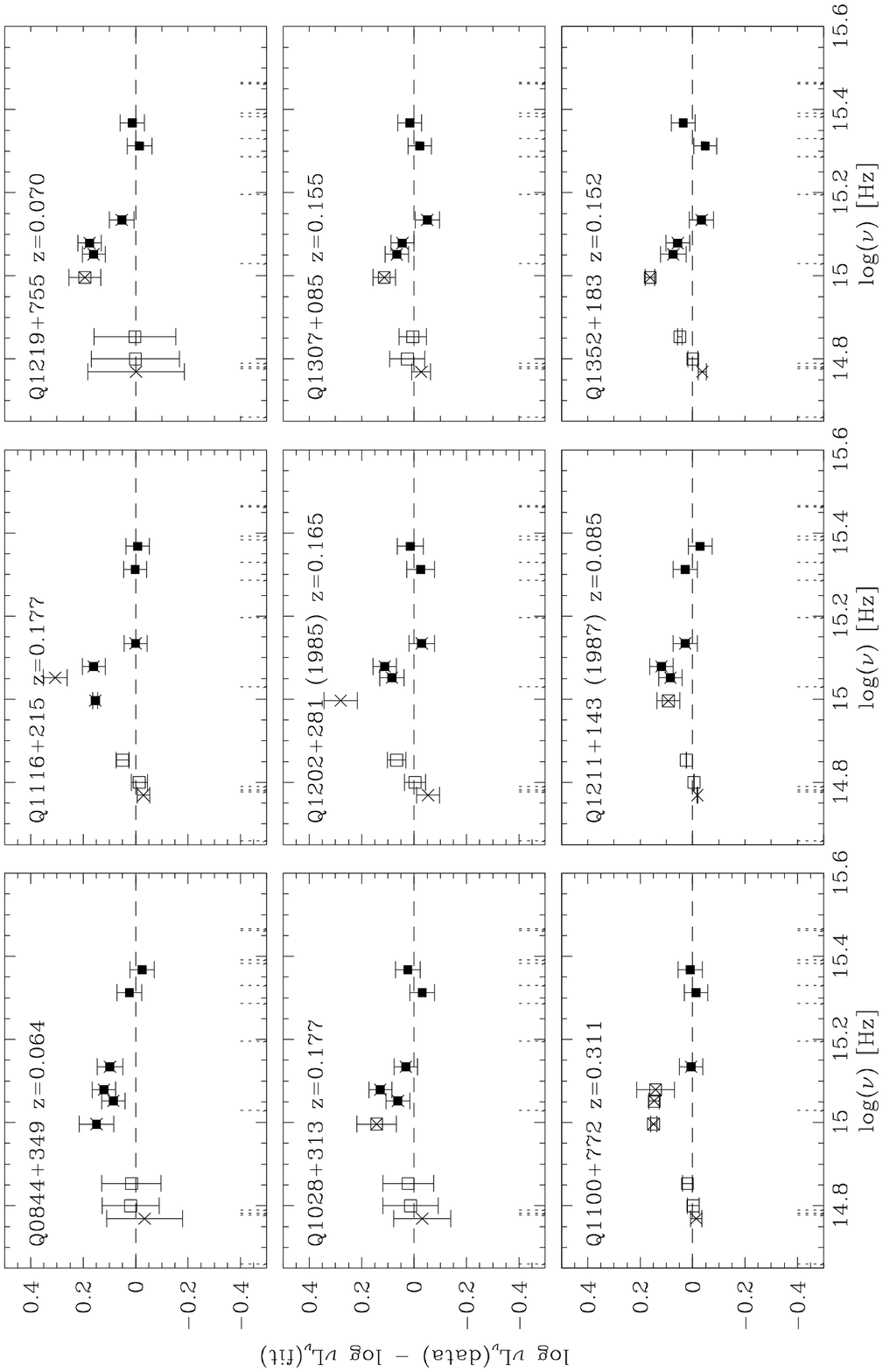}
\plotone{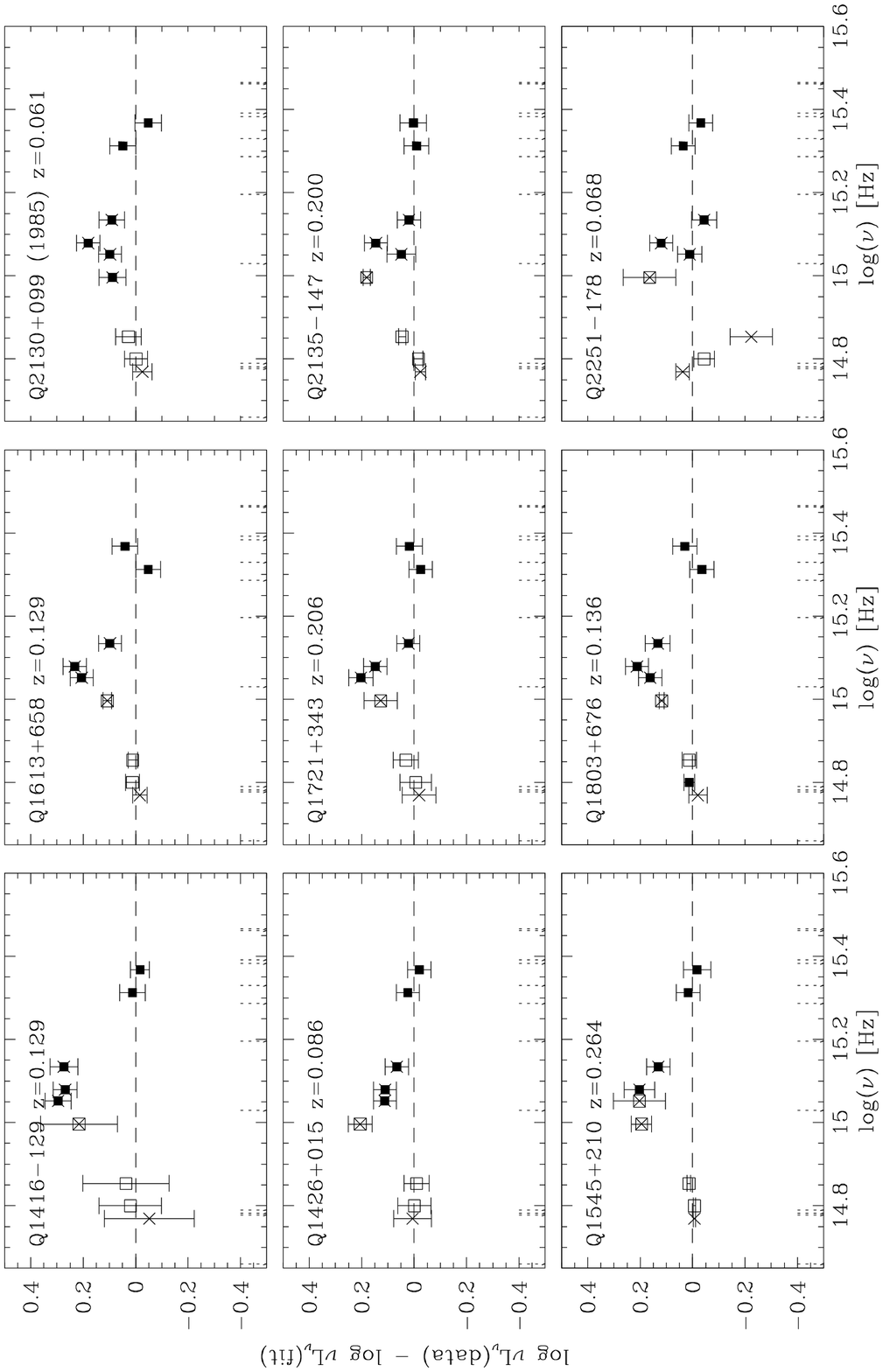}
\clearpage

\plotone{fig7.eps}
\clearpage

\plotone{fig8.eps}
\clearpage

\plotone{fig9.eps}
\clearpage

\plotone{fig10.eps}
\clearpage

\plotone{fig11.eps}
\clearpage

%
%
\include{table1} 
\include{table2} 
\include{table3}

\include{table4}

\include{table5}

\include{table6}

\include{table7}
\include{table8}
\include{table9}
\include{table10}
\include{table11}

\include{table12}
\include{table13}
\include{table14}
\include{table15}
\include{table16}
\include{table17}
\include{table18}
\include{table19}
\include{table20}
\include{table21}
\include{table22}
\include{table23}
\include{table24}


\end{document}

%% file: lmod.tex
\newcommand{\atoms}{\ifmmode{\rm ~atoms~cm^{-2}} \else ~atoms cm$^{-2}$\fi}
\newcommand{\rr}{\ifmmode {F_{PL} \over F_{disk}}\else${F_{PL} \over F_{disk}}$\fi}
\newcommand{\nufnu}{\ifmmode \nu f_{\nu} \else$\nu f_{\nu}$\fi}
\newcommand{\fnu}{\ifmmode f_{\nu} \else$f_{\nu}$\fi}
\newcommand{\Fnu}{\ifmmode F_{\nu} \else$F_{\nu}$\fi}
\newcommand{\aox}{\ifmmode{\alpha_{\tiny OX}} \else $\alpha_{\scriptsize OX}$\fi} 
\newcommand{\mdot}{\ifmmode {\dot{M}}\else${\dot{M}}$\fi}
\newcommand{\lstar}{\ifmmode {L^{*}}\else${L^{*}}$\fi}
\newcommand{\llstar}{\ifmmode {L/L^{*}}\else${L/L^{*}}$\fi}

\newcommand{\aouv}{\ifmmode{\alpha_{\tiny OUV}} \else $\alpha_{\scriptsize OUV}$\fi}
\newcommand{\costheta}{\ifmmode {\cos \theta}\else${\cos \theta}$\fi}
\newcommand{\astar}{\ifmmode {a_{*}}\else${a_{*}}$\fi}
\newcommand{\nh}{\ifmmode{\rm N_{H}} \else N$_{H}$\fi}
\newcommand{\asec}{\ifmmode ^{\prime\prime}\else$^{\prime\prime}$\fi}
\newcommand{\mv}{\ifmmode {m_{V}}\else${m_{V}}$\fi}
\newcommand{\Mv}{\ifmmode {M_{V}}\else${M_{V}}$\fi}
\newcommand{\msun}{\ifmmode {M_{\odot}}\else${M_{\odot}}$\fi}
\newcommand{\erg}{\ifmmode erg~cm^{-2}~s^{-1}\else erg cm$^{-2}$ s$^{-1}$\fi}
\newcommand{\amin}{\ifmmode ^{\prime}\else$^{\prime}$\fi}
\newcommand{\degs}{\ifmmode ^{\circ}\else$^{\circ}$\fi}
\newcommand{\nulnu}{\ifmmode {\log_{10}<\nu L_{\nu}>}\else${\log_{10}<\nu L_{\nu
}>}$\fi}
\newcommand{\lavgl}{\ifmmode {\log L}\else${\log L}$\fi}
\newcommand{\dotdot}{\ifmmode{\cdots} \else $\cdots$\fi}
\newcommand{\llstarz}{\ifmmode {L/L^{*}(z)}\else${L/L^{*}(z)}$\fi}

\newcommand{\ho}{\ifmmode {H_{o}}\else${H_{o}}$\fi}
\newcommand{\qo}{\ifmmode {q_{o}}\else${q_{o}}$\fi}
\newcommand{\ie}{{\it i.e.}}

\newcommand{\eg}{{\it e.g.}}
\newcommand{\etal}{{et al.}}
\newcommand{\twid}{\ifmmode\sim \else$^{\sim}$\fi}

%% file: table1.tex
\begin{deluxetable}{lllllrll} 
\tabletypesize{\scriptsize}
\tablecaption{High redshift sample \label{hizsample}}
\tablewidth{0pt}
\tablehead{
\colhead{Object} & 
\colhead{Other Names} & 
\colhead{class\tablenotemark{a}} & 
\colhead{z\tablenotemark{b}} & 
\colhead{V\tablenotemark{b}} & 
\colhead{$\alpha_{ox}$\tablenotemark{c}} & 
\colhead{selection\tablenotemark{d}} & 
\colhead{used?\tablenotemark{e}}
}
\startdata
Q$0000-263$ &      &RQ&     4.111        &17.5\tablenotemark{f} & 1.866 & OBJ & $\surd$ \\\tableline
Q$0014+813$ &S5    &RL&     3.3866\tablenotemark{g} &16.5  &1.344 & RAD & $\surd$  \\\tableline
Q$0055-269$ &      &RQ&     3.6625\tablenotemark{g} &17.1 &\dotdot & OBJ  & $\surd$  \\\tableline
Q$0114-089$ &UM670 &RQ&     3.1626\tablenotemark{g} &17.4  & \dotdot & OBJ  &$\surd$ \\\tableline
Q$0130-403$ &      &RQ&     3.03           &17.02 &1.713 &  OBJ  & X \\\tableline
Q$0347-383$ &      &RQ&     3.222\tablenotemark{i}  &17.3  &\dotdot  & OBJ  & $\surd$ \\\tableline
Q$0351-390$ &      &RQ&     3.01           &17.9  &\dotdot &  OBJ  & X \\\tableline
Q$0420-388$A &     &RL&     3.123\tablenotemark{j}&16.92 & 1.506&  OBJ & $\surd$  \\\tableline
Q$0636+680$ &S4    &RL&     3.1775\tablenotemark{g} &17.2\tablenotemark{m} & 1.679 & RAD &$\surd$ \\\tableline 
Q$0956+122$ &      &RQ&     3.306         &17.5 &\dotdot &  OBJ  &$\surd$  \\\tableline
Q$1107+487$ &SP    &RQ&     3.01\tablenotemark{h}     &16.9\tablenotemark{h} & 1.809 & OBJ  & X \\\tableline 
Q$1159+123$ &      &RQ&     3.502          &17.5  & $>1.688$ & OBJ &$\surd$  \\\tableline 
Q$1206+119$ &      &RQ&     3.106         &17.90  & \dotdot & OBJ  & X \\\tableline 
Q$1208+101$ &      &RQ&     3.811\tablenotemark{i} &17.5   & 1.818 & OBJ  &$\surd$  \\\tableline 
Q$1358+391$ & SP 1 &RQ&     3.3            &17.0   & \dotdot & OBJ  & X \\\tableline 
Q$1442+101$ &OQ172  &RL& 3.535            &17.78\tablenotemark{k}  &  \dotdot & RAD & X \\\tableline 
Q$1935-692$ & PKS  &RL& 3.170\tablenotemark{l}    &17.3\tablenotemark{l}    & \dotdot & RAD & $\surd$ \\\tableline 
Q$1946+769$ & HS1946+7658 &RQ& 3.02        &15.85  & 1.933 & OBJ  & $\surd$ \\\tableline 
Q$2000-330$ & PKS  &RL& 3.783\tablenotemark{g}     &17.5\tablenotemark{m} & 1.349 & RAD & $\surd$  \\\tableline 
Q$2126-158$ & PKS  &RL& 3.2663\tablenotemark{g}     &17.3 & 1.142 &  RAD & $\surd$  \\\tableline 
Q$2204-408$ &      &RL& 3.169\tablenotemark{i}     &17.57  & \dotdot & OBJ  & $\surd$  \\
\enddata
\tablenotetext{a}{Classification: RL = Radio Loud quasar; RQ = Radio quiet 
quasar. }
\tablenotetext{b}{Redshifts, $z$, and V magnitudes are taken from the catalog of Hewitt \& Burbidge (1993), unless otherwise noted.}
\tablenotetext{c}{$\alpha_{ox} \equiv -\log(l_{opt}/l_{x}) / \log(\nu_{opt}/\nu_{x})$ where $l_{opt}$ and $l_{x}$ are the 2500\AA\ and 2 keV fluxes measured in the rest-frame; from Bechtold \etal\ (1994).}
\tablenotetext{d}{Type of survey: radio (RAD) or objective prism (OBJ); in which the quasar was first discovered; from Hewitt \& Burbidge (1993).}
\tablenotetext{e}{`$\surd$'s indicate the quasars used in the final analysis and `X's indicate those omitted.} 
\tablenotetext{f}{The magnitude is an estimate at 6000\AA\ from Webb \etal\ (1988).} 
\tablenotetext{g}{redshift from Tytler \& Fan (1992)} 
\tablenotetext{h}{from our observations}
\tablenotetext{i}{redshift from Steidel (1990)} 
\tablenotetext{j}{redshift from Lanzetta \etal\ (1991)} 
\tablenotetext{k}{magnitude from Barthel, Tytler \& Thompson (1990)} 
\tablenotetext{l}{redshift and magnitude from Jauncey \etal\ (1989)} 
\tablenotetext{m}{magnitude from spectrophotometry of Bechtold \etal\ (1994)}
\end{deluxetable}

%% file: table2.tex
\begin{deluxetable}{lllllrll} 
\tabletypesize{\scriptsize}
\tablecaption{Low redshift sample \label{lowzsample}}
\tablewidth{0pt}
\tablehead{
\colhead{Object} & 
\colhead{Other Names}& 
\colhead{class\tablenotemark{a}}& 
\colhead{z\tablenotemark{b}}& 
\colhead{V\tablenotemark{b}} & 
\colhead{$\alpha_{ox}$\tablenotemark{c}} & 
\colhead{selection\tablenotemark{d}} & 
\colhead{used?\tablenotemark{e}} 
}
\startdata
Q$0007+106$ & PG & RL & 0.089 & 15.16 &1.33 & XRAY & $\surd$ \\\tableline 
Q$0026+129$ & PG & RQ  & 0.142\tablenotemark{f} &  15.41\tablenotemark{f} &1.35 &  UVX & $\surd$\\\tableline 
Q$0050+124$ & I Zw 1  & RQ  & 0.061 & 14.07 &1.44 & UVX & $\surd$ \\\tableline
Q$0052+251$ & PG, HEAO & RQ & 0.155 & 15.42 &1.39  &  UVX  & $\surd$ \\\tableline
Q$0054+144$ & PHL909 & RQ  & 0.171 & 16.70 & 1.34 & UVX & X\\\tableline 
Q$0121-590$ & {\em in} Fairall 9 & RQ & 0.045 & 13.23 &1.37& UVX & $\surd$  \\\tableline 
Q$0134+329$ & 3CR 48, 4C 32.08  & RL  & 0.367\tablenotemark{f} & 16.46\tablenotemark{f} & 1.28&RAD & $\surd$ \\\tableline 
Q$0205+024$ & NAB, Mkn 586 & RQ & 0.155\tablenotemark{f} & 15.39\tablenotemark{f}  & 1.53 &UVX  & $\surd$\\\tableline 
Q$0312-770$ & PKS & RL  & 0.223\tablenotemark{f} & 16.10\tablenotemark{f} & 1.16 & RAD  &  X \\\tableline 
Q$0804+761$ & PG  & RQ  & 0.1\tablenotemark{f} & 15.15\tablenotemark{f} & 1.44 & UVX   & $\surd$ \\\tableline 
Q$0837-120$ & 3C 206, PKS  & RL  & 0.198 & 15.76 & 1.15 & RAD  & $\surd$ \\\tableline 
Q$0844+349$ & PG & RQ   & 0.064 & 14.00 & 1.60 & UVX  & $\surd$ \\\tableline 
Q$1028+313$ & B2  & RL  & 0.177 & 16.71 & 1.30 & RAD & $\surd$ \\\tableline 
Q$1100+772$ & 3CR 249.1, 4C 77.09 & RL & 0.311 & 15.72 & 1.40 & RAD &  $\surd$\\ 
            & PG, S5 &    &       &       &      &     &         \\\tableline
Q$1116+215$ & PG, TON 1388 & RQ  & 0.177 & 15.17  & 1.62 & UVX & $\surd$ \\\tableline 
Q$1146-037$ & PKS          & RL  & 0.341 & 16.90  & 1.22 & RAD  & X \\\tableline 
Q$1202+281$ & PG, GQ Com & RQ  & 0.165 & 15.51  & 1.26 & VAR & $\surd$ \\\tableline 
Q$1211+143$ & PG  & RQ  & 0.085 & 14.63 &1.39 &  UVX  & $\surd$ \\\tableline 
Q$1219+755$ & Mkn 205 & RQ & 0.070 & 15.24 & 1.21&  UVX   & $\surd$ \\\tableline 
Q$1307+085$ & PG  & RQ & 0.155 & 15.28  & 1.42 &  UVX  & $\surd$ \\\tableline 
Q$1352+183$ &1E,PG & RQ & 0.152  & 15.5  & 1.46 & UVX  & $\surd$ \\\tableline 
Q$1416-129$ & PG & RQ & 0.129 & 15.40 & 1.26 & UVX & $\surd$ \\\tableline 
Q$1426+015$ & Mkn 1383 & RQ  & 0.086 & 15.05  & 1.41 &  UVX & $\surd$ \\\tableline 
Q$1545+210$ & 3CR 323.1, 4C 21.45 & RL & 0.264\tablenotemark{f} & 16.69\tablenotemark{f} &1.37 & RAD & $\surd$ \\ 
            & PKS, PG  &    &       &       &     &     &         \\\tableline
Q$1613+658$ & Mkn 876, PG & RQ & 0.129 & 15.37 & 1.36 & UVX  & $\surd$\\\tableline 
Q$1635+119$ &  & RL  & 0.146 & 16.50 & 1.31 & RAD  & X \\\tableline 
Q$1721+343$ & 4C 34.47  & RL  & 0.206 & 16.50 & 1.28 & RAD  & $\surd$ \\\tableline 
Q$1803+676$ &{\em in} Kazaryan 102& RQ &0.136\tablenotemark{f} & 15.78\tablenotemark{f} & 1.45 & UVX  & $\surd$ \\\tableline 
Q$2130+099$ &PG {\em in} II Zw 136 & RQ  &0.061 & 14.62  &1.52 &  UVX&$\surd$ \\\tableline 
Q$2135-147$ &PHL 1657, PKS & RL & 0.200\tablenotemark{f} & 15.19\tablenotemark{f} & 1.25 & UVX & $\surd$\\\tableline 
Q$2251-178$ & MR & RQ & 0.068 & 14.36 & 1.28 & XRAY & $\surd$\\  
\enddata
\tablenotetext{a}{Classification: RL = Radio Loud; RQ = Radio quiet; from Table 1 of Elvis \etal\ (1994a).}
\tablenotetext{b}{ Redshifts, and V magnitudes, from Table 1 of Elvis \etal\ (1994a), unless otherwise noted.}
\tablenotetext{c}{$\alpha_{ox} \equiv -\log(l_{opt}/l_{x}) / \log(\nu_{opt}/\nu_{x})$ where $l_{opt}$ and $l_{x}$ are the 2500\AA\ and 2 keV fluxes measured in the rest-frame; from Table 2 of Elvis \etal\ (1994a).}
\tablenotetext{d}{Primary selection technique based on: radio (RAD), color excess (UVX), X-ray (XRAY) or
variability (VAR) measurements. For the $z\geq0.1$ quasars these designations are from Table 1 of Hewitt \& Burbidge (1993) 
and for the $z<0.1$ ones not included in that version, from Table 1 of Hewitt \& Burbidge (1991). Three quasars not
in either version: Q0844+349, Q1211+143 and Q1426+015; appear in the Markarian or PG surveys so we list these
as UVX selected.
}
\tablenotetext{e}{ `$\surd$'s indicate the quasars used in the final analysis and `X's indicate those omitted.} 
\tablenotetext{f}{redshift and magnitude from Hewitt \& Burbidge (1993).} 
%
\end{deluxetable}

%% file: table3.tex
\begin{deluxetable}{lcccccc}
\tabletypesize{\small}
\tablecolumns{7}
\tablecaption{VLA Fluxes: $z>3$ Quasars \label{vlaobslog}}
\tablewidth{0pt}
\tablehead{
\colhead{} & \multicolumn{3}{c}{$\nu=4.95$ GHz} & \multicolumn{3}{c}{$\nu=1415$ MHz} \\
\cline{2-4} \cline{5-7} \\
\colhead{Quasar} & 
\colhead{$F_{max}$[mJy]\tablenotemark{a}} & \colhead{RMS flux [mJy]\tablenotemark{b}} & \colhead{SNR} &
\colhead{$F_{max}$[mJy]\tablenotemark{a}} & \colhead{RMS flux [mJy]\tablenotemark{b}} & \colhead{SNR} }
\startdata
Q$0000-263$ & 0.27 & 0.13 & 2.1 & 0.32 & 0.16 & 2.0 \\
Q$0055-269$ & 0.45 & 0.14 & 3.2 & 0.44 & 0.20 & 2.2 \\
Q$0114-089$ & 0.23 & 0.13 & 1.8 & 0.12 & 0.16 & 0.8 \\
Q$1107+48$  & 0.48 & 0.14 & 3.4 & 0.33 & 0.37 & 0.9 \\
Q$1159+123$ & 0.31 & 0.14 & 2.2 & 0.44 & 0.17 & 2.6 \\
Q$1206+119$ & 0.26 & 0.13 & 2.0 & 0.42 & 0.17 & 2.5 \\
Q$1208+101$ & 0.20 & 0.12 & 1.7 & 0.31 & 0.15 & 2.1 \\
Q$1358+391$ & 0.24 & 0.12 & 2.0 & 0.13 & 0.17 & 0.8 \\
Q$1946+7658$& 0.89 & 0.13 & 6.8 & 0.43 & 0.17 & 2.5 \\ 
\enddata
\tablenotetext{a}{$F_{max}$ is the flux measured at position of beam peak 
if no obvious source was visible.}
\tablenotetext{b}{RMS flux measured far from center of image}
\end{deluxetable}

%% file: table4.tex
\begin{deluxetable}{ll lc cc cc}
\tabletypesize{\tiny}
\tablecaption{Instrument/Observation Summary\label{obssum}}
\tablewidth{0pt}
\tablehead{
\colhead{Telescope/Run Dates} & 
\colhead{Instrument\tablenotemark{a}} &
\colhead{Detector} &
\colhead{$\lambda\lambda$} &
\colhead{W/D\tablenotemark{b}} &
\colhead{gpm\tablenotemark{c}} &
\colhead{R\tablenotemark{d}} &
\colhead{$t_{total}$\tablenotemark{e}} 
}
\startdata
\cutinhead{IR SPECTROSCOPY}
%
MMT      & Fspec &  NICMOS HgCdTe & H & $1.2\asec$  & 75 & 730 & 50 \\
Nov 93   &       & $256\times256$ & K &             & 75 & 670 & 40-90 \\\hline
KPNO 4-m & CRSP  &  SBRC 41 InSb  & J & $1.4\asec$ & 150 & 550 & 30-140 \\
Dec 93   &       & $256\times256$ & J &            & 300 & 1430& 20     \\
         &       &                & K &            & 150 & 700 & 50-80 \\
         &       &                & K &            & 300 & 1800& 25-50 \\\hline
 CTIO 4-m & OSIRIS  & NICMOS3 HgCdTe & IJHK\tablenotemark{f} & $1.2\asec$ & 120 & 550 & 50-200 \\
Sep 94, Dec 93 & & $256\times256$ & & & & & \\
\cutinhead{IR PHOTOMETRY}
%
MMT&  IR-phot & InSb & J,H,K \& L$^{\prime}$ & $5\farcs3$ or $8\farcs7$&\nodata&\nodata & 30-40 \\
Jun,Sep,Oct 93,Mar 94 &             &single- &  & & & &  \\
                       &             & channel&  & & & & \\
                       &             &        &  & & & & \\
                       &             &        &  & & & &  \\\hline
MMT   & IR-bolom & Ge diode & N & $5\farcs4$ &  \nodata & \nodata  & 13-27 \\
Nov 92 &          &          &   &            &    &    &   \\\hline
CTIO 4-m & OSIRIS  & NICMOS3        & H & 32 pxls (12\farcs7)& \nodata & \nodata &  4.5  \\
Sep 94, Dec 93 &  & HgCdTe & K & $0\farcs398$/pxl &         &         &  2.5 \\
\cutinhead{OPTICAL SPECTROPHOTOMETRY}
%
MMT                   &  Red channel & TI 5 CCD      & 4200-8400 & 5\arcsec & 150 & 60 & 5-20\\
May 91 \& Sep 91      &              & $800\times800$& 4200-8400 & 1\farcs5 & 150 & 250& \\
                      &              &               & 4000-6550 & 1\farcs5 & 300 & 450& \\
FLWO 1.5-m & FAST& Loral CCD       & $3700-7500$ & 5\arcsec & 300 & 500 & 45 \\
Feb 94     &                      & unthinned       &          &          &     &      & \\
           &                      & $512\times2688$ &          &          &     &      & \\
Steward Obs. 2.3-m& B\&C spec&Loral CCD       &           &        &   &   & 60\\
Sep 92            &          &$1200\times800$ &5500-9100 &4\farcs5 &400&160& \\
                  &          &                &3200-6500 &4\farcs5 &300&180& \\
May 90            &          &                &4500-7500 &4\farcs5 &300&400&\\

CTIO 1.5-m&RC spec&thick GEC CCD  &3500-8300  &4\arcsec &150 (\#13) &300& 60\\
Sep 93    &       &$576\times420$ &5000-9500  &4\arcsec &158 (\#11) &480& \\

\cutinhead{OPTICAL PHOTOMETRY}
%

FLWO 1.2-m & & Ford CCD  &UBVRI& 24,28 pxls (14\farcs4,16\farcs8)&\nodata  &\nodata   & 2-10 \\
Oct 93 \& Apr 94  &  & $2048\times2048$ &     & $0\farcs6$/pxl\tablenotemark{g} & & & \\\hline
CTIO 0.9-m        &  & Tek CCD          & BRI &24 pxls ($11\farcs9$) &\nodata  & \nodata  & 2-4 \\
Sep 93            &  & $1024\times1024$ &     &$0\farcs496$/pxl                 &  &   &  \\
\enddata
\tablenotetext{a}{Instrument references: Fspec | Williams \etal\ (1993); CRSP | Joyce, Fowler 
and Heim (1994); OSIRIS | DePoy \etal\ (1993); IR-photometer | Rieke (1984); IR-bolometer | 
Keller, Sabol and Rieke (1990); FAST spectrograph | Fabricant \etal\ (1998)}
\tablenotetext{b}{Aperture: In the case of spectroscopic observations, the projected slit width 
on the sky in arcseconds. All spectroscopy was done using long slits which included object and 
sky. In the case of photometric observations, the diameter of the aperture through which the
object was observed (MMT IR photometer and bolometer) or that chosen in the data reduction 
(OSIRIS).}
\tablenotetext{c}{gratings per mm of the grating used.}
\tablenotetext{d}{R = $\lambda/\Delta\lambda$, where $\Delta\lambda$ is the FWHM which 
corresponds to the 2-pixel dispersion for the IR spectra. }
\tablenotetext{e}{The range of total on-source integration times (min) per object per filter or setting.}
\tablenotetext{f}{Cross-dispersed, spectral range $\sim 1.1\mu$m to 2.4$\mu$m (IJHK)}
\tablenotetext{g}{The pixels were binned $2\times2$, and the plate scale given here is for the
$2\times2$ binning.}
\end{deluxetable}

%% file: table5.tex
\begin{deluxetable}{llllllll}
\tabletypesize{\scriptsize}
\tablecaption{Log of IR spectroscopic observations\label{irspec}}
\tablewidth{0pt}
\tablehead{
\colhead{Quasar}&
\colhead{Instrument}& 
\colhead{bands}& 
\colhead{slit$\times$grating\tablenotemark{a}} & 
\colhead{R\tablenotemark{b}} & 
\colhead{t\tablenotemark{c}} & 
\colhead{date (UT)} & 
\colhead{comments}
}
\startdata
Q$0000-263$ & OSIRIS & JHK & 1.2x120 &550 & 80 & 1994 Jul 27  & \\ \tableline
Q$0014+813$ & CRSP & J & 1.4x150 &550& 27 & 1993 Dec 6  & \\ 
  & CRSP & J & 1.4x300 &1430& 20 & 1993 Dec 7  & \\ 
  & Fspec & K & 1.2x75 &670 & 88 & 1993 Nov 26 & \\ \tableline
Q$0055-269$ & CRSP & K & 1.4x150 &1450& 50 & 1993 Dec 6  & $0\farcs8$ seeing \\ 
          & CRSP & J & 1.4x150 &550& 93 & 1993 Dec 7  & {variable seeing, cirrus} \\ 
        & OSIRIS & JHK & 1.2x120 &550 &102 & 1993 Dec 5  & \\ \tableline
Q$0114-089$ & CRSP & J & 1.4x150 &550 & 53 & 1993 Dec 4  & {$1\farcs4$ seeing}\\ 
  & Fspec & K & 1.2x75 &670 & 56 & 1993 Nov 28 & {thin cirrus} \\ \tableline
Q$0347-383$ & OSIRIS & JHK & 1.2x120 &550 &198 & 1993 Dec 8  & \\ \tableline
Q$0420-388$A & OSIRIS & JHK & 1.2x120 &550 & 48 & 1994 Sep 26 & \\ \tableline
Q$0636+680$ & CRSP & J & 1.4x150 &550 & 80 & 1993 Dec 4  &{$\gtrsim1$ \asec\ seeing}\\ 
  & CRSP & K & 1.4x150 &670 & 47 & 1993 Dec 5  & {cirrus}\\ 
  & CRSP & K & 1.4x300 &1720 & 53 & 1993 Dec 6  & \\ 
  & Fspec & K & 1.2x75 &670 & 88 & 1993 Nov 27 & \\ 
  & Fspec & H & 1.2x75 &730 & 48 & 1993 Nov 29 & {$1\farcs8$ seeing; clouds} \\ \tableline 
Q$0956+122$ & CRSP & J & 1.4x150 &550 &137 & 1993 Dec 4  &{$1\farcs1$ seeing}\\ 
  & CRSP & K & 1.4x300 &1930 & 26 & 1993 Dec 6  & \\ 
  & Fspec & K & 1.2x75 &670 & 88 & 1993 Nov 27 & {thin clouds} \\ \tableline
Q$1159+123$ & CRSP & J & 1.4x150 &550 & 53 & 1993 Dec 5  &{1\asec\ seeing; some cirrus}\\ 
  & Fspec & K & 1.2x75 &670 & 72 & 1993 Nov 29 & \\ \tableline
Q$1208+101$ & CRSP & K & 1.4x150 &700 & 80 & 1993 Dec 6  &  \\ \tableline
Q$1935-692$ & OSIRIS & JHK & 1.2x120 &550 & 80 & 1994 Sep 25 & \\ \tableline
Q$1946+7658$& CRSP & J & 1.4x150 &550 & 57 & 1993 Dec 6  & $0\farcs8$ seeing\\  
  & Fspec & K & 1.2x75 &670 & 64 & 1993 Nov 26 & {$\sim2$\asec seeing}\\ \tableline
Q$2000-330$ & OSIRIS & JHK & 1.2x120 &550 &104 & 1994 Sep 22 & \\ \tableline
Q$2126-158$ & CRSP & J & 1.4x150 &550 & 53 & 1993 Dec 5  & {$1\farcs4$ seeing}\\ 
  & Fspec & K & 1.2x75 &670 & 40 & 1993 Nov 26 &{$\sim2$\asec\ seeing} \\  
  & OSIRIS & JHK & 1.2x120 &550 &128 & 1994 Sep 24 & \\ \tableline
Q$2204-408$ & OSIRIS & JHK & 1.2x120 &550 & 64 & 1994 Sep 22 & \\   \tableline
\enddata
\tablenotetext{a}{The slit width (in arcsec on the sky) x grating (gpm).}
\tablenotetext{b}{$R = \lambda / \Delta \lambda$, where $\Delta \lambda$ is the FWHM and 
corresponds to the 2-pixel dispersion.}
\tablenotetext{c}{Total exposure time in minutes.}
\end{deluxetable}



%% file: table6.tex
\begin{deluxetable}{lllllllll}
\tabletypesize{\tiny}
\tablecaption{IR Photometry\label{irphot}}
\tablewidth{0pt}
\tablehead{
\colhead{Quasar} & 
\colhead{Instrument} & 
\colhead{date (UT)}  & 
\colhead{J} & 
\colhead{H} & 
\colhead{K} & 
\colhead{L\tablenotemark{a}} & 
\colhead{N\tablenotemark{a}} &  
\colhead{comments\tablenotemark{b}}}
\startdata
Q0000-263 & OSIRIS & 1993 Dec 5 &\nodata & \nodata &$15.07\pm0.05$ & \nodata & \nodata &   \\ 
 & MMT-IRphot & 1993 Oct 30 & $16.09\pm0.05$ & $15.48\pm0.05$ &$14.85\pm0.05 $ & \nodata & \nodata & $c$\\ 
\tableline
Q0014+813 & MMT-IRphot & 1993 Jun 4 & \nodata & \nodata &$13.60\pm0.02$ &$>13.18$& \nodata &5\farcs3 \\ 
 & MMT-IRphot & 1993 Jun 8 & $14.88\pm0.01$ & $14.21\pm0.17$ &$13.61\pm0.02$ &$12.28\pm0.52$& \nodata & 5\farcs3 \\ 
 & MMT-IRbol & 1992 Nov 8 & \nodata & \nodata & \nodata & \nodata & $>8.81 $ & 5\farcs4 \\ 
\tableline
Q0055-269 & OSIRIS & 1993 Dec 5 & \nodata & \nodata &$15.48\pm0.06$ & \nodata & \nodata  & \\ 
 & OSIRIS & 1994 Sep 26 & \nodata & $15.92\pm0.09$ & \nodata & \nodata & \nodata  &\\ 
&MMT-IRphot& 1993 Sep 27 & \nodata & $16.06\pm0.06$ & $15.12 \pm 0.07$ &$>12.89$ & \nodata & 8\farcs7, $d$\\
\enddata
\tablenotetext{a}{Lower limits to the magnitudes were determined from the 3$\sigma$ flux upper limits following: $M_{3\sigma}=-2.5log_{10}[3\sigma(F)/F_{0}$] where $F_{0}=309$ Jy at L$^{\prime}$ (Cutri 1994, private communication) and $F_{0}=36$ Jy at N (Rieke, Lebofsky and Low 1985)}
\tablenotetext{b}{Comments include aperture used ($\arcsec$) for MMT IR photometer or bolometer observations and the additional notes:
{c. Photometry obtained and reduced by R. Cutri}; {d. Photometry obtained and reduced by S. 
Kenyon}; {e. A 10\arcsec, rather than 15\arcsec, throw used.}}

\end{deluxetable}

%% file: table7.tex
\begin{deluxetable}{lllllll}
\tabletypesize{\footnotesize}
\tablecaption{Log of Optical Spectrophotometric Observations\label{optspec}}
\tablewidth{0pt}
\tablehead{
\colhead{Quasar} & 
\colhead{Telescope} & 
\colhead{$\lambda$ (\AA)} & 
\colhead{slit/grating$^{a}$} & 
\colhead{R$^{b}$} &
\colhead{t$^{c}$} & 
\colhead{date (UT)}}
\startdata
Q$0000-263$ & CTIO 1.5-m & 5090-9580 & 4/158 &500 & 60 & 1993 Sep 16 \\
  & MMT  & 3685-8780 & 5/150   &60  & 15  & 1991 Sep 16  \\
  & MMT  & 3685-8780 & 1.5/150 &250 & 20  & 1991 Sep 16  \\\tableline
Q$0014+813$ & MMT  & 3685-8780 & 5/150   &60  & 5   & 1991 Sep 16 \\
          & MMT  & 3685-8780 & 1.5/150 &250 & 10  & 1991 Sep 16  \\\tableline
Q$0055-269$ & CTIO 1.5-m & 3500-8300 & 4/150 &400 & 60 & 1993 Sep 15 \\
          & CTIO 1.5-m & 5000-9500 & 4/158 &500 & 60 & 1993 Sep 16 \\\tableline
Q$0114-089$ & CTIO 1.5-m & 5000-9500 & 4/158 &500 & 60  & 1993 Sep 17 \\
\enddata
\tablenotetext{a}{The slit width (in arcsec on the sky) and grating (gpm).}
\tablenotetext{b}{$R = \lambda / \Delta \lambda$, where $\Delta \lambda$ is the FWHM.}
\tablenotetext{c}{The total exposure time in minutes.}
\tablenotetext{d}{from Bechtold \etal\ 1994.}
\end{deluxetable}

%% file: table8.tex
\begin{deluxetable}{lllllllll}
\tabletypesize{\scriptsize}
\tablecaption{Optical Photometric Observations \label{optphot}}
\tablewidth{0pt}
\tablehead{
\colhead{Quasar} & 
\colhead{telescope} & 
\colhead{date (UT)} & 
\colhead{U} & 
\colhead{B} & 
\colhead{V} & 
\colhead{R} & 
\colhead{I} \\
}
\startdata
Q0000-263  &CTIO 0.9-m & 93 Sep 12 &\dotdot &\dotdot  & \dotdot& $17.33\pm0.03$ & $16.89\pm0.03$ \\ 
           &CTIO 0.9-m & 93 Sep 12 &\dotdot &\dotdot &\dotdot & $17.34\pm0.03$ &\dotdot           \\ 
           &CTIO 0.9-m & 93 Sep 13 &\dotdot & $20.21\pm0.03$  &\dotdot &   \dotdot     &\dotdot    \\ \tableline
Q0014+813  &FLWO 1.2-m&93 Oct 27 &\dotdot &$17.56\pm0.04$ & \dotdot &$16.14\pm0.02$ & $15.66\pm0.02$ \\ 
           &FLWO 1.2-m&93 Oct 27 &\dotdot &$17.56\pm0.07$ & \dotdot & \dotdot & \dotdot  \\ 
           &FLWO 1.2-m&93 Oct 27 &\dotdot &$17.53\pm0.04$ & \dotdot & \dotdot & \dotdot  \\ \tableline
Q0055-269   &CTIO 0.9-m & 93 Sep 12 &\dotdot & \dotdot  &\dotdot & $17.57\pm0.03$ & $17.46\pm0.03$ \\ 
             &CTIO 0.9-m & 93 Sep 13 &\dotdot & $18.68\pm0.03$  &\dotdot &  \dotdot      & \dotdot  \\
\enddata
\tablecomments{As mentioned in the text, the CTIO observations were made in photometric conditions. During the runs at FLWO, however, we did not have strictly 
photometric weather, so some observations were repeated.}
\end{deluxetable}

%% file: table9.tex
\begin{deluxetable}{crrrrcc} 
\tabletypesize{\footnotesize}
\tablecolumns{7}
\tablecaption{{Adopted calibration data for IR and optical photometry}\label{flux0mag}}
\tablewidth{0pt}
\tablehead{
\multicolumn{1}{c}{Filter} &
\multicolumn{1}{c}{$\lambda_{0}$} &
\multicolumn{1}{c}{$\lambda_{-}$} &
\multicolumn{1}{c}{$\lambda_{+}$} &
\multicolumn{1}{c}{$F_{\nu}(\lambda_{0}$; Vega)}  &
\multicolumn{1}{c}{$F_{\nu}(\lambda_{0}$; PL)}\tablenotemark{a} &
\multicolumn{1}{c}{$\Delta m$ \tablenotemark{b}} \\
\colhead{} &
\colhead{[$\mu$m]} &
\colhead{[$\mu$m]} &
\colhead{[$\mu$m]} &
\colhead{[Jy]} &
\colhead{[Jy]} &
\colhead{}
}
\startdata
 U & 0.365 & 0.337 & 0.392 & 1353 & 1759 & $-0.28$\phn\phn \\ 
 B & 0.434 & 0.384 & 0.484 & 4390\tablenotemark{c} & 4970 & $-0.13$\phn\phn \\ 
 V & 0.542 & 0.495 & 0.589 & 3630 & 3610 & \phs0.006\phn \\ 
 R & 0.628 & 0.566 & 0.690 & 3080 & 3040 & \phs0.014\phn \\ 
 I & 0.836 & 0.725 & 0.948 & 2265 & 2265 & \phs0.0\phn\phn \\    
 J\tablenotemark{d} & 1.25 & 1.15 & 1.35 & 1603 & \nodata & \nodata \\    
 H\tablenotemark{d} & 1.60 & 1.40 & 1.80 & 1075 & \nodata & \nodata \\    
 K\tablenotemark{d} & 2.22 & 1.97 & 2.47 & 667 & \nodata &\nodata \\
\enddata
\tablenotetext{a}{
$F_{\nu}$($\lambda_{0};PL$) is the flux within a band that would be 
expected for a power law as opposed to a stellar spectrum. We compute
it by scaling the zero magnitude Vega flux by the ratio of the 
normalized convolution of a power law spectrum ($F_{\nu}(PL)$) with 
the filter ($\phi_{f}$) and detector ($\phi_{d}$) response functions to 
the likewise convolved Vega spectrum ($F_{\nu}(Vega)$; 
from Hayes 1985).
($F_{\nu}$($\lambda_{0};PL$)=$F_{\nu}$($\lambda_{0};Vega$)$\times[{\int{F_{\nu}(Vega)\phi_{f}\phi_{d} d\nu}/F_{\nu}(\lambda_{0}; Vega)}$] \\
$\times[{{F_{\nu}(\lambda_{0};PL)/\int{F_{\nu}(PL)\phi_{f}\phi_{d}d\nu}}}$].)
We computed $F_{\nu}$($\lambda_{0};PL$) assuming
both $\alpha = -0.3$ (Francis \etal\ 1991) and $\alpha = -1.0$ 
(Elvis \etal\ 1994a), and this column gives the average of these results, which did not differ 
by much. These fluxes, rather than those of Vega, were used to flux calibrate the 
optical photometry (\S \ref{ophot}) but not the IR (\S \ref{iphot}).}
\tablenotetext{b}{Synthetic color correction for spectra of a power law form: \\
$\Delta~m$=$-2.5\log(F_{\nu,PL}/F_{\nu,Vega})$.}
\tablenotetext{c}{Vega spectrum extrapolated over H$\gamma$ absorption.}
\tablenotetext{d}{Central wavelengths and bandpasses for the near-IR bands from:
G. Rieke and M. Rieke, private communication. 
References for zero-magnitude fluxes: Campins, Rieke \& Lebofsky 1985 (J, H and K),
Cutri 1994 (private communication, L$^{\prime}$), and Rieke, Lebofsky \& Low
1985 (N).
}
\end{deluxetable}

%% file: table10.tex
\begin{deluxetable}{llllll}
\tablewidth{0pt}
\tabletypesize{\scriptsize}
\tablecaption{Scaling factors applied to spectra\label{scalfac}}
\tablehead{
\colhead{Quasar/$\lambda\lambda$} & 
\colhead{source of spectrum} &
\colhead{$\lambda$ range} &
\colhead{$g$} & 
\colhead{${\sigma{(F_{p})} \over F_{p}}$\tablenotemark{a}} & 
\colhead{${\sigma{<A>} \over <A>}$}
}
\startdata
Q$0000-263$ & &&&&\\
  optical   & MMT Red & 4000-8800 & 1.1791\tablenotemark{g} & 0.023 & 0.036 \\*
            & CTIO    & 5100-9600 & 1.0494\tablenotemark{g} & 0.023 &  0.043 \\* 
   J        & OSIRIS & 1.20-1.33, 1.23-1.35 & 1.7425\tablenotemark{d} & 0.046 & 0.015 \\*
   H        & OSIRIS & 1.41-1.46, 1.48-1.83 & 1.7425\tablenotemark{d} & 0.046 & 0.015 \\*
   K        & OSIRIS & 1.94-2.33 & 1.7425\tablenotemark{d} & 0.046 & 0.015 \\  \tableline     
Q$0014+813$ & &&&&\\* 
  optical   & MMT Red & 3700-8800 & 1.0585\tablenotemark{h} & 0.014 & 0.010 \\
\enddata
\tablenotetext{a}{{$\sigma{(F_{p})}/F_{p}$ is equal to $0.4 \ln(10) \sigma{(m)}$ where the values $\sigma{(m)}$ are given in the logs of optical and IR photometry}}
\tablenotetext{b}{Normalized to match MMT J band photometry} 
\tablenotetext{c}{Normalized to match MMT H band photometry}
\tablenotetext{d}{Normalized to match MMT K band photometry}
\tablenotetext{e}{Normalized to match OSIRIS H band photometry}
\tablenotetext{f}{Normalized to match OSIRIS K band photometry}
\tablenotetext{g}{Normalized to match CTIO R band photometry} 
\tablenotetext{h}{Normalized to match 48\asec R band photometry}
\tablenotetext{i}{Normalized to match the average over the OSIRIS J band spectrum}
\end{deluxetable}

%% file: table11.tex
\begin{deluxetable}{lllcccc}
\tabletypesize{\footnotesize}
\tablecaption{Galactic and Ly$\alpha$ column densities toward the high redshift quasars\label{galext}}
\tablewidth{0pt}
\tablehead{
\colhead{Quasar}& 
\colhead{$\alpha$(1950)} & 
\colhead{$\delta$(1950)} & 
\colhead{$z_{QSO}$} & 
\colhead{N$_{HI}$(Gal)$^{a}$} &
\colhead{z$_{d Ly\alpha}$} &
\colhead{N$_{HI}(Ly\alpha)^{a}$}
}
\startdata
Q$0000-263$& 00 00 49.5&$-$26 20 01 &4.111\phn & \phn1.43\tablenotemark{b}&3.395\tablenotemark{c}&20.00\tablenotemark{c} \\*
Q$0014+813$&00 14 04.50&+81 18 29\tablenotemark{d}  &3.3866    &14.30\tablenotemark{b} &\nodata &\nodata \\*
Q$0055-269$&00 55 32.6&$-$26 59 25\tablenotemark{d} &3.6625  & \phn1.77\tablenotemark{b} &\nodata&\nodata \\*
Q$0114-089$&01 14 53& $-$08 57 20\tablenotemark{d}  &3.1626  & \phn2.44\tablenotemark{e} & \nodata & \nodata \\*
Q$0347-383$&03 47 54.20&$-$38 19 34\tablenotemark{d}&3.222\phn & \phn1.98\tablenotemark{e} & 3.0244\tablenotemark{f}  & \phn6.31\tablenotemark{f} \\*
Q$0420-388$A&04 20 29.90&$-$38 51 51\tablenotemark{d}&3.123\phn & \phn1.91\tablenotemark{g} &\nodata&\nodata \\*
Q$0636+680$&06 36 47.60 & +68 01 26\tablenotemark{d}&3.1775  & \phn5.11\tablenotemark{b} &\nodata &\nodata \\* 
Q$0956+122$&09 56 11.10&+12 17 07\tablenotemark{d}  &3.306\phn    & \phn2.88\tablenotemark{e} &\nodata &\nodata \\
\enddata
\tablenotetext{a}{Column densities in the Galaxy, N$_{HI}$(Gal) and the damped Ly$\alpha$
systems, N$_{HI}$(Ly$\alpha$), are in units of $10^{20}$ cm$^{-2}$.}
\tablenotetext{b}{from Table 1(e) of Murphy et al. 1996. $\tau \ll 1$ is
assumed (and so the peak observed brightness temperature, $T_{peak}=T_{spin}$).
The differences between the column densities they determine for $\tau \ll 1$ and
for $T_{spin}=150$K are less 5\% for the above quasars. Accurate to 
$<10^{19}$ cm$^{-2}$}
\tablenotetext{c}{N$_{HI}$(Ly$\alpha$) and $z_{abs}$ from Lanzetta \etal\ (1991).}
\tablenotetext{d}{positions from Schneider \etal\ 1992 }
\tablenotetext{e}{N$_{HI}$ from Stark \etal\ (1992)}
\tablenotetext{f}{from Williger \etal\ (1989).}
\tablenotetext{g}{from Table 1 of Elvis, Lockman \& Wilkes (1989). All values are for 
$T_{spin}=250$K and are accurate to $<10^{19}$ cm$^{-2}$}
\tablenotetext{h}{The redshift and column density of the damped Ly$\alpha$ system in the
foreground of HS1946+7658 are from Hagen \etal\ (1992; redshift) and Dobrzycki \& 
Bechtold (1996; redshift and column density).}
\end{deluxetable}

%% file: table12.tex
\begin{deluxetable}{lcccclll}
\tabletypesize{\scriptsize}
\tablecolumns{8}
\tablecaption{Parameters used in correcting for `Ly$\alpha$ forest' absorption\label{lymatch}}
\tablewidth{0pt}
\tablehead{
\colhead{Quasar} &
\multicolumn{1}{c}{ref\tablenotemark{a}} &
\multicolumn{1}{c}{source of\tablenotemark{b}} &
\colhead{smooth\tablenotemark{c} } &
\colhead{shift\tablenotemark{d} } &
\multicolumn{1}{c}{$\lambda$\tablenotemark{e}} &
\multicolumn{1}{c}{fit\tablenotemark{f}} &
\multicolumn{1}{c}{valid fit\tablenotemark{g}} \\
\colhead{} &
\colhead{high R} &
\colhead{spectrophot} &
\colhead{} &
\colhead{} &
\colhead{range} &
\colhead{order } &
\colhead{range (\AA)} 
}
\startdata
Q0000-263 & 3 & 1 & 2 & 7  & 5850-6250 & 3 & 5850-6250 \\*
Q0000-263 & 3 & 2 & 1 & 16 & 5850-6250 & 3 & 5850-6250 \\*\tableline
%
Q0014+813 & 1 & 1 & 3 & 22 & 4580-5425 & 5 & 4600-5400 \\*\tableline
%
Q0420-388(1) & 5 &  2   & 1 & 13  & 3770-3910 & 4 (fit to 1 only)& 3775- \\*
Q0420-388(2) & 5 &      & 1 & 13  & 3965-4140 & 8 (fit to 1-6)&\\*
Q0420-388(3) & 5 &      & 1 & 13  & 4195-4390 & 8 (fit to 1-6)&\\*
Q0420-388(4) & 5 &      & 1 & 22  & 4435-4640 & 8 (fit to 1-6)&\\*
Q0420-388(5) & 5 &      & 1 & 24  & 4695-4920 & 3 (fit to 5 only)& \\
\enddata
\end{deluxetable}

%% file: table13.tex
\begin{deluxetable}{lllllcl}
\tablewidth{0pt}
\tablecolumns{7}
\tabletypesize{\scriptsize}
\tablecaption{Construction of quasi-simultaneous and simultaneous low $z$ SEDs \label{simultan}}
\tablehead{
\colhead{Object(yr)} & 
\colhead{IUE obs} & 
\colhead{IUE date} & 
\colhead{optical spec\tablenotemark{a}} & 
\colhead{optical phot\tablenotemark{a}} & 
\colhead{Comments\tablenotemark{b}} &
\colhead{used?\tablenotemark{c}}
}\startdata
Q0007+106& SWP01806 & 78 Jun 18      & 1980 Nov 15 & 1986 Sep 28& N & $\surd$\\* 
         & LWR01687 & 78 Jun 18      & && & \\* 
         & SWP23248 & 84 Jun 12      & && & \\* 
         & LWP03560 & 84 Jun 12      & && & \\ \tableline
Q0026+129& SWP10953 & 81 Jan 4 & 1980 Nov 14 & 1981 Feb 5& S & $\surd$\\* 
         & LWR09629 & 81 Jan 4 &      &      & & \\ \tableline
Q0050+124& SWP18557 & 82 Nov 15 & 1986 & 1986 Sep 28& N & $\surd$\\* 
         & LWR14626 & 82 Nov 14& & & &\\
\enddata
\tablenotetext{a}{References to the sources of the optical spectrophotometry and
photometry are given in Elvis \etal\ (1994)}
\tablenotetext{b}{Notes on the spectra: S=simultaneous (data taken within 1 year); 
Q=quasi-simultaneous (data taken within 2 years); N=non-simultaneous (time span
greater than 2 years); R=was rejected from the sample, as noted in Table 2}
\tablenotetext{c}{used ($\surd$) or not used (X) in data analysis}
\tablenotetext{d}{used in analysis, except when its luminosity at $\lambda_{rest}=1460$
\AA\ is needed, \eg\ 1285-1460\AA\ spectral index}
\end{deluxetable}

%% file: table14.tex
\begin{deluxetable}{clcl}
\tabletypesize{\small}
\tablecaption{Narrow Bands\label{windows}}
\tablewidth{0pt}
\tablehead{
\colhead{central wavelength} & 
\colhead{spectral range} & 
\colhead{continuum?} & 
\colhead{comments} 
}
\startdata
 1115\AA & 1100 - 1130\AA &  yes & close to a continuum window used by \\* 
         &                &      & Zheng et al. 1997 (HST quasar composite)\\* 
 1160\AA & 1150 - 1170\AA &  yes & just blueward of Ly$\alpha$ \\*  
 1285\AA & 1275 - 1295\AA &  yes & between Ly$\alpha$ and SiIV/OIV]; \\* 
         &                &      & from Francis et al. (1991)  \\* 
 1460\AA & 1450 - 1470\AA &  yes & between SiIV/OIV] and CIV  \\* 
 2200\AA & 2180 - 2220\AA &  yes & from Francis et al. (1991) \\* 
 2500\AA & 2480 - 2520\AA &  no  & FeII multiplets 1-6,33-36 \\* 
         &                &      & and 64 (Joly 1993) \\* 
 2660\AA & 2640 - 2680\AA &  yes & blueward of MgII \\* 
 3023\AA & 3001 - 3041\AA &  yes & redward of MgII \\* 
 4200\AA & 4180 - 4220\AA &  yes & for low $z$, between IUE and \\* 
	 &                &      & optical data; for high $z$, between \\* 
	 &                &      & optical and IR; Francis et al. (1991) \\* 
 4750\AA & 4730 - 4770\AA &  yes & blueward of H$\beta$/[OIII] \\* 
 5100\AA & 5080 - 5120\AA &  yes & redward of H$\beta$/[OIII]; \\* 
 7500\AA & 7480 - 7520\AA &  yes & long wavelength constraint\\*  \tableline
\enddata
\end{deluxetable}

%% file: table15.tex
\begin{deluxetable}{l | lllcc | lllcc | lllcc}
\tabletypesize{\footnotesize}
\tablewidth{0pt}
\tablecolumns{16}
\tablecaption{Narrow band luminosities (High $z$ Sample)\label{hiz_cfluxes}}
\tablehead{}
\tablehead{
\colhead{$\lambda_{o}$} &
\colhead{$\log \nu L_{\nu}$\tablenotemark{a,b}} &
\colhead{$\sigma_{avg}$} &
\colhead{$\sigma_{var}$} &
\colhead{N\tablenotemark{c}} &
\colhead{Q\tablenotemark{d}} &
\colhead{$\log \nu L_{\nu}$} &
\colhead{$\sigma_{avg}$} &
\colhead{$\sigma_{var}$} &
\colhead{N} &
\colhead{Q} &
\colhead{$\log \nu L_{\nu}$} &
\colhead{$\sigma_{avg}$} &
\colhead{$\sigma_{var}$} &
\colhead{N} &
\colhead{Q}
}
\startdata
\colhead{} & \multicolumn{5}{|c|}{Q$0000-263$}& \multicolumn{5}{|c|}{Q$0014+813$}& \multicolumn{5}{|c}{Q$0055-269$} \\\tableline
1285\AA & 47.72 & 0.01 & 0.02 & 29    & 3  & 48.06 & 0.01 & 0.01 & 13    & 3  & 47.21 & 0.02 & 0.03 & 11    & 3 \\*
1460\AA & 47.66 & 0.01 & 0.02 & 29    & 3  & 48.01 & 0.01 & 0.01 & 14    & 3  & 47.13 & 0.02 & 0.04 & 12    & 3 \\*
2200\AA & 47.55 & 0.11 & 0.18 & 24,33 & 2  & 47.74 & 0.02 & 0.02 & 21,34 & 1  & 47.13 & 0.06 & 0.09 & 22,30 & 2 \\*
2500\AA & 47.52 & 0.03 & 0.09 & 18    & 3  & 47.78 & 0.01 & 0.01 & 12    & 3  & 47.12 & 0.02 & 0.04 & 18    & 3 \\*
2660\AA & 47.49 & 0.10 & 0.10 & 1     & 3  & 47.74 & 0.00 & 0.01 & 25    & 3  & 47.13 & 0.02 & 0.08 & 41    & 3 \\*
3023\AA & 47.45 & 0.03 & 0.09 & 15    & 3  & 47.70 & 0.01 & 0.01 & 16    & 3  & 47.06 & 0.05 & 0.07 & 21,41 & 2 \\*
4200\AA & 47.34 & 0.03 & 0.06 & 10    & 3  & 47.63 & 0.01 & 0.01 & 13,1  & 2  & 46.85 & 0.06 & 0.09 & 13,16 & 1 \\
\enddata
\tablenotetext{a}{Values of $\lavgl$ are $\nulnu$ where the average is taken over
the indicated spectral range.  When no datapoints lie within the band, the value $\nu
L_{\nu}$ is determined, as described in the text, by interpolating between the averaged
fluxes, $<\nu L_{\nu}>$, where the averages are taken over the first 50\AA\ longwards
and shortwards of the nearest datapoint on the long and short wavelength sides of the
band, respectively.}
\tablenotetext{b}{N is the number of datapoints within the band.  When no datapoints
lie within the band, the column for N contains the number of datapoints within the
50\AA\ bands on the long and short wavelength sides (N$_{long}$, N$_{short}$).}
\tablenotetext{c}{When no datapoints lie within the band and the band lies beyond the
extent of our dataset, a value for $\nu L_{\nu}$ could not be computed.}
\tablenotetext{d}{The numbers in this column indicate the confidence that the computed
luminosity represents the continuum level: 3 signifies that the luminosity was computed
directly from points within the band and represents the continuum level well; 2 that it
was interpolated but looks reasonable; 1 that it does not represent the continuum level
well; and 0 that it could not be computed because of insufficient data coverage to the
red.}
\end{deluxetable}

%% file: table16.tex
\begin{deluxetable}{l | lllcc | lllcc | lllcc}
\tabletypesize{\footnotesize}
\tablewidth{0pt}
\tablecolumns{16}
\tablecaption{Narrow band luminosities (Low $z$ Sample)\label{lowz_cfluxes}}
\tablehead{
\colhead{$\lambda_{o}$} &
\colhead{$\log \nu L_{\nu}$}\tablenotemark{a} &
\colhead{$\sigma_{avg}$} &
\colhead{$\sigma_{var}$} &
\colhead{N\tablenotemark{b}} &
\colhead{Q\tablenotemark{c}} &
\colhead{$\log \nu L_{\nu}$} &
\colhead{$\sigma_{avg}$} &
\colhead{$\sigma_{var}$} &
\colhead{N} &
\colhead{Q} &
\colhead{$\log \nu L_{\nu}$} &
\colhead{$\sigma_{avg}$} &
\colhead{$\sigma_{var}$} &
\colhead{N} &
\colhead{Q}
}
\startdata
\colhead{} & \multicolumn{5}{|c|}{Q$0007+106$} & \multicolumn{5}{|c|}{Q$0026+129$} & \multicolumn{5}{|c}{Q$0050+124$} \\\tableline
1285\AA & 44.91 & 0.03 & 0.09 & 33  & 3  & 45.27 & 0.05 & 0.10 & 15 &  3  & 44.44 & 0.04 & 0.12 & 55 & 3 \\*
1460\AA & 44.86 & 0.04 & 0.14 & 35  & 3  & 45.19 & 0.04 & 0.06 & 17 &  3  & 44.45 & 0.04 & 0.10 & 50 & 3 \\*
2200\AA & 44.88 & 0.03 & 0.10 & 44  & 3  & 45.22 & 0.05 & 0.11 & 12 &  3  & 44.59 & 0.04 & 0.11 & 67 & 3 \\*
2500\AA & 44.95 & 0.03 & 0.07 & 46  & 3  & 45.14 & 0.05 & 0.10 & 21 &  3  & 44.75 & 0.03 & 0.05 & 66 & 3 \\*
2660\AA & 44.90 & 0.03 & 0.06 & 46  & 3  & 45.18 & 0.05 & 0.11 & 10 &  1  & 44.69 & 0.03 & 0.06 & 69 & 3 \\*
3023\AA & 44.79 & 0.04 & 0.06 & 1,33& 2  & 45.02 & 0.02 & 0.02 & 1,1& 2  & 44.75 & 0.04 & 0.12 & 21 & 3 \\*
4200\AA & 44.53 & 0.04 & 0.04 & 1,1 & 2  & 44.89 & 0.00 & 0.00 & 1,1& 2  & 44.56 & 0.02 & 0.02 & 1  & 3 \\
\enddata
\tablenotetext{a}{Values of $\lavgl$ are $\nulnu$ where the average is taken over
the indicated spectral range. When no datapoints lie within the band, the value
$\nu L_{\nu}$ is determined, as described in the text, by interpolating between the
averaged fluxes, $<\nu L_{\nu}>$, where the averages are taken over the first
50\AA\ longwards and shortwards of the nearest datapoint on the long and short
wavelength sides of the band, respectively.}
\tablenotetext{b}{N is the number of datapoints within the band.  When no datapoints
lie within the band, the column for N contains the number of datapoints within the
50\AA\ bands on the long and short wavelength sides (N$_{long}$, N$_{short}$).}
\tablenotetext{c}{The numbers in this column indicate the confidence that the computed
luminosity represents the continuum level: 3 signifies that the luminosity was computed
directly from points within the band and represents the continuum level well; 2 that it
was interpolated but looks reasonable; 1 that it does not represent the continuum level
well; and 0 that it could not be computed because of insufficient data coverage to the
red.}
\tablenotetext{d}{At the redshift of Q0837-120, the 2660\AA\ band falls on long
wavelength edge of the LWP spectrum (LWR04214) which is (noisy and) rising to longer
wavelengths, possibly because it is close to MgII$\lambda2798$. Hence, the average
at 2660\AA\ overestimates the continuum level.}
\tablenotetext{e}{The 3023\AA\ average in Q1202+281 (1985) is interpolated from the
IUE spectrum and the B band photometry. The edge of the LWP spectrum (LWP04904)
rises toward longer wavelengths, probably as it approaches MgII$\lambda2978$.  Thus
the interpolated value overestimates the continuum.}
\end{deluxetable}

%% file: table17.tex
%
%
%
\begin{deluxetable}{lllllllll}
\tabletypesize{\scriptsize}
\tablewidth{0pt}
\tablecolumns{9}
\tableheadfrac{0.01}
\tablecaption{UV luminosities \label{uvfluxes}}
\tablehead{
\colhead{} &
\multicolumn{4}{c}{1100-1130\AA} &
\multicolumn{4}{c}{1150-1170\AA} \\
\cline{2-5} \cline{6-9} \\
\colhead{Quasar} &
\colhead{log($\nu L_{\nu}$)} &
\colhead{$\sigma_{avg}$} &
\colhead{$\sigma_{var}$} &
\colhead{N} &
\colhead{log($\nu L_{\nu}$)} &
\colhead{$\sigma_{avg}$} &
\colhead{$\sigma_{var}$} &
\colhead{N}
}
\startdata
Q0000-263\tablenotemark{a} & 47.48 & 0.02 & 0.12 &44 &47.46 & 0.03 & 0.12 &29 \\* 
Q0000-263(CTIO) & \nodata & \nodata & \nodata & \nodata &47.78 & 0.04 & 0.04 &185 \\* 
Q0000-263(MMT) & \nodata & \nodata & \nodata & \nodata &47.79 & 0.05 & 0.05 &185 \\* 
Q0014+813\tablenotemark{a} & 47.99 & 0.01 & 0.03 &21 &47.94 & 0.01 & 0.05 &13 \\* 
Q0014+813 & 48.08 & 0.04 & 0.05 &526 &48.09 & 0.04 & 0.05 &350 \\* 
Q0055-269\tablenotemark{a} & 47.06 & 0.03 & 0.10 &18 &47.07 & 0.03 & 0.10 &11 \\* 
Q0114-089 & \nodata & \nodata & \nodata & \nodata & \nodata & \nodata & \nodata & \nodata \\* 
Q0347-383\tablenotemark{a} & 47.09 & 0.02 & 0.07 &16 &46.74 & 0.10 & 0.31 &11 \\
\enddata
\tablenotetext{a}{Data not corrected for Ly$\alpha$ forest absorption.}
\end{deluxetable}

%% file: table18.tex
\begin{deluxetable}{lllrllrr}
\tabletypesize{\scriptsize}
\tablecolumns{8}
\tablecaption{Linear Least Square fits from $1285-5100$\AA \label{aouv}}
\tablewidth{0pt}
\tablehead{
\colhead{} & 
\multicolumn{3}{c}{All 9 points from 1285-5100\AA} &
\multicolumn{3}{c}{5 points, excluding `3000\AA\ bump'} &
\colhead{} \\
\cline{2-4} \cline{5-7} \\
\colhead{Quasar} & 
\colhead{$\chi^{2}_{r}$ (dof)\tablenotemark{a}} & \colhead{P($\chi^{2}$)\tablenotemark{b}} & \colhead{$\alpha_{ouv}$} & 
\colhead{$\chi^{2}_{r}$ (dof)\tablenotemark{a}} & \colhead{P($\chi^{2}$)\tablenotemark{b}} & \colhead{$\alpha_{ouv}$} & 
\colhead{$\Delta$\tablenotemark{c}}}
\startdata
\cutinhead{High redshift sample}
Q0000-263 & \phn\phn1.7 (5) & 0.8777 & $-0.30 \pm 0.02$  & \phn\phn4.3 (1) & 0.9609 & $-0.29 \pm 0.02$ 	& $0.01$ \\* 
Q0014+813 & \phn16.4 (6) & 1.0 & $-0.14 \pm 0.03$  & \phn\phn2.9 (3) & 0.9674 & $-0.13 \pm 0.01$ 	& $0.01$ \\* 
Q0055-269 & \phn\phn8.1 (6) & 1.0 & $-0.40 \pm 0.10$  & \phn\phn3.8 (2) & 0.9765 & $-0.40 \pm 0.04$ 	& $0.00$ \\* 
Q0114-089 & \phn\phn1.6 (6) & 0.8707 & $-0.27 \pm 0.02$  & \phn\phn2.0 (2) & 0.8672 & $-0.26 \pm 0.01$ 	& $0.00$ \\* 
Q0347-383 & \phn\phn5.8 (7) & 1.0 & $-0.23 \pm 0.07$  & \phn\phn3.6 (3) & 0.9868 & $-0.25 \pm 0.04$ 	& $-0.02$ \\* 
Q0420-388 & \phn\phn1.7 (7) & 0.8941 & $-0.61 \pm 0.02$  & \phn\phn1.8 (3) & 0.8565 & $-0.61 \pm 0.01$ 	& $-0.00$ \\* 
Q0636+680 & \phn\phn9.0 (7) & 1.0 & $-0.57 \pm 0.04$  & \phn16.8 (3) & 1.0 & $-0.56 \pm 0.05$ 	& $0.01$ \\* 
Q0956+122 & \phn\phn4.3 (7) & 0.9999 & $-0.22 \pm 0.05$  & \phn\phn6.4 (3) & 0.9997 & $-0.22 \pm 0.06$	& $-0.00$ \\
\enddata
\tablenotetext{a}{Reduced $\chi^{2}$ of the power law fit. The number of degrees of freedom
is shown in parenthesis.} 
\tablenotetext{b}{Probability of randomly obtaining a value of $\chi^{2}_{r}$ better than 
that computed from the fit. These are all quite high, indicating either the fits are poor, or
the errors are underestimated.}
\tablenotetext{c}{$\Delta$ is the difference between the spectral indices
computed using all points and using all but the 3000\AA\ bump points.}
\end{deluxetable}

%% file: table19.tex
\begin{deluxetable}{llrrcc}
\tabletypesize{\normalsize}
\tablecaption{Statistics on the 1285-5100\AA\ optical/UV spectral indices, $\alpha_{ouv}$ \label{astats}}
\tablewidth{0pt}
\tablehead{
\colhead{Sample\tablenotemark{a}}&
\colhead{N$_{obj}$}&
\colhead{median}&
\colhead{mean}&
\colhead{$\sigma_{\tiny{N-1}}$}&
\colhead{${\sigma_{\tiny{N-1}} \over \sqrt(N)}$}
}
\startdata
low $z$ (E)&27 & $-0.40$ &$-0.38$ &$0.35$ &$0.07$ \\* 
high $z$(E)&15 & $-0.29$ &$-0.32$ &$0.28$ &$0.07$ \\*   
low+high $z$(E)&42 & $-0.40$ &$-0.36$ &$0.32$ &$0.05$ \\* 
low $z$ (A)&27 & $-0.41$ &$-0.36$ &$0.36$ &$0.07$ \\* 
high $z$(A)&15 & $-0.30$ &$-0.32$ &$0.28$ &$0.07$ \\* 
low+high $z$(A)&42 & $-0.38$ &$-0.34$ &$0.33$ &$0.05$ \\* 
\enddata
\tablenotetext{a}{
(A): Power law fits were made to all good luminosities (marked 3 or 2 in tables
\ref{hiz_cfluxes} and \ref{lowz_cfluxes}) between 1285\AA\ and 5100\AA;
(E): Luminosities between 2000\AA\ and 4000\AA\ were excluded from the power law fits.
}
 
\end{deluxetable}

%% file: table20.tex
\begin{deluxetable}{lrrrrrrr}
\tabletypesize{\scriptsize}
\tablecaption{UV and optical point-to-point spectral indices\label{uvao}}
\tablewidth{0pt}
\tablehead{
\colhead{Object} &
\colhead{$\alpha_{1115\AA}^{1285\AA}$} &
\colhead{$\alpha_{1160\AA}^{1285\AA}$} &
\colhead{$\alpha_{1285\AA}^{1460\AA}$} &
\colhead{$\alpha_{1285\AA}^{2200\AA}$} &
\colhead{$\alpha_{1285\AA}^{3023\AA}$} &
\colhead{$\alpha_{4200\AA}^{5100\AA}$} &
\colhead{$\alpha_{3023\AA}^{5100\AA}$} 
}
\startdata
\cutinhead{High $z$ sample}
Q$0000-263$& \nodata&$0.51\pm1.03$\tablenotemark{a}& $ 0.12\pm0.37$&$-0.27\pm0.47$&$ -0.28\pm 0.09$&\nodata&\nodata \\
Q$0014+813$& $-0.71\pm0.72$&$-0.46\pm1.00$& $-0.09\pm0.20$&$ 0.36\pm0.08$&$ -0.03\pm 0.03$&$   0.15\pm0.15$&$  -0.23\pm 0.05$\\
Q$0055-269$& \nodata&\nodata& $ 0.44\pm0.48$&$-0.66\pm0.27$&$ -0.60\pm 0.14$&$  -0.95\pm0.96$&$  -0.06\pm 0.31$\\
Q$0114-089$& \nodata&\nodata& $-0.28\pm0.30$&$-0.17\pm0.08$&$ -0.24\pm 0.18$&$  -0.12\pm0.80$&$  -0.28\pm 0.36$\\
Q$0347-383$& \nodata&\nodata& $ 0.47\pm0.31$&$-0.48\pm0.13$&$ -0.36\pm 0.13$&$  -0.27\pm0.86$&$   0.09\pm 0.33$\\
Q$0420-388$& $-0.97\pm0.74$&$-0.55\pm1.02$& $-0.80\pm0.34$&$-0.59\pm0.08$&$ -0.65\pm 0.10$&$  -0.64\pm0.57$&$  -0.54\pm 0.22$\\
Q$0636+680$& $-0.69\pm0.71$&$ 0.55\pm0.99$& $ 0.22\pm0.13$&$-0.44\pm0.07$&$ -0.62\pm 0.06$&$   0.19\pm0.35$&$  -0.31\pm 0.11$\\
Q$0956+122$& \nodata&\nodata& $ 0.57\pm0.33$&$-0.16\pm0.18$&$ -0.25\pm 0.07$&$   1.01\pm0.46$&$   0.14\pm 0.13$\\
\enddata
\tablenotetext{a}{Used the Ly$\alpha$ forest corrected spectrophotometry from CTIO. The spectral index that is measured from
the corrected MMT spectrophotometry is similar: $\alpha(1160,1285) = 0.57\pm1.06$.}
\tablenotetext{b}{Used the Ly$\alpha$ forest corrected shorter wavelength spectrophotometry from the MMT. The spectral index that 
is measured from the longer wavelength set is $\alpha(1160,1285) = -2.01\pm1.10$.}
\end{deluxetable}

%% file: table21.tex
\begin{deluxetable}{llrrrrl}
\tabletypesize{\footnotesize}
\tablecaption{Linear-regression fits\label{uvcorr}}
\tablewidth{0pt}
\tablehead{
\colhead{$\alpha$} &
\colhead{$\alpha_{o}$} &
\colhead{$\log(L_{\nu})$} &
\colhead{$\log(1+z)$} &
\colhead{$p_{K-T}$\tablenotemark{a}} &
\colhead{N$_{obj}$} &
\colhead{reference} 
}
\startdata
\cutinhead{$\alpha_{uv}$ vs. $L_{uv}$: $H_{o}=75$km/s/Mpc, $q_{o}=0.1$}
$\alpha$(1285\AA,1460\AA)\tablenotemark{b} & $-6.0\pm5.0$ & $0.13\pm0.11$ & \nodata & 58& 41 & this work\\
$\alpha$(1285\AA,2200\AA) & $-7.12\pm2.49$ & $0.14\pm0.05$ & \nodata & 99& 42 & this work\\
$\alpha$(1285\AA,1460\AA)\tablenotemark{b} & $-0.34\pm0.18$ & \nodata & $0.45\pm0.46$ & 82& 41 & this work\\
$\alpha$(1285\AA,2200\AA) & $-0.71\pm0.09$ & \nodata & $0.48\pm0.23$ & 98& 42 & this work\\
\cutinhead{$\alpha_{uv}$ vs. $L_{uv}$: Comparison with O'Brien \etal\ (1988; $H_{o}=50$km/s/Mpc, $q_{o}=0$)}
$\alpha_{UV,OGW88}$ & $-1.58$ & $0.02\pm0.05$ & $0.71\pm0.20$ & $93,<80$ \tablenotemark{d} & 68 & O'Brien \etal\ 1988 \\
$\alpha_{UV,OGW88}$\tablenotemark{c} & $-5.64\pm1.43$ & $0.16\pm0.05$ & \nodata & $>99$ & 68 & O'Brien \etal\ 1988 \\
$\alpha$(1285\AA,1460\AA)\tablenotemark{b} & $-3.8\pm3.2$ & $0.12\pm0.10$ & \nodata & 54& 41 & this work\\
$\alpha$(1285\AA,2200\AA) & $-4.7\pm1.6$ & $0.13\pm0.05$ & \nodata & 99.0& 42 & this work\\
\enddata
\tablenotetext{a}{One minus the percentage probability of a chance correlation, 
based on a Kendall-tau correlation test.}
\tablenotetext{b}{Q0134+329 was not used in the fits because its 1460\AA\ luminosity, hence
also its 1285-1460\AA\ slope, are discrepant.}
\tablenotetext{c}{O'Brien \etal\ measure the UV spectral index from Ly$\alpha$ to 1900\AA.}
\tablenotetext{d}{The first number is the significance of the $\alpha-\log(1+z)$
partial correlation, and the second, that of the $\alpha-\log~L$ one.}
\tablenotetext{e}{Cheng \etal\ (1991) have fit to normalized variables, hence the quite 
different intercept values.}
\tablenotetext{f}{One-tailed significance of the $\alpha$-$z$ partial correlation 
when log$L$ is included in the fit.}
\tablenotetext{g}{No number given, but Cheng \etal\ state the relation between $\alpha$ and 
$L$ is not significant.}
\tablenotetext{h}{Significance not given.}
\end{deluxetable}

%% file: table22.tex
\begin{deluxetable}{llrrrrrrr}
\tabletypesize{\normalsize}
\tablewidth{0pt}
\tablecolumns{9}
\tablecaption{UV slope change\label{dauv}}
\tablehead{
\colhead{sample} & 
\colhead{$\Delta$}\tablenotemark{a} &
\colhead{total\tablenotemark{b}} &
\colhead{average} &
\colhead{$\sigma_{N-1}$}&
\colhead{$\sigma_{avg}$}&
\colhead{+\tablenotemark{c}} &
\colhead{0\tablenotemark{c}} & 
\colhead{$-$\tablenotemark{c}}}
\startdata
All &$\Delta_{02}^{23}$\tablenotemark{d} & 23 & 1.28 & 1.27 & 0.26 & 10 & 13 & 0 \\*  
    &$\Delta_{12}^{23}$\tablenotemark{d} & 32 & 0.28 & 1.79 & 0.32 & 7 & 20 & 5 \\* 
    &$\Delta_{02}^{24}$ & 23 & 0.71 & 1.07 & 0.22 & 9 & 12 & 2 \\* 
    &$\Delta_{12}^{24}$ & 32 & $-0.033$ & 1.37 & 0.24 & 4 & 21 & 7 \\* \tableline 

low z &$\Delta_{02}^{23}$\tablenotemark{d} & 18 & 1.38 & 1.39 & 0.33 & 7 & 11 & 0 \\* 
      &$\Delta_{12}^{23}$\tablenotemark{d} & 25 & 0.23 & 1.98 & 0.40 & 5 & 15 & 5 \\* 
      &$\Delta_{02}^{24}$ & 18 & 0.72 & 1.20 & 0.28 & 6 & 10 & 2 \\* 
      &$\Delta_{12}^{24}$ & 25 & $-0.13$ & 1.46 & 0.29 & 3 & 16 & 6 \\
\enddata
\tablenotetext{a}{
$\Delta_{02}^{23} \equiv \alpha_{1115\AA}^{1285\AA} - \alpha_{1285\AA}^{1460\AA}$;
$\Delta_{12}^{23} \equiv \alpha_{1160\AA}^{1285\AA} - \alpha_{1285\AA}^{1460\AA}$;
$\Delta_{02}^{24} \equiv \alpha_{1115\AA}^{1285\AA} - \alpha_{1285\AA}^{2200\AA}$;
and
$\Delta_{12}^{24} \equiv \alpha_{1160\AA}^{1285\AA} - \alpha_{1285\AA}^{2200\AA}$
}
\tablenotetext{b}{Number of sample objects with measured spectral indices}
\tablenotetext{c}{Number of objects showing a $>1\sigma$ turndown ($+$), turnup ($-$) or
whose spectral slopes on either side of Ly$\alpha$ differ by less than $1\sigma$ ($0$).}
\tablenotetext{d}{Statistics involving $\alpha_{23}$ exclude Q$0134+329$ from the sample.}
\end{deluxetable}

%% file: table23.tex
\begin{deluxetable}{lrrrrc}
\tabletypesize{\scriptsize}
\tablecaption{Measures of the $\lambda\sim3000$\AA\ excess\label{fe2}}
\tablewidth{0pt}
\tablehead{
\colhead{Quasar} & 
\colhead{R(2500\AA)\tablenotemark{a}} & 
\colhead{R(2660\AA)\tablenotemark{a}} & 
\colhead{excess\tablenotemark{b}} &
\colhead{`EW(2200-3023)'\tablenotemark{c}} &
\colhead{quality\tablenotemark{d}}
}
\startdata
\cutinhead{High redshift sample}
Q$0000-263$   &   $0.016  \pm  0.029$   &   $0.011  \pm  0.103$   &   $0.03\pm0.09$  & $23\pm77$ & 2333\\* 
Q$0014+813$   &   $-0.035  \pm  0.007$  &   $-0.045  \pm  0.004$  &  ($-0.12\pm0.01$) & ($-93\pm6$) & 1333\\* 
Q$0055-269$   &   $0.100  \pm  0.025$   &   $0.133  \pm  0.022$   &   $0.28\pm0.04$ & $232\pm44$ & 2332 \\* 
Q$0114-089$   &   ($0.005 \pm 0.082$)   &   ($0.002\pm0.077$)     &   $-0.01\pm0.08$  & $-8\pm70$ & 3222 \\* 
Q$0347-383$   &   ($0.070 \pm 0.099$)   &   ($0.112\pm0.090$)     &   $0.21\pm0.10$  & $172\pm97$ & 3223 \\* 
Q$0420-388$A   &  ($0.034\pm0.049$)    &   ($0.025\pm0.051$)      &   $0.04\pm0.05$  & $35\pm46$ & 3223 \\* 
Q$0636+680$   &   ($0.003\pm0.019$)    &   $0.005  \pm  0.020$    &  $0.02\pm0.02$   & $19\pm19$ &  2233 \\* 
Q$0956+122$   &   ($0.027\pm0.028$)    &   $0.036  \pm  0.020$    &  $0.06\pm0.03$   & $51\pm27$ & 2233 \\
\enddata
\tablenotetext{a}{Values in parentheses were not plotted or used
in the analysis, since they rely solely on interpolated fluxes or were determined using a discrepant flux point.}
\tablenotetext{b}{The excess is defined as $(L_{total} - L_{c})/L_{c}$ between 2200\AA\ and 3023\AA, where
$L_{c}$ is determined from the power law fit through the average fluxes in the narrow continuum bands from
1285\AA - 5100\AA\ but not including those between 2000\AA\ and 4000\AA.}
\tablenotetext{c}{`EW(2200-3023)' is computed from the 4 narrow band average fluxes at 2200\AA, 2500\AA, 2660\AA,
and 3023\AA\ and the continuum level as determined by the power law fits described in the previous note.}
\tablenotetext{d}{The digits, left to right, represent the quality of the average narrow band fluxes at
2200\AA, 2500\AA, 2660\AA\ and 3023\AA, as follows: 3 | directly computed; 2 | had to be interlopated; and
1 | look discrepant.}
\end{deluxetable}

%% file: table24.tex
\begin{deluxetable}{lrrrrrrrr}
\tabletypesize{\normalsize}
\tablecolumns{8}
\tablecaption{Optical change of slope\label{optstats}}
\tablehead{
\multicolumn{1}{l}{} &
\multicolumn{4}{c}{Short $\lambda$ (UV) spectral index} &
\multicolumn{4}{c}{Long $\lambda$ (optical) spectral index} \\
\cline{2-5} \cline{6-9} \\
\colhead{Sample} & 
\colhead{$N_{obj}$} &
\colhead{avg} &
\colhead{$\sigma_{N-1}$}&
\colhead{$\sigma_{avg}$}&
\colhead{$N_{obj}$} &
\colhead{avg} &
\colhead{$\sigma_{N-1}$}&
\colhead{$\sigma_{avg}$}}
\startdata
\cutinhead{\hspace{2.5in} $\alpha(1285\AA,3023\AA)$ \hspace{0.8in} $\alpha(3023\AA,5100\AA)$}
All & 39 & $-0.676$ & 0.437 & 0.070 &39& 0.176 & 0.464 & 0.074 \\*
high z & 12&$-0.437$& 0.342 & 0.099 &12& $-0.206$& 0.365 & 0.105 \\*
low z & 27 &$-0.782$& 0.438 & 0.084 &27& 0.345 & 0.402 & 0.077 \\*
\cutinhead{\hspace{2.5in} $\alpha(1285\AA,2200\AA)$ \hspace{0.8in} $\alpha(4200\AA,5100\AA)$ }
All & 39 & $-0.609$ & 0.449 & 0.072 &39& $-0.091$ & 0.833 & 0.133 \\*
high z&12& $-0.451$ & 0.404 & 0.117 &12& $-0.289$ & 0.551 & 0.159 \\*
low z& 27& $-0.679$ & 0.457 & 0.088 &27& $-0.004$ & 0.927 & 0.178 \\*
\cutinhead{\hspace{2.5in}$\alpha(1285\AA,1460\AA)$\tablenotemark{a} \hspace{0.8in} $\alpha(4200\AA,5100\AA)$}
All & 39 & $-0.330$ & 1.091 & 0.175 &39& $-0.091$ & 0.833 & 0.133 \\*
high z&12& $-0.036$ & 0.465 & 0.134 &12& $-0.289$ & 0.551 & 0.159 \\
\enddata
\tablenotetext{a}{Statistics involving $\alpha_{23}$ exclude Q$0134+329$ from the sample.}
\end{deluxetable}

%% file: ms.bbl
\begin{thebibliography}{}

\bibitem[Barthel, Tytler \& Thomson 1990]{btt} Barthel, P. D., Tytler, D. R. \& Thomson, B. \aaps, 82, 339.

\bibitem[Barvainis 1993]{barv93} Barvainis, R. 1993, \apj, 412, 513.

\bibitem[Bechtold 1994]{bechtold94} Bechtold, J. 1994, \apjs, 91, 1.

\bibitem[Bechtold \etal\ 1994]{brosat94} Bechtold, J., et al. 1994, \aj, 108, 374.

\bibitem[Bohlin \etal\ 1980]{iue80} Bohlin, R. C., Holm, A. V., Savage, B. D., Snijders, M. A. J. \& Sparks, W. M. 1980 \aap, 85, 1.


\bibitem[Bouchet \etal\ 1985]{bouchet} Bouchet, P., Lequeux, J., Maurice, E., Pr\'{e}vot, L. \& Pr\'{e}vot-Burnichon, M. L. 1985, \aap, 149, 330.

\bibitem[Boyle, Shanks \& Peterson 1988]{bsp88}  Boyle, B. J., Shanks, T. \& Peterson, B. A. 1988, \mnras, 235, 935.

\bibitem[Boyle 1990]{compboyle}  Boyle, B. J. 1990, \mnras, 243, 231.

\bibitem[Boyle \etal\ 1991]{bjsmzz91}  Boyle B. J., Jones L.R., Shanks T., Marano B., Zitelli V., \&  Zamorani G., 1991, in ASP Conf. Ser. 21, The Space Distribution of Quasars, ed. D. Crampton, (San Francisco: ASP), 191.

\bibitem[Boyle \etal\ 2000]{2dfevol} Boyle, B. J., Shanks, T., Croom, S. M., Smith, R. J., Miller, L., Loaring, N. \&  Heymans, C. 2000, \mnras, 317, 1014.

\bibitem[Campins, Rieke \&  Lebofsky 1985]{campins85} Campins, H., Rieke, G. H. \&  Lebofsky, M. J. 1985, \aj, 90, 896. 

\bibitem[Carballo \etal\ 1999]{carballo99} Carballo, R.,  Gonz\'alez-Serrano J. I., Benn, C. R., 
S\'anchez, S. F.  \&  Vigotti, M. 1999, MNRAS, 306, 137.

\bibitem[Carrili \etal\ 1998]{carilli} Carilli, C. L., Menten, K. M., Reid, M. J., Rupen, M. P. \&  Yun, M. S. 1998, \apj, 494, 175. 

\bibitem[Casali \&  Hawarden 1992]{UKIRTfs} Casali, M. M. \&  Hawarden, T. G. 1992, UKIRT Newsletter, 4, 33.

\bibitem[Cavaliere \etal\ 1988]{cavetal88} Cavaliere, A., Giallongo, E., Padovani, P. \&  Vagnetti, F. 1988, in ASP Conf. Ser. 2, Proceedings of a Workshop on Optical Surveys for Quasars, ed. P. 
Osmer, A. Porter, R. Green \&  C. Foltz (San Francisco: ASP), 335.

\bibitem[Cavaliere \&  Vittorini 1998]{cv98} Cavaliere, A. \&  Vittorini, V. 1998, in ASP Conf. Ser. 146, 
The Young Universe: Galaxy Formation and Evolution at Intermediate and High Redshift, eds. S. D'Odorico, 
A. Fontana \& E. Giallongo, (San Francisco: ASP), 26.

\bibitem[Cheng, Gaskell \&  Koratkar 1991]{cgk91} Cheng, F. H., Gaskell, C. M. \&  Koratkar, A. P. 1991, \apj, 370, 487.

\bibitem[Choi, Yang \&  Yi 1999]{choi99} Choi, Y., Yang, J. \&  Yi, I. 1999, ApJ, 518, L77

\bibitem[Choi, Yang \&  Yi 2001]{choi00} Choi, Y., Yang, J. \&  Yi, I. 2001, ApJ, 555, 673.


\bibitem[Collin \& Joly]{cj01} Collin, S. and Joly, M. 2001, New Astronomy Reviews, 44, 531.

\bibitem[Cooper, Bui \&  Bailey 1993]{nicmos3} Cooper, D. E, Bui, D. \&  Bailey, R. B. 1993, \procspie, 1946, 170.

\bibitem[Czerny \&  Elvis 1987]{czernyelvis87} Czerny, B. \&  Elvis, M. 1987, \apj, 321, 305.

\bibitem[Czerny \etal\ 1995]{czerny95} Czerny, B., Loska, Z., Szczerba, R., Cukierska, J. \&  Madejski, G. 1995, Acta Astron., 45, 623.

\bibitem[DePoy \etal\ 1993]{osiris} DePoy, D. L., Atwood, B., Byard, P., Frogel, J. \&  O'Brien, T. 1993, \procspie, 1946, 667. 

\bibitem[Dobrzycki \&  Bechtold 1996]{dobrz96} Dobrzycki, A. \&  Bechtold, J 1996 \apj, 457, 102.

\bibitem[Elias \etal\ 1982]{elias82} Elias, J. H., Frogel, J. A., Matthews, K. \&  Neugebauer, G. 1982, \aj, 87, 1029.

\bibitem[Elston, Thompson \&  Hill 1994]{eth94} Elston, R., Thompson, K. L. \&  Hill, G. J. 1994, \nat, 367, 250.

\bibitem[Elvis, Lockman \&  Wilkes 1989]{ewl89} Elvis, M., Lockman, F. J. \&  Wilkes, B. J. 1989, \aj, 97, 777.

\bibitem[Elvis 1992]{sedwind92} Elvis, M. 1992, in ``Frontiers of X-ray Astronomy'', eds.  Y. Tanaka 
\&  K. Koyama, Universal Academic Press, p.567.

\bibitem[Elvis \etal\ 1994a]{atlas} Elvis, M., Wilkes, B. J., McDowell, J. C., Green, R. F., Bechtold, J., Willner, S. P., Cutri, R., Oey, M. S. \&  Polomski, E., 1994a, \apjs, 95, 1.

\bibitem[Elvis \etal\ 1994b]{highzabs} Elvis, M., Fiore, F., Wilkes, B. J., McDowell, J. C. 1994b, \apj, 422, 60.

\bibitem[Elvis \etal\ 1998]{wgacatII} Elvis, M., Fiore, F., Giommi, P. \&  Padovani, P. 1998, \apj, 492, 91.

\bibitem[Fabricant \etal\ 1998]{fast94} Fabricant, D., Cheimets, P., Caldwell, N. \&  Geary, J. 1998, \pasp, 110, 79.

\bibitem[Fan et al. 2000]{fan58} Fan, X., \etal\ 2000, \aj, 120, 1167.

\bibitem[Fiore \etal\ 1998]{wgacatI} Fiore, F., Elvis, M., Giommi, P. \& Padovani, P. 1998, \apj, 492, 79.

\bibitem[Francis \etal~1991]{complbqs}  Francis, P. J., Hewett, P. C., Foltz, C. B., Chaffee, F. H, Weymann, R. J. \&  Morris, S. L. 1991, \apj, 373, 465.

\bibitem[Frank, King \&  Raine 1992]{fkr}  Frank, J., King, A. R. \&  Raine, D. J. 1992, Accretion Power in Astrophysics, 2nd Edition, (Cambridge: Cambridge University Press). 

\bibitem[Giallongo \etal\ 1999]{glens99}Giallongo, E., Fontana, A., Cristiani, S. \&  D'Odorico, S. 1999, \apj, 510, 605. 

\bibitem[Giveon \etal\ 1999]{pgvar99} Giveon, U., Maoz, D., Kaspi, S., Netzer, H. \&  Smith, P. S. 1999, \mnras, 306, 637.

\bibitem[Goldschmidt \etal\ 1992]{gold92} Goldschmidt, P., Miller, L., LaFranca, F. \&  Cristiani, S. 1992, \mnras, 256, 65P.

\bibitem[Goldschmidt \&  Miller 1998]{gold98} Goldschmidt, P. \&  Miller, L. 1998, \mnras, 293, 107.


\bibitem[Green \etal\ 1995]{green95} Green, P. J., et al. 1995, \apj, 450, 51.

\bibitem[Gregory \&  Condon 1991]{87gb} Gregory, P. C. \&  Condon, J. J. 1991, \apjs, 75, 1011.

\bibitem[Haehnelt \&  Rees 1993]{harees93} Haehnelt, M. G. \&  Rees, M. J. 1993, \mnras, 263, 168. 

\bibitem[Haehnelt, Natarajan \&  Rees 1998]{hnr98} Haehnelt, M. G. , Natarajan, P. \&  Rees, M. J. 1998, MNRAS, 300, 817.

\bibitem[Hagen \etal\ 1992 ]{hagen92} Hagen, H.-J., Cordis, L., Engels, D., Groote, D., Haug, U., Heber, U., K\"ohler, Th.,
Wisotzki, L. \& Riemers, D. 1992, \aap, 253, L5.

\bibitem[Haiman \&  Menou 2000]{hm2000} Haiman, Z. \&  Menou, K. 2000, \apj, 531, 42.


\bibitem[Hamuy \etal\ 1992]{hamuy92} Hamuy, M., Walker, A. R., Suntzeff, N. B., Gigoux, P., Heathcote, S. R. \&  Phillips, M. M. 1992, \pasp, 104, 533.

\bibitem[Hawarden \etal\ 2001]{UKIRTfs2} Hawarden, T. G., Leggett, S. K., Letawsky, M. B., Ballantyne, D. R. \&  Casali, M. M. 2001, \mnras, 325, 563.

\bibitem[Hawkins \&  V\'{e}ron 1993]{varsel93} Hawkins, M. R. S. \&  V\'{e}ron, P. 1993, \mnras, 260, 202.

\bibitem[Hawkins \&  V\'{e}ron 1995]{varsel95} Hawkins, M. R. S. \&  V\'{e}ron, P. 1995, \mnras, 275, 1102.

\bibitem[Hayes 1985]{hayes85} Hayes, D. S. 1985, in ``Calibration of Fundamental Stellar Quantities: proceedings of the 111th Symposium of the IAU'', eds. D. S. Hayes, L. E. Pasinetti \&  A. G. Davis Phillip, [Boston: D. Reidel Publishing Co.], pg 225. 

\bibitem[Hewett, Foltz \&  Chaffee 1993]{lbqsevol93} Hewett, P. C., Foltz, C. B. \&  Chaffee, F. H. 1993, \apj, 406, L43.

\bibitem[Hewett, Foltz \&  Chaffee 1995]{lbqs95} Hewett, P. C., Foltz, C. B. \&  Chaffee, F. H. 1995, \aj, 109, 1498.

\bibitem[Hewitt \&  Burbidge 1991]{hb91} Hewitt, A. \&  Burbidge, G. 1991, \apjs, 63, 1.

\bibitem[Hewitt \&  Burbidge 1993]{hb93} Hewitt, A. \&  Burbidge, G. 1993, \apjs, 87, 451.

\bibitem[Jauncey \etal\ 1989]{jaun} Jauncey, D. L, Savage, A., Morabito, D. D., Preston, R. A., 
Nicolson, G. D. \&  Tzioumis, A. K. 1989, \aj, 98, 54.

\bibitem[Johnson 1966]{johnson66} Johnson, H. L. 1966, \araa, 4, 193.

\bibitem[Joly 1993]{joly93} Joly, M. 1993, Annales de Physique, 18(3), p. 241.


\bibitem[Joyce, Fowler \&  Heim 1994]{crspref} Joyce, R. R., Fowler, A. M. \&  Heim, G. B. 1994, in Proc. SPIE, 2198, 725.

\bibitem[Kauffmann \&  Haehnelt 2000]{kh2000} Kauffmann, G. \&  Haehnelt, M. 2000, \mnras, 311, 576.

\bibitem[Keller, Sabol \&  Rieke 1990]{irbolo} Keller, L. D., Sabol B. A. \&  Rieke, G. H. 1990, \procspie, 1235, 160.

\bibitem[Kennefick, Djorgovski \&  de Carvalho 1995]{kdc} Kennefick, J. D., Djorgovski, S. G. \&  de Carvalho, R. R. 1995, \apj, 110, 2553.

\bibitem[K\"ohler \etal\ 1997]{kohler97} K\"ohler, T., Groote, D., Reimers, D. \&  Wisotzki, L. 1997, \aap, 325, 502.

\bibitem[Koo \&  Kron 1988]{kookron88} Koo, D. C. \&  Kron, R. G. 1988, \apj, 325, 92.

\bibitem[Krolik \&  Kallman 1988]{krolikk88} Krolik, J. H. \&  Kallman, T. R. 1988, \apj, 324, 714.

\bibitem[Kuhn \etal\ 1995]{kuhn95} Kuhn, O., Bechtold, J., Cutri, R., Elvis, M. \&  Rieke, M. 1995, \apj, 438, 643.

\bibitem[Kurpiewski, Kuraszkiewicz, \&  Czerny 1997]{kkc97} Kurpiewski, A., Kuraszkiewicz, J. \&  Czerny, B. 1997, \mnras, 285, 725.

\bibitem[Kwan \etal\ 1995]{kwan95} Kwan, J. Cheng, F.-Z., Fang, L.-Z., Zheng, W. and Ge, J. 1995, \apj, 440, 628. 

\bibitem[LaFranca \&  Cristiani 1995]{lafranca95} LaFranca, F. \&  Cristiani, S. 1997, \aj, 113, 1517.

\bibitem[Landolt 1992]{landolt92} Landolt, A. 1992, \aj, 104, 340.

\bibitem[Lanzetta \etal\ 1991]{lanz} Lanzetta, K. M., Wolfe, A. M., Turnshek, D. A., Lu, L., McMahon, R. G. \&  Hazard, C. 1991, \apjs, 77, 1.

\bibitem[Laor \&  Netzer 1989]{ln89} Laor, A. \&  Netzer, H. 1989, \mnras, 238, 897.

\bibitem[Laor et al. 1997]{laor97} Laor, A., Fiore, F., Elvis, M., Wilkes, B. J. \& McDowell, J. C. 1997, 
\apj, 477, 93.

\bibitem[Lawrence \etal\ 1997]{lawrence97} Lawrence, A., Elvis, M., Wilkes, B. J., McHardy, I. \&  Brandt, N. 1997, \mnras, 285, 879. 

\bibitem[L\'ehar \etal\ 2000]{lehar2000} L\'ehar, J., Falco, E. E., Kochanek, C. S., McLeod, B. A., Mu\~noz, J. A., Impey, C. D., Rix, H.-W., Keeton, C. R. \&  Peng, C. Y. 2000, \apj, 536, 584.

\bibitem[Londish, Boyle \&  Schade 2000]{lde2000} Londish, D., Boyle, B. J. \& 
Schade, D. J. 2000, \mnras, 318, 411.

\bibitem[Magorrian \etal\ 1998]{mag98} Magorrian, J., et al. 1998, \apj, 115, 2285.

\bibitem[Malkan \&  Sargent 1982]{ms82} Malkan, M. A. \&  Sargent, W. L. W. 1982, \apj, 254, 22.


\bibitem[Malkan 1988]{malkan88} Malkan, M. A. 1988, Adv. Space Res., 8, 49.

\bibitem[Malkan 1991]{malkan91} Malkan, M. A. 1991, in ``Structure and Emission Properties of Accretion Disks'', Proc. of the Sixth IAP Astrophysics Meeting/IAU Colloquium No. 129, eds. C. Bertout, S. Collin-Souffrin, J. P. Lasota \&  J. Tran Thanh Van, (Gif sur Yvette Cedex, France: Editions Fronti\`{e}res), 165.

\bibitem[Maoz \etal\ 1992]{maoz92} Maoz, D., Bahcall, J. N., Schneider, D. P., Doxsey, R., Bahcall,
N. A., Filippenko, A. V., Goss, W. M., Lahav, O. \&  Yanny, B. 1992, \apj, 386, L1.



\bibitem[Marshall 1985]{marshall85} Marshall, H. L. 1985, \apj, 299, 109. 

\bibitem[Massey \etal\ 1988]{massey88} Massey, P., Strobel, K., Barnes, J. V. \&  Anderson, E. 1988, \apj, 328, 315.

\bibitem[Massey \&  Gronwall 1990]{massey90} Massey, P., \&  Gronwall, C. 1990, \apj, 358, 344.

\bibitem[Mathur 2000]{mathur00} Mathur, S. 2000, \mnras, 314, L17.

\bibitem[Matt, Fabian \&  Ross 1993]{mfr93} Matt, G., Fabian, A. C. \&  Ross, R. R. 1993, \mnras, 264, 839.

\bibitem[McDowell \etal\ 1991]{incdisk} McDowell, J., Kuhn, O., Elvis, M. \&  Wilkes, B. 1991, 
in ``Structure and Emission Properties of Accretion Disks'', Proc. of the Sixth IAP Astrophysics 
Meeting/IAU Colloquium No. 129, eds. C. Bertout, S. Collin-Souffrin, J. P. Lasota \&  J. Tran Thanh Van, 
(Gif sur Yvette Cedex, France: Editions Fronti\`{e}res), 473.

\bibitem[Murayama \etal\ 1998]{mura0636} Murayama, T., Taniguchi, Y., Evans, A. S., Sanders, D. B., Ohyama, Y., Kawara, K. \&  Arimoto, N. 1998, \aj, 115, 2237. 

\bibitem[Murphy \etal\ 1996]{mlle96} Murphy, E. M., Lockman, F. J., Laor, A. \&   Elvis, M. 1996, \apjs, 105, 369.

\bibitem[Mushotzky \&  Wandel 1989]{mw89} Mushotzky, R. F. \&  Wandel, A. 1989, \apj, 339, 674.

\bibitem[Natali \etal\ 1998]{natali98} Natali, F., Giallongo, E., Cristiani, S. \&  LaFranca, F. 1998, \aj, 115, 397. 

\bibitem[Neugebauer \etal\ 1987]{n87} Neugebauer, G., Green, R. F., Matthews, K., Schmidt, M., Soifer, B. T. \&  Bennett, J. 1987, \apjs, 63, 615.

\bibitem[O'Brien, Gondhalekar \&  Wilson 1988]{obrien88} O'Brien, P. T., Gondhalekar, P. M. \&  Wilson, R. 1988 \mnras, 233, 801.

\bibitem[Oke 1974]{oke74} Oke, J. B. 1974, \apjs, 27, 210.

\bibitem[Oke, Shields \&  Korycansky 1984]{oke84} Oke, J. B., Shields, G. A. \&  Korycansky, D. G. 1984, \apj, 277, 64.

\bibitem[Peebles 1993]{peebles}  Peebles, P. J. E. 1993, Principles of Physical Cosmology, (Princeton: Princeton University Press).

\bibitem[Pei, Fall \&  Bechtold 1991]{pfb91}  Pei, Y. C., Fall, S. M. \&  Bechtold, J. B. 1991, \apj, 378, 6.

\bibitem[Pei 1995a]{pei_ple95}  Pei, Y. C. 1995a, \apj, 438, 623. 

\bibitem[Pei 1995b]{pei_gl95}  Pei, Y. C. 1995b, \apj, 440, 485. 

\bibitem[Pr\'{e}vot \etal\ 1984]{prevot} Pr\'{e}vot, M. L., Lequeux, J., Maurice, E., Pr\'{e}vot, L. \&  Rocca-Volmerange, B. 1984, \aap, 132, 389.

\bibitem[Puchnarewicz \etal\ 1996]{rixosc96} Puchnarewicz, E. M., et al. 1996, \mnras, 315, 1.

\bibitem[Reeves \etal\ 1997]{reeves97} Reeves, J. N., Turner, M. J. L., Ohashi, T. \&  Kii, T. 1997, \mnras, 292, 468.

\bibitem[Richstone \etal\ 1998]{richstone98} Richstone, D., et al. 1998, \nat, 395, A14.

\bibitem[Richstone \& Schmidt 1980]{rs80} Richstone, D. O. \& Schmidt, M. 1980, \apj, 235, 361.

\bibitem[Rieke 1984]{irphot} Rieke, G. H. 1984, in MMTO Visiting Astronomer Information, Multiple Mirror Telescope Observatory Technical Report No. 13.

\bibitem[Rieke \&  Lebofsky 1985]{rlext85} Rieke, G. H. \&  Lebofsky, M. J. 1985, \apj, 288, 618.

\bibitem[Rieke, Lebofsky \&  Low 1985]{rll85} Rieke, G. H., Lebofsky, M. J. \&  Low, F. J.  1985, , \aj, 90, 900.


\bibitem[Rowan-Robinson 1995]{rr95} Rowan-Robinson, M. 1995, \mnras, 272, 737.

\bibitem[Salucci \etal\ 1999]{salucci99} Salucci, P., Szuszkiewicz, E., Monaco, P. \&  Danese, L. 1999, \mnras, 307, 637.

\bibitem[Savage \&  Mathis 1979]{smred79} Savage, B. D. \&  Mathis, J. S. 1979, \araa, 17, 73.

\bibitem[Schmidt 1968]{schmidt68} Schmidt, M. 1968, \apj, 151, 393.

\bibitem[Schmidt \&  Green 1983]{sg83} Schmidt, M. \&  Green, R. F. 1983, \apj, 269, 352.

\bibitem[Schmidt, Schneider \&  Gunn 1995]{ssg95} Schmidt, M., Schneider, D. P. \&  Gunn, J. E. 1995, \aj, 110, 68.

\bibitem[Schneider et al. 1992]{accpos} Schneider, D. P., et al. 1992, \pasp, 104, 678.

\bibitem[Shields 1979]{shields79} Shields, G. A. 1979, \nat, 272, 706.

\bibitem[Siemiginowska \&  Elvis 1997]{se97} Siemiginowska, A. \&  Elvis, M. 1997, \apj, 482, L9.


\bibitem[Sincell \&  Krolik 1997]{sk97} Sincell, M. W. \&  Krolik, J. H. 1997, \apj, 476, 605.

\bibitem[Stark \etal\ 1992]{stark92} Stark, A. A., Gammie, C. F., Wilson, R. W., Bally, J., Linke, R. A., Heiles, C. \&  Hurwitz, M. 1992, \apjs, 79, 77.

\bibitem[Steidel 1990]{ste90} Steidel, C. C. 1990, \apjs, 72, 1.

\bibitem[Sun \&  Malkan 1989]{sm89} Sun, W.-H. \&  Malkan, M. A. 1989, \apj, 346, 68.

\bibitem[Thompson, Hill \&  Elston 1999] {the99} Thompson, K. L., Hill, G. J. \&  Elston, R. 1999, ApJ, 515, 487.

\bibitem[Tokunaga 1984]{tokunaga84} Tokunaga, A. T. 1984, \aj, 89, 172.

\bibitem[Tytler \&  Fan 1992]{tf92}  Tytler, D. \&  Fan, X.-M. 1992, \apjs, 79, 1.

\bibitem[Tytler 1999]{tytler99}  Tytler, D. 1999, in ASP Conf. Ser. 162, Quasars and Cosmology, ed. G. Ferland
and J. Baldwin (San Francisco: ASP), 449.

\bibitem[V\'{e}ron-Cetty \&  V\'{e}ron 1993]{vcv93} V\'{e}ron-Cetty, M.-P. \&  V\'{e}ron, P. 1993, ``A Catalogue of Quasars and Active Nuclei (6th edition)'', ESO Scientific Report, No. 13.

\bibitem[Wandel 1987]{wandel87} Wandel, A. 1987, \apj, 316, L55.

\bibitem[Wang, Brinkmann \& Bergeron 1996]{wbb96} Wang, T., Brinkmann, W. \& Bergeron, J. 1996, \aap, 309, 81.

\bibitem[Warren, Hewett \&  Osmer 1994]{war94} Warren, S. J., Hewett, P. C. \& Osmer, P. J., 1994, \apj, 421, 412.

\bibitem[Webb \etal\ 1988]{q0000} Webb, J. K., Parnell, H. C., Carswell, R. F., McMahon, R. G., Irwin, M. J., Hazard, C., Ferlet, R. \&  Vidae-Madjar, A. 1988, The Messenger, 51, 15.

\bibitem[Wilkes \&  Elvis 1987]{we87} Wilkes, B. J. \&  Elvis, M. 1987, \apj, 323, 243.

\bibitem[Wilkes \&  McDowell 1995]{tiger} Wilkes, B. J. \&  McDowell, J. C. 1995, in ASP Conf.
Ser. 61, Astronomical Data Analysis and Software Systems III, ed. D. Crabtree, R. Hanisch and 
J. Barnes (San Francisco: ASP), 423.

\bibitem[Wilkes et al. 1999]{wlines} Wilkes, B. J., Kuraszkiewicz, J., Green, P. J., Mathur, S. and McDowell, 
J. C. 1999, \apj, 513, 76.

\bibitem[Williams \etal\ 1993]{williams93} Williams, D. M., Thompson, C. L., Rieke, G. H. \&  Montgomery, E. F. 1993, \procspie, 1308, 482.

\bibitem[Williger \etal\ 1989]{gwilliger89} Williger, G. M., Carswell, R. M., Webb, J. K., Boksenberg, A. \&  Smith, M. G. 1989, \mnras, 237, 635.

\bibitem[Willner \etal\ 1985]{will85} Willner, S. P., Elvis, M., Fabbiano, G., Lawrence, A. \&  Ward, M. J. 1985, \apj, 299, 443.

\bibitem[Wills, Netzer \&  Wills 1985]{wnw85} Wills, B. J., Netzer, H. \&  Wills, D. 1985, \apj, 288, 94.

\bibitem[Wisotzki \etal\ 2000]{wisotzki00} Wisotzki, L., Christlieb, N., Bade, N., Beckmann, V., K\"ohler, T., Vanelle, C. \&  Reimers, D. 2000, \aap, 358, 77.

\bibitem[Wisotzki 2000]{wisotzki00} Wisotzki, L. 2000, \aap, 288, 94.

\bibitem[Zheng \&  Malkan 1993]{zm93} Zheng, W. \&  Malkan, M. A. 1993, \apj, 415, 517.

\bibitem[Zheng \etal\ 1997]{hstcomp} Zheng, W., Kriss, G. A., Telfer, R. C., Grimes, J. P. \&  Davidsen, A. F. 1997, \apj, 475, 469.

\end{thebibliography}
